\documentclass[twocolumn]{aastex63}
\usepackage{bm}
\usepackage{amsmath,amssymb}
\usepackage{booktabs}
\usepackage{graphicx}
\usepackage{comment}
\usepackage{diagbox}
\usepackage{tabularx}
\usepackage{threeparttable}
\usepackage{xspace}
\usepackage{placeins}   
\usepackage{isotope}
\usepackage{layouts}
\def\real{{\mathbb R}}

\def\bh{{\mathbf h}}
\def\bx{{\mathbf x}}
\def\by{{\mathbf y}}
\def\bz{{\mathbf z}}

\def\bX{{\mathbf X}}
\def\bY{{\mathbf Y}}
\def\bZ{{\mathbf Z}}

\def\ltsima{$\; \buildrel < \over \sim \;$}
\def\simlt{\lower.5ex\hbox{\ltsima}}
\def\gtsima{$\; \buildrel > \over \sim \;$}
\def\simgt{\lower.5ex\hbox{\gtsima}}

\usepackage{color}

\submitjournal{AAS Journals}

\shorttitle{The \posydon{} binary population synthesis code}
\shortauthors{Fragos et al.}

\newcommand{\Teff}{\ifmmode {T_{\rm eff}}\else${T_{\rm eff}}$\fi}
\newcommand{\Msun}{\ensuremath{M_\odot}}
\newcommand{\Renv}{\ensuremath{R_{\rm{conv.reg.}}}\xspace}
\newcommand{\Denv}{\ensuremath{D_{\rm{conv.reg.}}}\xspace}
\newcommand{\Rsun}{\ensuremath{R_\odot}}
\newcommand{\Lsun}{\ensuremath{L_\odot}}
\newcommand{\posydon}{\texttt{POSYDON}\xspace}

\newcommand{\mesa}{\texttt{MESA}\xspace}

\newcommand{\change}[2]{\textcolor{red!50!gray}{\sout{#1}} \textbf{#2}}

\definecolor{turquoise}{HTML}{249494}
\definecolor{juanga}{rgb}{0.56, 0.84, 0.60}
\definecolor{forestgreen}{rgb}{0.13, 0.55, 0.13}

\begin{document}

\title{{\tt POSYDON}: A General-Purpose Population Synthesis Code with Detailed Binary-Evolution Simulations}

\correspondingauthor{T.\ Fragos}
\email{Anastasios.Fragkos@unige.ch}

\author[0000-0003-1474-1523]{Tassos\,Fragos}
\affiliation{Département d’Astronomie, Université de Genève, Chemin Pegasi 51, CH-1290 Versoix, Switzerland}

\author[0000-0001-5261-3923]{Jeff\, J.\,Andrews}
\affiliation{Center for Interdisciplinary Exploration and Research in Astrophysics (CIERA), Northwestern University, 2145 Sheridan Road, Evanston, IL 60208, USA}
\affiliation{Department of Physics, University of Florida, 2001 Museum Rd, Gainesville, FL 32611, USA}

\author[0000-0002-3439-0321]{Simone\,S.\,Bavera}
\affiliation{Département d’Astronomie, Université de Genève, Chemin Pegasi 51, CH-1290 Versoix, Switzerland}

\author[0000-0003-3870-7215]{Christopher\,P.L.\,Berry}
\affiliation{Center for Interdisciplinary Exploration and Research in Astrophysics (CIERA), Northwestern University, 2145 Sheridan Road,
Evanston, IL 60208, USA}
\affiliation{ Institute for Gravitational Research, University of Glasgow, Kelvin Building, University Avenue, Glasgow, G12 8QQ, Scotland}

\author[0000-0002-0403-4211]{Scott\,Coughlin}
\affiliation{Center for Interdisciplinary Exploration and Research in Astrophysics (CIERA), Northwestern University, 2145 Sheridan Road,
Evanston, IL 60208, USA}

\author[0000-0002-4442-5700]{Aaron\,Dotter}
\affiliation{Center for Interdisciplinary Exploration and Research in Astrophysics (CIERA), Northwestern University, 2145 Sheridan Road,
Evanston, IL 60208, USA}

\author[0000-0001-9058-7228]{Prabin\,Giri}
\affiliation{Department of Electrical and Computer Engineering, Iowa State University, 2520 Osborn Dr, Ames, IA 50011, USA}

\author[0000-0001-9236-5469]{Vicky\,Kalogera}
\affiliation{Center for Interdisciplinary Exploration and Research in Astrophysics (CIERA), Northwestern University, 2145 Sheridan Road,
Evanston, IL 60208, USA}
\affiliation{Department of Physics and Astronomy, Northwestern University, 2145 Sheridan Road, Evanston, IL 60208, USA}

\author[0000-0003-4554-0070]{Aggelos\,Katsaggelos}
\affiliation{Center for Interdisciplinary Exploration and Research in Astrophysics (CIERA), Northwestern University, 2145 Sheridan Road,
Evanston, IL 60208, USA}
\affiliation{Electrical and Computer Engineering, Northwestern University, 2145 Sheridan Road, Evanston, IL 60208, USA}

\author[0000-0003-3684-964X]{Konstantinos\,Kovlakas}
\affiliation{Département d’Astronomie, Université de Genève, Chemin Pegasi 51, CH-1290 Versoix, Switzerland}

\author{Shamal\,Lalvani}
\affiliation{Electrical and Computer Engineering, Northwestern University, 2145 Sheridan Road, Evanston, IL 60208, USA}

\author[0000-0003-4260-960X]{Devina\,Misra}
\affiliation{Département d’Astronomie, Université de Genève, Chemin Pegasi 51, CH-1290 Versoix, Switzerland}

\author{Philipp\,M.\,Srivastava}
\affiliation{Center for Interdisciplinary Exploration and Research in Astrophysics (CIERA), Northwestern University, 2145 Sheridan Road,
Evanston, IL 60208, USA}
\affiliation{Electrical and Computer Engineering, Northwestern University, 2145 Sheridan Road, Evanston, IL 60208, USA}

\author[0000-0002-2956-8367]{Ying\,Qin}
\affiliation{Department of Physics, Anhui Normal University, Wuhu, Anhui 241000, China}

\author[0000-0003-4474-6528]{Kyle\,A.\,Rocha}
\affiliation{Center for Interdisciplinary Exploration and Research in Astrophysics (CIERA), Northwestern University, 2145 Sheridan Road,
Evanston, IL 60208, USA}
\affiliation{Department of Physics and Astronomy, Northwestern University, 2145 Sheridan Road, Evanston, IL 60208, USA}

\author[0000-0002-5962-4796]{Jaime\,Rom\'an-Garza}
\affiliation{Universidad de Monterrey, Ave. Morones Prieto 4500 Pte.,  C.P. 66283, San Pedro Garza Garc\'ia, Nuevo Le\'on, M\'exico.}

\author{Juan\,Gabriel\,Serra}
\affiliation{Center for Interdisciplinary Exploration and Research in Astrophysics (CIERA), Northwestern University, 2145 Sheridan Road,
Evanston, IL 60208, USA}
\affiliation{Electrical and Computer Engineering, Northwestern University, 2145 Sheridan Road, Evanston, IL 60208, USA}

\author[0000-0003-0161-8032]{Petter\,Stahle}
\affiliation{Département d’Informatique, Université de Genève, Route de Drize 7, CH-1227 Carouge, Switzerland}

\author[0000-0001-9037-6180]{Meng\,Sun}
\affiliation{Center for Interdisciplinary Exploration and Research in Astrophysics (CIERA), Northwestern University, 2145 Sheridan Road,
Evanston, IL 60208, USA}

\author[0000-0002-5169-1425]{Xu\,Teng}
\affiliation{Department of Electrical and Computer Engineering, Iowa State University, 2520 Osborn Dr, Ames, IA 50011, USA}

\author[0000-0002-8839-6278]{Goce\,Trajcevski}
\affiliation{Department of Electrical and Computer Engineering, Iowa State University, 2520 Osborn Dr, Ames, IA 50011, USA}

\author[0000-0001-6252-3606]{Nam\,Hai\,Tran}
\affiliation{DARK, Niels Bohr Institute, University of Copenhagen,
Jagtvej 128, 2200 Copenhagen, Denmark}

\author[0000-0002-0031-3029]{Zepei\,Xing}
\affiliation{Département d’Astronomie, Université de Genève, Chemin Pegasi 51, CH-1290 Versoix, Switzerland}

\author[0000-0002-7464-498X]{Emmanouil\,Zapartas}
\affiliation{Département d’Astronomie, Université de Genève, Chemin Pegasi 51, CH-1290 Versoix, Switzerland}

\author[0000-0002-0147-0835]{Michael Zevin}
\affiliation{Kavli Institute for Cosmological Physics, The University of Chicago, 5640 South Ellis Avenue, Chicago, Illinois 60637, USA}
\affiliation{Enrico Fermi Institute, The University of Chicago, 933 East 56th Street, Chicago, Illinois 60637, USA}

\begin{abstract}

Most massive stars are members of a binary or a higher-order stellar systems, where the presence of a binary companion can decisively alter their evolution via binary interactions. 
Interacting binaries are also important astrophysical laboratories for the study of compact objects. 
Binary population synthesis studies have been used extensively over the last two decades to interpret observations of compact-object binaries and to decipher the physical processes that lead to their formation. 
Here, we present \posydon{}, a novel, publicly available, binary population synthesis code that incorporates full stellar-structure and binary-evolution modeling, using the \mesa{} code, throughout the whole evolution of the binaries. 
The use of \posydon{} enables the self-consistent treatment of physical processes in stellar and binary evolution, including: realistic mass-transfer calculations and assessment of stability, internal angular-momentum transport and tides, stellar core sizes, mass-transfer rates and orbital periods.  
This paper describes the detailed methodology and implementation of \posydon{}, including the assumed physics of stellar- and binary-evolution, the extensive grids of detailed single- and binary-star models, the post-processing, classification and interpolation methods we developed for use with the grids, and the treatment of evolutionary phases that are not based on pre-calculated grids. The first version of \posydon targets binaries with massive primary stars (potential progenitors of neutron stars or black holes) at solar metallicity.

\end{abstract}



\section{Introduction \label{sec:intro}}

Throughout their lives, stars affect their surroundings via the immense energy radiated across the electromagnetic spectrum \citep[e.g.,][]{2013ARA&A..51..393C,2022arXiv220201413E} and the nuclear-processed material emitted as a stellar wind \citep[e.g.,][]{2000ARA&A..38..613K, 2014ARA&A..52..487S}. 
The deaths of massive ($\gtrsim 8\,\rm M_{\odot}$) stars, even more than their lives, transform their environments as their cores run out of nuclear fuel and collapse to form neutron stars (NSs) and black holes (BHs). 
The formation of these compact objects (COs) is often accompanied by a supernova (SN) or a $\gamma$-ray burst that release more energy in $10\,\mathrm{s}$ than our Sun in $10^{10}\,\mathrm{yr}$ \citep[e.g.,][]{2012ARNPS..62..407J, 2021Natur.589...29B, 2006ARA&A..44..507W}. 
These explosive events enrich their environments with heavier elements while also regulating any ongoing star-formation \citep[e.g.,][]{2013ARA&A..51..457N,2011MNRAS.417..950H,2012MNRAS.421.3522H}.

It is now established that most massive stars are members of a binary or a higher-order stellar system \citep[][]{2013A&A...550A.107S, 2017ApJS..230...15M}.
More often than not, the presence of a binary companion decisively alters the evolution and final fate of both binary components via binary-interaction processes such as tidal dissipation, mass-transfer phases, and stellar mergers \citep{2012Sci...337..444S, 2017PASA...34....1D}. 
Furthermore, interacting binaries are arguably some of the most important astrophysical laboratories available for the study of COs.  
Accretion of matter from a binary companion gives rise to X-ray emission, bringing the system to the X-ray binary (XRB) phase \citep{1991PhR...203....1B, 1992ApJ...391..246P}, while gravitational waves (GWs) enable us to witness the last moments of the lives of coalescing binary COs \citep{2016PhRvL.116f1102A,2021arXiv211103606T}. 

Over the last two decades, multi-wavelength surveys of the Milky Way and its neighborhood, as well as numerous nearby and more distant galaxies, have amassed large datasets of binary stellar systems. 
These datasets range from targeted, high-resolution observations of galaxies in the local Universe \citep[e.g., the PHAT and LEGUS surveys;][]{2012ApJS..200...18D, 2015AJ....149...51C}, to serendipitous  \citep[e.g., the Chandra Source Catalogue 2.0;][]{2019HEAD...1711401E}, all-sky  \citep[e.g., {\it Gaia};][]{2018A&A...616A...1G} and  transient surveys \citep[e.g., Pan-STARRS, Zwicky Transient Facility;][]{2002SPIE.4836..154K, 2019PASP..131g8001G}. 
In addition to electromagnetic surveys, there is also the global GW observatory network of LIGO \citep{2015CQGra..32g4001L}, Virgo \citep{2015CQGra..32b4001A} and KAGRA \citep{2019NatAs...3...35K}, which have detected nearly a hundred binary CO mergers \citep{2021arXiv211103606T}. 
Combined, these surveys are revolutionizing our view of binary stellar systems, including CO binaries, and their environments.

Aspects of the astrophysics of all these different types of stellar binaries can be obtained from observations and modeling of present-day properties of individual, well-studied systems. 
However, more comprehensive insight requires understanding the statistical properties of their entire populations.
For these studies, binary population synthesis (BPS) modeling is often employed. 
BPS modeling first generates initial binary populations, whose properties are randomly sampled from probability distributions that can be observationally constrained. 
Then, this initial population is evolved with a computationally efficient simulation tool using our best understanding of the physics dictating binary star interactions, to produce observable properties of the target population. 
If the number of binaries evolved is large enough to provide a statistically significant description of a population of interest, then BPS can provide valuable insights about the expected rate and distribution of the target population's properties, the different evolutionary pathways that lead to formation of these systems, and the effect that different physical processes have on their evolution. 

Over the last two decades, many general purpose BPS codes have been developed, e.g., {\tt binary\_c} \citep{2004MNRAS.350..407I,2006AA...460..565I,2009AA...508.1359I}, {\tt BPASS} \citep{2017PASA...34...58E},  the {\tt Brussels} code \citep{1998NewA....3..443V, 1998A&ARv...9...63V}, {\tt BSE} \citep{2002MNRAS.329..897H}, {\tt ComBinE} \citep{2018MNRAS.481.1908K}, {\tt COMPAS} \citep{2017NatCo...814906S,2021arXiv210910352T}, {\tt COSMIC} \citep{2020ApJ...898...71B}, {\tt MOBSE} \citep{2018MNRAS.474.2959G}, the {\tt Scenario Machine} \citep{1996smbs.book.....L,2009ARep...53..915L}, {\tt SEVN} \citep{2015MNRAS.451.4086S}, {\tt SeBa} \citep{1996AA...309..179P, 2012AA...546A..70T}, {\tt StarTrack} \citep{2002ApJ...572..407B, 2008ApJS..174..223B}, and {\tt TRES} \citep{2016ComAC...3....6T}. These have been used in studies of a wide variety of binary populations. 
A hard requirement for BPS is computational efficiency, as for most studies one would need to model the evolution of many millions of binaries in a reasonable computational time.

BPS codes stand in stark contrast to detailed stellar-structure and binary-evolution codes, e.g., {\tt BEC} \citep{2000ApJ...528..368H, 2000ApJ...544.1016H}, {\tt BINSTAR} \citep{2013A&A...550A.100S}, the {\tt Cambridge STARS} code \citep[][]{1971MNRAS.151..351E, 1995MNRAS.274..964P,2004MNRAS.353...87E,2009MNRAS.396.1699S}, {\tt MESA} \citep{2015ApJS..220...15P}, and the {\tt TWIN} code \citep{2001ApJ...552..664N, 2002ApJ...575..461E}, which self-consistently solve the stellar structure equations of a binary's component stars along with the orbital evolution. Many studies have used detailed binary-evolution calculations \citep[e.g.,][]{2001ApJ...552..664N, 2002ApJ...565.1107P, 2007A&A...467.1181D, 2017A&A...604A..55M,  2018A&A...616A..28Q, 2019ApJ...870L..18Q,2020A&A...638A..39L,2020A&A...642A.174M,2020A&A...637A...6L,2021A&A...656A..58L} to generate grids of models, varying the masses of the two stars and the binary's orbital period. However, in all those cases, the grids of detailed binary tracks either cover a limited part of the initial parameter space, or focus on a specific evolutionary phase. This limitation is principally caused by the computational demands of detailed grids; each simulation typically requires $\sim 10$--$100$ CPU hours for the modeling of a single system \citep[e.g.,][]{2019ApJS..243...10P}.

As a result of this computational expense, a common thread among the vast majority of current BPS codes is that they approximate each star's evolution, employing either fitting formulae \citep[e.g., {\tt SSE};][]{2000MNRAS.315..543H} or look-up tables \citep[e.g., {\tt COMBINE};][]{2018MNRAS.481.1908K} for the properties of \emph{single stars}, based on grids of pre-calculated detailed, single-star models.
Then, the effects of binary interactions (e.g., Roche-lobe overflow or tides) are modeled using approximate prescriptions and parametrizations.
This modeling approach is often called rapid or parametric BPS; throughout the remainder of this work, we choose to use the term \emph{parametric BPS} (pBPS) modeling, to make a distinction between computational efficiency and modeling accuracy. 
A notable exception among BPS codes is {\tt BPASS} \citep{2017PASA...34...58E}, which uses extensive grids of detailed binary evolution models computed with a custom version of the {\tt Cambridge STARS} binary evolution code \citep[][]{2009MNRAS.396.1699S}. In the grids of binary-star models employed in {\tt BPASS}, both the primary and the secondary stars are followed in detail, but only one at a time (for computational-cost reasons). 
During the primary's evolution, the properties of the secondary star are approximated by formulae based on single-star models \citep[][]{2000MNRAS.315..543H}. Subsequently, once the modeling of the primary's evolution is completed, the secondary star's evolution is re-computed, accounting for mass-transfer and rejuvenation effects.

Studies using pBPS techniques have allowed us to make advances in our understanding of the formation pathways leading to different types of binary systems, and to interpret observations of binary populations. 
Such examples are studies on white-dwarf binaries \citep[e.g.,][]{2001A&A...368..939N, 2009ApJ...699.2026R, 2010ApJ...717.1006R,  2012A&A...546A..70T, 2018ApJ...854L...1B,2020A&A...638A.153K}, XRBs \citep[e.g.,][]{2000A&A...358..462V, 2004ApJ...601L.147B, 2008ApJ...683..346F, 2013ApJ...764...41F, 2013ApJ...776L..31F, 2012ApJ...749..130L, 2013ApJ...774..136T,2013ApJ...766...19T,  2014MNRAS.437.1187Z, 2014MNRAS.442.1980Z,   2015A&A...579A..33V, 2019IAUS..346..332A,2019ApJ...875...53W,2020ApJ...898..143S}, the Galactic population of double NSs \citep[DNSs; e.g.,][]{2011MNRAS.413..461O, 2015ApJ...801...32A, 2017AcA....67...37C,2018MNRAS.481.4009V,2020MNRAS.494.1587C}, GW sources observable by ground-based observatories \citep[e.g.,][]{2012ApJ...759...52D,2013ApJ...779...72D,2015ApJ...806..263D,2014A&A...564A.134M,2016A&A...589A..64M,2016Natur.534..512B,2018A&A...619A..77K,2018MNRAS.474.2959G,2018MNRAS.479.4391M,2019MNRAS.490.3740N,2019MNRAS.485..889S,2020MNRAS.498.3946K,2020ApJ...898...71B,2020ApJ...899L...1Z,2021MNRAS.508.5028B} and SNe in binary systems \citep[e.g.,][]{2003NewA....8..817D, 2013A&A...552A.105V, 2014A&A...563A..83C, 2017A&A...601A..29Z, 2019A&A...631A...5Z, 2021A&A...645A...6Z}. 

However, the implicit assumption in pPBS codes that the binary components have properties identical to single stars of the same mass in thermal equilibrium (e.g., abundance profiles, core sizes, mass-radius relations and response to mass-loss), as well as the lack of information about the star's internal structure at different critical evolutionary phases (e.g., the onset of a dynamically unstable mass-transfer, the end of stable and unstable mass-transfer phases or the core-collapse), may introduce systematic uncertainties and inaccuracies. 
Current pBPS codes therefore rely on approximate prescriptions for modeling binary interactions and  difficult-to-calibrate additional model parameters. 
These complications could be avoided by instead employing detailed stellar-structure and binary-evolution simulations (hereafter \emph{detailed models}). 
Focusing on aspects that are relevant to the formation of CO binaries, detailed models \textit{(i)} allow for a self-consistent estimation of the mass-transfer rate, especially during thermal-timescale mass-transfer phases, and therefore an accurate assessment of mass-transfer stability; \textit{(ii)} allow for a more accurate description of the type and properties of the formed CO as well as any potential associated transient events since the internal structure of pre-core-collapse stars is known, \textit{(iii)} account for the transport of angular momentum between and within the binary components, including its back-reaction on the structure and evolution of each star (e.g., rotational mixing), and \textit{(iv)} allow for the self-consistent modeling of the end of a mass transfer phase (e.g., accounting for a potential partial stripping of the envelope).

In this work, we build upon the combined experience gained from the large body of BPS studies to date, to create \posydon{} (POpulation SYnthesis with Detailed binary-evolution simulatiONs), a general-purpose code that can generate entire populations of binaries, underpinned by detailed, self-consistent models of stellar binaries.\footnote{\posydon{} \change{will become}{is} publicly available at  \href{https://posydon.org}{https://posydon.org} \change{upon publication of the paper}{}.} 
With \posydon{} we aim to address many of the caveats of pBPS codes, while at the same time maintaining much of their flexibility. 
In its first release (v1.0), \posydon{} is limited to stars of solar metallicity, and binaries where the primary star is massive enough to form a BH or a NS. 
Future releases, which are already in development, will lift these limitations. 
In Section~\ref{sec:overview}, we introduce \posydon{} and the approach it takes to modeling binary populations.
In Sections~\ref{sec:single_physics} and \ref{sec:binary_physics} we provide the physics adopted for our detailed models of single and binary stars, respectively. 
We describe the pre-calculated grids of single and binary stellar evolution models in Section~\ref{sec:grids}, the way they are post-processed in Section~\ref{sec:postprocessing}, and our classification and interpolation methods for their optimal use in Section~\ref{sec:machine_learning}. In Section~\ref{sec:other_physics} we detail our treatment of evolutionary phases which are not based on pre-calculated grids, such as the core-collapse and the common-envelope (CE) phase, while in Section~\ref{sec:flow} we describe how all the aforementioned pieces come together to model the entire evolution of a binary. 
In Section~\ref{sec:populations} we outline our assumptions and methods in modeling populations of binary systems and present some example results.   
We conclude in Section~\ref{sec:future}, where we present an outlook of future development directions of the \posydon{} code. 
The definitions of all the symbols used throughout this paper can be found in Table~\ref{tab:varialbe_defs}.

\startlongtable
\begin{center}\label{tab:varialbe_defs}
\begin{deluxetable}{p{1.2cm} p{4.5cm} p{1cm}} 
\tabletypesize{\footnotesize}
\tablecaption{List of variables used throughout the paper.}
\tablehead{\colhead{Name} & \colhead{Description} & \colhead{First appears}}
\startdata
$a$ & Orbital separation & \ref{sec:Tides}\\
$a_i$ & Orbital separation before orbital kick & \ref{sec:SN_kicks} \\
$a_f$ & Orbital separation after orbital kick & \ref{sec:SN_kicks} \\
$a_\mathrm{pre,CE}$ & Orbital separation pre common envelope & \ref{sec:common_env} \\
$a_\mathrm{post,CE}$ & Orbital separation post common envelope & \ref{sec:common_env}\\
$\dot{a}$ &  Rate of change of orbital separation & \ref{sec:method:ODE} \\
$\dot{a}_\mathrm{wind}$ & Rate of change of orbital separation due to wind mass loss & \ref{sec:method:ODE} \\
$\dot{a}_\mathrm{tides}$ & Rate of change of orbital separation due to tides & \ref{sec:method:ODE} \\
$\dot{a}_\mathrm{GR}$ & Rate of change of orbital separation due to gravitational-wave radiation & \ref{sec:method:ODE} \\
$a_{\rm spin}$ & Non-dimensional spin & \ref{sec:MT_degenerate}\\
$c$ & Speed of light & \ref{sec:MT_non_degenerate}\\
$D_{\rm conv.reg.}$ & Depth of a convective region   & \ref{sec:sanitization}\\
$e$ & Orbital eccentricity &\ref{sec:method:ODE}\\
$e_f$ & Orbital eccentricity after orbital kick & \ref{sec:SN_kicks} \\
$\dot{e}$ &  Rate of change of orbital eccentricity & \ref{sec:method:ODE} \\
$\dot{e}_\mathrm{tides}$ &Rate of change of orbital eccentricity due to tides & \ref{sec:method:ODE} \\
$\dot{e}_\mathrm{GR}$ & Rate of change of orbital eccentricity due to gravitational-wave radiation & \ref{sec:method:ODE} \\
$E$ & Eccentric anomaly & \ref{sec:SN_kicks}\\
$E_{\rm 2}$ & Second-order tidal torque coefficient & \ref{sec:Tides}\\
$f_{\rm conv}$ & Dimensionless factor accounting for slow convective shells that cannot contribute to the tidal viscosity within an orbital timescale & \ref{sec:Tides}\\
$f_\mathrm{fb}$ & Fallback mass fraction & \ref{sec:rembar} \\
$f_{\rm ov}$ & Convective exponential overshooting parameter & \ref{sec:mixing}\\
$g$ & Local gravitational acceleration & \ref{sec:surface_boundary}\\
$G$ & Gravitational constant & \ref{sec:winds}\\
$I$ & Moment of inertia & \ref{sec:Tides}\\
$\dot{I}$ & Moment of inertia rate of change & \ref{sec:method:ODE}\\
$j$ & Specific angular momentum  & \ref{sec:CO-HeMS}\\
$j_\mathrm{ISCO}$ & Specific angular momentum of the ISCO  & \ref{sec:birth_spin}\\
$J$ & Angular momentum  & \ref{sec:CO-HeMS}\\
$J_\mathrm{shell}$ & Stellar shell's angular momentum  & \ref{sec:birth_spin}\\
$J_\mathrm{direct}$ & Stellar shell's angular momentum of directly collapsing material & \ref{sec:birth_spin}\\
$J_\mathrm{disk}$ & Stellar shell's angular momentum of disk forming material & \ref{sec:birth_spin}\\
$J_\mathrm{BH}$ & Black hole angular momentum & \ref{sec:birth_spin}\\
$k$ & Apsidal motion constant  & \ref{sec:Tides}\\
$L$ & Star's luminosity & \ref{sec:winds}\\
$L_{\rm Edd}$ & Eddington luminosity & \ref{sec:winds}\\
$L_{2}$ & Second lagrange point & \ref{sec:common_env}\\
$M_{1}$ & Mass of the initially more massive star  & \ref{sec:HMS-HMS}\\
$M_{2}$ & Mass of the initially less massive star  & \ref{sec:HMS-HMS}\\
$M_{\rm acc}$ & Mass of the accretor & \ref{sec:MT_non_degenerate}\\
$M_{\rm don}$ & Mass of the donor & \ref{sec:common_env} \\
$M_{\rm CO}$ & Mass of the compact object  & \ref{sec:CO-HMS_RLO}\\
$M_{\rm{C/O\mbox{-}core}}$ & Mass of the C/O core & \ref{sec:sanitization}\\
$M_{\rm conv.reg.}$ & Mass of a convective region & \ref{sec:Tides}\\
$M_{\rm comp}$ & Mass of the binary companion star  & \ref{sec:Tides}\\
$M_{\rm disk}$ & Mass of accretion disk & \ref{sec:birth_spin}\\
$M_{\rm{env}}$ & Mass of the stellar envelope & \ref{sec:sanitization}\\
$M_{\rm{grav}}$ & Remnant's gravitational mass & \ref{sec:grab_mass} \\
$M_{\rm{He\mbox{-}core}}$ & Mass of the He core & \ref{sec:sanitization}\\
$M_{\rm{rembar}}$ & Remnant's baryonic mass & \ref{sec:sanitization}\\
$M^\mathrm{max}_{\rm{NS}}$ & Maximum neutron star mass & \ref{sec:grab_mass}\\
$\dot{M}_{\rm Edd}$ & Mass-accretion rate corresponding to the Eddington limit & \ref{sec:MT_non_degenerate}\\
$\dot{M}_{\rm w}$ & Wind mass-loss rate & \ref{sec:winds}\\
$M_\mathrm{tot}^\mathrm{i}$ & Binary stellar mass before core collapse & \ref{sec:SN_kicks} \\
$M_\mathrm{tot}^\mathrm{f}$ & Binary stellar mass after core collapse & \ref{sec:SN_kicks} \\
$m_{\rm{shell}}$ & Stellar shell's mass & \ref{sec:birth_spin}\\
$P$ & Pressure & \ref{sec:surface_boundary}\\
$P_{\rm orb}$ & Orbital period & \ref{sec:HMS-HMS}\\
$q$ & Binary mass ratio & \ref{sec:Tides}\\
$R$ & Stellar radius & \ref{sec:winds}\\
$R_{\rm acc}$ & Radius of the accretor & \ref{sec:MT_non_degenerate}\\
$R_{\rm b,conv.reg.}$ & Radial coordinate of a convective region's bottom boundary & \ref{sec:Tides}\\
$R_{\rm{core}}$ & Radius of the stellar core & \ref{sec:sanitization}\\
$R_{\rm{C/O\mbox{-}core}}$ & Radius of the C/O core & \ref{sec:sanitization}\\
$R_{\rm conv}$ & Radius of the convective core & \ref{sec:Tides}\\
$R_{\rm conv.reg.}$ & Radial coordinate of a convective region's center & \ref{sec:sanitization}\\
$R_{\rm don}$ & Radius of the donor & \ref{sec:common_env} \\
$R_{\rm{He-core}}$ & Radius of the He core & \ref{sec:sanitization}\\
$R_{\rm L}$ & Roche lobe radius & \ref{sec:MT_non_degenerate}\\
$R_{\rm L,acc}$ & Roche lobe radius of the accretor & \ref{sec:MT_non_degenerate}\\
$R_{\rm t,conv.reg.}$ & Radial coordinate of a convective region's top boundary  & \ref{sec:Tides}\\
$r$ & Stellar shell's radius & \ref{sec:birth_spin}\\
$r_i$ & Instantaneous orbital separation before orbital kick & \ref{sec:SN_kicks}\\
$T_{\rm eff}$ & Effective temperature & \ref{sec:surface_boundary}\\
$T$ & Timescale for orbital changes due to tides  & \ref{sec:Tides}\\
$v_k$ & Magnitude of velocity kick & \ref{sec:SN_kicks}\\
$v_r$ & Orbital velocity of the collapsing star & \ref{sec:SN_kicks}\\
$X$ & Hydrogen mass function & \ref{sec:MT_non_degenerate}\\
$X_\mathrm{center}$ & Center hydrogen mass function & \ref{sec:matching} \\
$X_\mathrm{surf}$ & Surface hydrogen mass function & \ref{sec:matching} \\
$Y$ & Helium mass fraction & \ref{sec:microphysics}\\
$Y_\mathrm{center}$ & Center helium mass fraction & \ref{sec:matching} \\
$Y_\mathrm{surf}$ & Surface helium mass fraction & \ref{sec:interpolation_accuracy}\\
$Z$ & Metallicity (mass fraction of elements heavier than $^4$He)   & \ref{sec:microphysics}\\
$\alpha_{\rm CE}$ & Fraction of the orbital energy that contributes to the unbinding of the CE & \ref{sec:common_envelope}\\
$\alpha_{\rm MLT}$ & Convective mixing length parameter & \ref{sec:mixing}\\
$\alpha_{\rm th}$ & Thermohaline mixing parameter & \ref{sec:mixing}\\
$\eta$ & Dimensionless factor denoting the radiative efficiency of the accretion process  & \ref{sec:MT_non_degenerate}\\
$\theta$ & Stellar profile's polar angle & \ref{sec:birth_spin} \\
$\theta_\mathrm{disk}$ & Stellar profile's polar angle of disk formation & \ref{sec:birth_spin} \\
$\kappa$ & Opacity & \ref{sec:surface_boundary}\\
$\lambda_{\rm CE}$ & Parametrization of the CE's binding energy & \ref{sec:common_envelope}\\
$\tau$ & Optical depth & \ref{sec:surface_boundary}\\
$\tau_{\rm sync}$ & Tidal synchronization timescale & \ref{sec:Tides}\\
$\tau_{\rm conv}$ & Convective timescale & \ref{sec:Tides}\\
$\tau_{\rm mb}$ & Magnetic braking torque & \ref{sec:method:ODE}\\
$\nu$ & Reduced mass & \ref{sec:method:ODE} \\
$\sigma$ & Maxwellian distribution dispersion & \ref{sec:SN_kicks}\\
$\sigma_\mathrm{CCSN}$ & Maxwellian distribution dispersion for CCSN kicks& \ref{sec:SN_kicks}\\
$\sigma_\mathrm{ECSN}$ & Maxwellian distribution dispersion for ECSN kicks& \ref{sec:SN_kicks}\\
$\psi$ & Binary orbital inclination with respect to before the kick   & \ref{sec:SN_kicks} \\
$\omega_{\rm s}$ & Surface angular velocity & \ref{sec:winds} \\
$\omega_{\rm s,crit}$ & Critical surface angular velocity & \\
$\Omega_\mathrm{orb}$ & Orbital angular velocity & \ref{sec:method:ODE} \\
$\Omega_{\rm{shell}}$ & Stellar shell's angular velocity & \ref{sec:birth_spin}\\
$\Omega$ & Stellar angular velocity & \ref{sec:method:ODE} \\
$\dot{\Omega}$ &  Stellar angular velocity rate of change & \ref{sec:method:ODE}\\
$\dot{\Omega}_\mathrm{wind}$ & Stellar angular velocity rate of change due to winds & \ref{sec:method:ODE} \\
$\dot{\Omega}_\mathrm{inertia}$ & Stellar angular velocity rate of change due to changes of star's moment of inertia & \ref{sec:method:ODE} \\
$\dot{\Omega}_\mathrm{tides}$ & Stellar angular velocity rate of change due to tides & \ref{sec:method:ODE} \\
$\dot{\Omega}_\mathrm{mb}$ & Stellar angular velocity rate of change due to magnetic breaking& \ref{sec:method:ODE} 
\ref{sec:winds} \\
\enddata
\end{deluxetable}
\end{center}

\section{Overview of the structure of \posydon{} \label{sec:overview}}

At its core, a BPS code requires two elements: a method to generate random binaries at zero-age main sequence (ZAMS) and a mechanism to evolve those binaries. 
The former is described in Section \ref{sec:initialization}. 
Regarding the evolution of each binary, our primary goal with \posydon is to self-consistently evolve the internal structures of the two stars comprising a binary along with the binary's orbit. 
To achieve this goal, we have opted to employ the stellar-structure code Modules for Experiments in Stellar Astrophysics \citep[\mesa{};][]{2011ApJS..192....3P, 2013ApJS..208....4P,2015ApJS..220...15P, 2018ApJS..234...34P,2019ApJS..243...10P}. 
However, calculating the evolutionary track of a binary  using \mesa can take in excess of $100$ CPU hours, a six-orders-of-magnitude increase in computational cost compared to a typical pBPS code such as {\tt COSMIC} \citep{2020ApJ...898...71B}. 
This  means that, even with modern computing resources, we cannot feasibly run more than $\sim10^5$--$10^6$ binaries, far too few to accurately model a Milky Way sized population with $\sim10^{11}$ stellar binaries. 
As a further complication, codes like \mesa{} cannot run individual binaries from start to finish; key physics, including CE phases and supernovae require the code to be stopped and restarted.\footnote{In principle these phases could be run within a stellar-structure code like \mesa{}; for instance, more updated versions of \mesa{} than the one we use can handle the evolution of a binary through a CE \citep{2021A&A...650A.107M}.}

\begin{figure*}
    \centering
    \includegraphics[width=\textwidth]{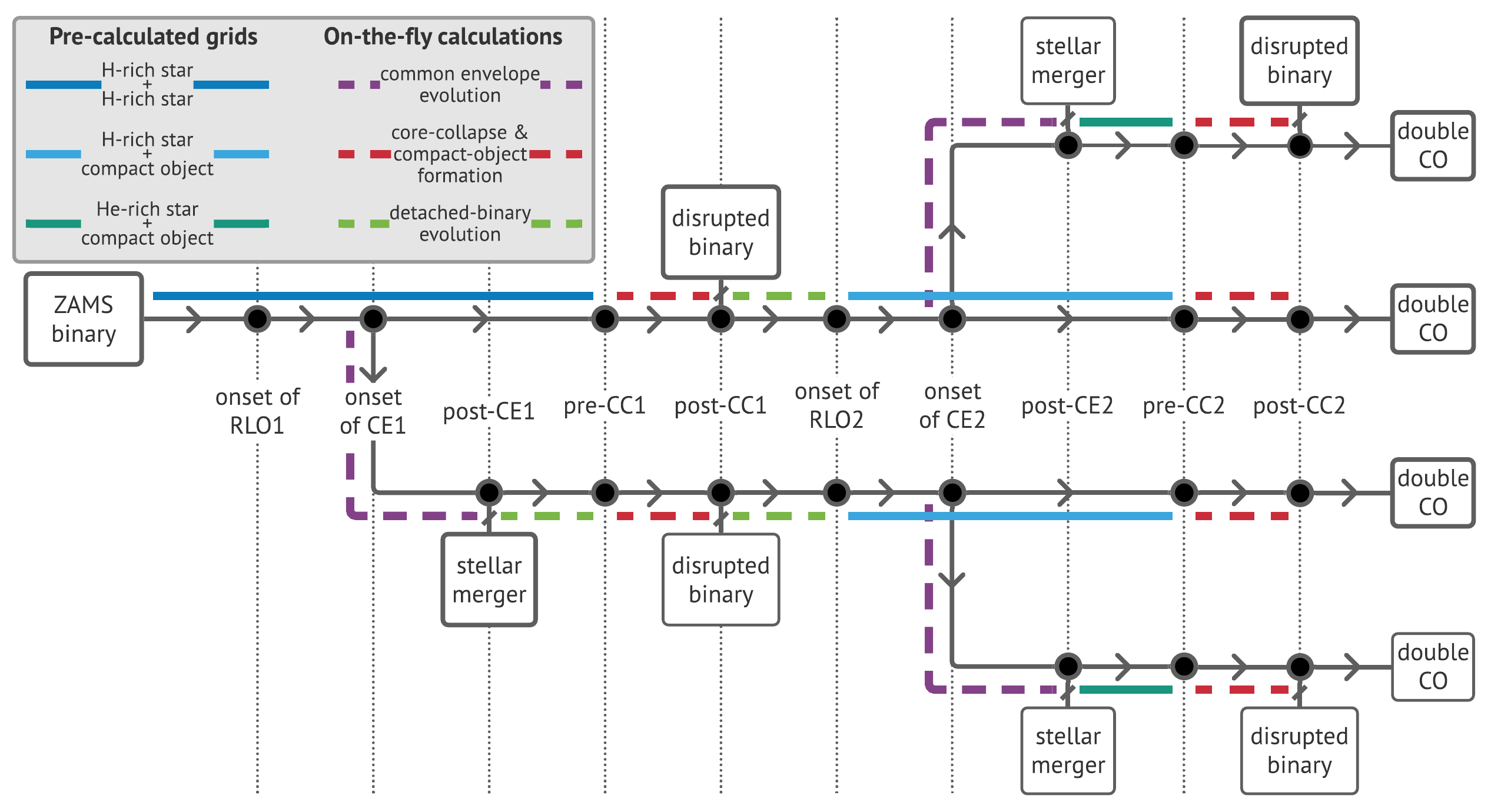}
    \caption{The structure of \posydon{} (v1.0) for modeling the evolution of a binary star. 
    Rectangles represent the initial and possible final outcomes of the evolution, and black circles represent events in the evolution of a binary.
    Colored lines showcase the different evolutionary steps that \posydon{} follows. 
    Evolutionary steps that are based on pre-calculated grids of detailed binary-evolution tracks are designated with solid lines, while those that are based on computations performed on-the-fly for each modeled binary are shown with dashed lines. 
    }
    \label{fig:structure}
\end{figure*}

\posydon{} solves the problems associated with stellar-structure codes by using extensive, pre-calculated grids of single and binary stellar-evolution models, covering the parameter space relevant for the formation of high-mass binary stars, with separate grids being calculated for each phase of binary evolution. In v1.0 our grids contain a combined total of nearly 120,000 separate detailed binary simulations.
To compute these grids, \posydon{} has an infrastructure specifically designed for high-performance computing environments that streamlines the process of producing large grids with consistent physics inputs. 
Using this infrastructure, we have generated five separate grids of single and binary stellar evolution models.
We computed three grids of interacting binary stars initially composed  \textbf{of} two hydrogen (H)-rich ZAMS stars (Section \ref{sec:HMS-HMS}); a CO and a H-rich star at the onset of Roche-lobe overflow (RLO; Section \ref{sec:CO-HMS_RLO}), and a helium (He)-rich ZAMS star with a CO companion (Section \ref{sec:CO-HeMS}). 
We further computed two grids of single H-rich and He-rich stars (Sections \ref{sec:HMS} and \ref{sec:HeMS}, respectively), which we use for the modeling of detached, non-interacting binaries. 
These five grids are then post-processed, so that their data size is reduced (Section \ref{sec:postprocessing}). 
We additionally apply classification and interpolation algorithms on the outputs of these extensive grids (Section \ref{sec:machine_learning}), allowing us to effectively interpolate between \mesa{} simulations to estimate the evolution of any arbitrary binary within some bounded region of the parameter space. 
As a simpler alternative, we also provide functionality to evolve individual binaries using nearest-neighbor matching, and in that case no classification or interpolation methods are required.

The second major component of \posydon{} is the code infrastructure to follow the entire evolution of a binary from start to end. 
To achieve this, we combined the aforementioned grids (and classification and interpolation methods) with physics dictating a binary's evolution through key phases, including core collapse and CO formation (Section~\ref{sec:core_collapse}) and CE (Section~\ref{sec:common_envelope}). 
These latter phases are not modeled based on pre-calculated grids, but rather with on-the-fly calculations. 
Similarly, the evolution of detached, potentially eccentric, post-core-collapse binaries is also modeled with on-the-fly calculations, where we use the single-star grids coupled to binary evolution routines, i.e.\ orbital evolution due to tides, stellar winds, magnetic breaking and gravitational-wave emission (Section~\ref{sec:detached}).

In the \posydon{} approach, each separate evolutionary phase (step) has its own dedicated function which determines the binary's state resulting from that step, the quantitative values characterizing that binary (e.g., masses of the two stars), and the event describing how that state ended (e.g., onset of RLO).
Once a step is completed, the \posydon{} framework uses the resulting binary state and event as well as each component stars' states to determine an individual binary's next evolutionary step. The process is repeated until a binary's evolution is complete, resulting in a disrupted binary, a binary merger, or a double CO. At this point, the next binary is run. 
The modular nature of \posydon{} allows a user to also provide their own prescriptions to model each phase of evolution, or even their own breakdown of the binary-evolution tree. 
As default in \posydon{}, we provide a complete set of evolutionary steps which we visually summarize in Figure~\ref{fig:structure} and present in detail in the following sections.

\section{Adopted stellar physics \label{sec:single_physics}}
\label{sec:stellar_physics}
All stellar-evolution models described in this paper were computed with the state-of-the-art, open source stellar-structure and evolution code \mesa{} \citep[][]{2011ApJS..192....3P, 2013ApJS..208....4P,2015ApJS..220...15P, 2018ApJS..234...34P,2019ApJS..243...10P} revision 11701 together with the 20190503 version of the \mesa\ software development kit \citep[SDK;][]{MESASDK}.\footnote{We made one minor bug fix in the \mesa\ source code, which involves replacing the mass of the proton with the atomic mass unit where it appears in the code that evaluates the \citet{2010CoPP...50...82P} equation of state (E.\ Bauer, 2020, private communication). 
This change is included in later \mesa{} releases.} 
\mesa{} solves the one-dimensional stellar-structure and composition equations. 
Mixing and burning processes are solved simultaneously; mixing is treated as a diffusive process.  
Discussion of specific elements of stellar physics are described in the following subsections, which are split into microphysical and macrophysical processes. 
We implement all physics that are not readily available in \mesa\ using the functionality provided by \texttt{run\_star\_extras} and \texttt{run\_binary\_extras}.

\subsection{Microphysics \label{sec:microphysics}}

We adopt the \citet{2009ARA&A..47..481A} protosolar abundances as our initial composition, with $Z=0.0142$ and $Y=0.2703$.
The equation of state is the standard \mesa\ amalgamation of the SCVH \citep{1995ApJS...99..713S}, OPAL \citep{2002ApJ...576.1064R}, HELM \citep{2000ApJS..126..501T} and PC \citep{2010CoPP...50...82P} equations of state \citep{2019ApJS..243...10P}. 
Radiative opacities are taken from \citet{ferguson2005} and \citet{iglesias1996} for the \citet{2009ARA&A..47..481A} mixture, along with electron conduction opacities from \citet{cassisi:2007}. 
Nuclear reaction rates are drawn from the JINA Reaclib database \citep{cyburt2010}.  
All models were computed using the \texttt{approx21} nuclear reaction network that consists of 21 species: $^1$H, $^3$He, $^4$He, $^{12}$C, $^{14}$N, $^{16}$O, $^{20}$Ne, $^{24}$Mg, $^{28}$Si, $^{32}$S, $^{36}$Ar, $^{40}$Ca, $^{44}$Ti, $^{48}$Cr, $^{56}$Cr, $^{52}$Fe, $^{54}$Fe, $^{56}$Fe and $^{56}$Ni, plus protons and neutrons (for the purpose of photo-disintegration).

\subsection{Macrophysics}

\subsubsection{Surface Boundary Conditions  \label{sec:surface_boundary}}

The stellar surface boundary condition is satisfied by the \texttt{simple\_photosphere} option, which sets the photosphere temperature using the \citet{Eddington1926} $T_{\rm eff}(\tau)$ relation, and the photosphere pressure via $P=\tau g/ \kappa$ with enhancement due to radiation pressure at the photosphere \citep[e.g.,][]{2011ApJS..192....3P}.

\subsubsection{Stellar Winds  \label{sec:winds}}

Stellar winds are a complex subject due to the varied physical mechanisms (known and unknown) that drive them and their dependence on the evolutionary state of their parent star. 
With this in mind, we have kept the wind prescription as simple as possible, while still capturing the key phenomenology of massive stellar evolution, but avoiding fine-tuning to reproduce any single subset of observations. In \posydon{}, changing the wind prescription would  require the computation of new set of single- and binary-star model grids.

For stars with initial masses above $8\,\Msun$ we use the \mesa\ {\tt Dutch} scheme, which consists of \citet{dejager1988} for $\Teff < 10,000$\,K and \citet{vink2001} for $\Teff > 11,000$\,K.
In cases where $\Teff > 11,000$\,K and the surface $^1$H mass fraction is below $0.4$, the \citet{vink2001} wind is replaced with the Wolf--Rayet wind of \citet{nugis2000}. 
Between 10,000\,K and 11,000\,K there is a linear ramp (as a function of $\Teff$) between the two wind prescriptions. 
We do not explicitly include any luminous blue variable (LBV) type winds; however, our stellar models at solar metallicity (e.g., Figure~\ref{fig:HRD}) do not enter the regime where LBV-type winds are typically applied in other studies \citep[e.g.,][]{2010ApJ...714.1217B}.

For stars with initial masses below $8\,\Msun$, we again use the {\tt Dutch} scheme for stars with $\Teff$ hotter than 12,000\,K. 
For stars with $\Teff$ less than 8,000\,K, we use the \citet{reimers1975} wind with scaling factor $\eta_R=0.1$ for stars on the first ascent of the giant branch, and the \citet{blocker1995} wind with scaling factor $\eta_B=0.2$ for stars in the thermally-pulsating phase.  
For the case of 8,000\,K $< \Teff <$ 12,000\,K, we calculate the wind for both the hot and cool schemes, and linearly interpolate between the two.

For mass-loss rates that have an explicit dependence on metallicity $Z$, we rescale wind mass loss based on the \emph{initial} metallicity, not the current surface $Z$, as winds are driven predominantly by iron-group elements that remain almost constant throughout stellar evolution \citep[e.g.,][]{2005A&A...442..587V}. 
The primary motivation for this approach is to avoid the dredge-up of carbon and oxygen to the surface layers in the later phases of evolution, which can cause surface $Z$ to approach $1$, from unduly influencing the mass-loss rate. 
The only exception here is the wind prescription by \citet{nugis2000} for Wolf--Rayet stars, which is specifically calibrated to the total surface metal content, including carbon and oxygen: in this case, we use the current, surface $Z$ value of the He-rich star.

We further boost stellar winds to limit a star's rotation  below its critical threshold $(\omega_{\rm s}/\omega_{\rm s,crit} \leq 1)$, so that the sum of the centrifugal force and the photon pressure never exceeds gravity on the surface of the star. 
The impact of rotation on the mass loss rate is considered as indicated in  \citep{1998A&A...334..210H,1998A&A...329..551L},
    \begin{equation}\label{ml}
    \centering
    \dot{M}_{\rm w}(\omega)= \dot{M}_{\rm w}(0)\left(\frac{1}{1-\omega_{\rm s}/\omega_{\rm s,crit}}\right)^\xi,
    \end{equation}
where $\dot{M}_{\rm w}$ is the star's wind mass-loss rate, and $\omega_{\rm s}$ and $\omega_{\rm s,crit}$  are the angular velocity and critical angular velocity at the surface, respectively. The default value of the exponent $\xi = 0.43$ is taken from \citet{1998A&A...329..551L}. The critical angular velocity is given the expression $\omega_{\rm s,crit}^2 = (1- L/L_{\rm Edd})GM/R^3$, where $L_{\rm Edd}$ is the Eddington luminosity and its expression is given in Eq.~\eqref{Ledd}. 
This explicit boost to the wind is supplemented by an implicit numerical scheme implemented in \mesa{} which ensures that the rotation of a star never exceeds its critical value.

\subsubsection{Convection, Rotation, and Mixing Processes}\label{sec:mixing}

Convective energy transport is modeled using mixing length theory \citep[MLT;][]{bohm-vitense58} except in superadiabatic, radiation-dominated regions where we employ the \texttt{MLT++} modifications introduced in \mesa{} \citep{2013ApJS..208....4P} that reduce the superadiabaticity in radiation-dominated convective regions, to improve numerical convergence.  
For the condition of convective neutrality we use the Ledoux criterion, and we use the convective premixing scheme as described by \citet{2019ApJS..243...10P}.  
We adopt a solar-calibrated mixing length parameter, $\alpha_{\rm MLT}=1.93$, based on results from the \texttt{MIST} project (Dotter et al., in preparation).

Rotation is implemented in \mesa\ as described in \citet{2013ApJS..208....4P,2019ApJS..243...10P}. 
Rotational mixing and angular-momentum transport follow the \texttt{MIST} project \citep{2016ApJ...823..102C}. 
It has been suggested that magnetic angular-momentum transport processes are main candidates for efficient coupling between the stellar core and its envelope during the post-MS (post-main sequence). 
Here, we adopt the Spruit--Tayler (ST) dynamo \citep{2002A&A...381..923S} that can be produced by differential rotation in the radiative layers and amplify a seed magnetic field. 
Stellar models with ST dynamo can reproduce the flat profile of the Sun \citep{2005A&A...440L...9E} and observations of the final spins of both white dwarfs and neutron stars \citep{2005ApJ...626..350H,2008A&A...481L..87S}, but struggles to explain the slow rotation rates of cores in red giants \citep{2012A&A...544L...4E,2014ApJ...788...93C, 2019MNRAS.485.3661F}.

\mesa\ treats mixing processes in the diffusive approximation with  MLT providing the basic description. 
In addition to MLT convection, we consider thermohaline mixing with the parameter $\alpha_{\rm th}=17.5$ \citep[][Eq.\ 14]{2013ApJS..208....4P}, also referred to as  $C_{\rm t}$ \citep[][Eq.\ 4]{2007A&A...467L..15C}, corresponding to an aspect ratio $\sim1$ of the instability fingers \citep{1980A&A....91..175K}. 
Thermohaline mixing is important during mass accretion from an evolved primary star onto an unevolved secondary star because the accreting material typically has a higher mean molecular weight than the material near the surface of the secondary star \citep[e.g.,][]{1980A&A....91..175K}.

Overshoot mixing is treated in the exponential decay formalism \citep{2000A&A...360..952H,2011ApJS..192....3P}.  
For the parameter $f_\mathrm{ov}$ describing the extent of the overshoot mixing in this formalism, we adopt an initial-mass-dependent relation. For lower mass stars (initial masses less than 4 \Msun)
we adopt a value taken from the \texttt{MIST} project, $f_\mathrm{ov}=0.016$, which is calibrated using the Sun, as well as open clusters \citep{2016ApJ...823..102C}. 
In the high-mass regime (initial masses greater than 8 \Msun), we adopt a value of $f_\mathrm{ov}=0.0415$ motivated by the work of \citet{2011A&A...530A.115B}, who used the step overshoot formalism. Both of these values of $f_\mathrm{ov}$ are measured from a distance of $0.008$ the local pressure scale height 
into the convection zone from the formal convective-radiative boundary. This is the same approach adopted in the \texttt{MIST} models \citep{2016ApJ...823..102C}.
In order to translate between the step and exponential-decay versions of overshoot mixing, we rely on the work of \citet{2017ApJ...849...18C} which shows that the free parameter in the step formalism is a factor of $\sim 10$ larger than $f_\mathrm{ov}$ (their Figure 3).  
For stars with initial masses between $4\,\Msun$ and $8\,\Msun$ we smoothly ramp between the two values of $f_\mathrm{ov}$. 
The mass range was chosen to be roughly consistent with the ranges considered in the two studies. 

We include no extra mixing due to semiconvection \citep[in the sense of ][]{1983A&A...126..207l}, as this process is implicitly accounted for in the convective premixing scheme \citep{2019ApJS..243...10P}.

\section{Adopted binary-star evolution physics \label{sec:binary_physics}}

\posydon{} is predominantly a BPS tool that simulates the evolution of an ensemble of binary systems through various stages of their life. 
In the \posydon{} framework, we base the evolution of binary systems on three extended \mesa{} binary grids, as shown in Figure~\ref{fig:structure}. 
One grid consists of initially detached binary systems of two H-rich stars starting from ZAMS, where we follow the internal evolution of both stars with detailed models (Section~\ref{sec:HMS-HMS}). 
A second grid consists of H-rich stars in a semi-detached system with a CO companion (Section~\ref{sec:CO-HMS_RLO}), and a third grid consisting of naked helium stars in an initially detached system with a CO companion (Section~\ref{sec:CO-HeMS}). 
For the non-CO components in these binaries, we follow the same prescriptions for stellar structure and evolution as described in Section~\ref{sec:single_physics}. 
However, the internal structures of stars can be affected by the presence of a companion, principally through tidal interactions and mass transfer. 
In this section we describe how we use the {\tt binary} module within \mesa{} to model each binary's orbit, while self-consistently accounting for the impact on each star's structure.

\subsection{Tides}
\label{sec:Tides}

Tidal forces take place in binary systems, as each star tends to be deformed by the gravitational pull of its companion. 
Invoked by this gravitational deformation, frictional forces inside a star drive a binary toward circularization and stellar spin - orbit synchronization.
In our \mesa{} binary grids, we assume that the initial orbit is circularized and the stellar spins are synchronized with the orbit.
This assumption should be valid especially for close orbits of massive stars, where tides are strong \citep{1996AA...309..179P, 2002MNRAS.329..897H}. 
As binaries evolve, their orbital periods change as well as the individual stars' rotation periods, potentially driving them out of synchronization. Therefore, it is only the process of spin-orbit coupling that is relevant for our grids of detailed binary-star models. In Section~\ref{sec:detached} we discuss our treatment of eccentric, detached binaries.

We follow the linear approach to tides, which defines a timescale for synchronization \citep{1981A&A....99..126H}. 
In this approach, a torque is applied to non-degenerate stars in a binary corresponding to the difference between the orbital and the spin angular velocity $\Delta \Omega$, divided by the synchronization timescale $\tau_{\rm sync}$:
\begin{equation}
    \delta \Omega = \frac{\Delta \Omega}{\tau_{\rm sync}} \delta \tau,
    \label{eq:omega_tides}
\end{equation}
where $\delta \tau$ is the timestep and $\delta \Omega$ is the change in spin angular velocity over a particular timestep. 
Since every layer of a star rotates with its own angular frequency, Eq.~\eqref{eq:omega_tides} is separately applied to every layer. 
The torque applied to each layer of the star is added up and an opposite torque is applied to the binary's orbit to ensure angular momentum conservation. 
Winds and mass transfer somewhat complicate the picture, and \citet{2015ApJS..220...15P} gives a detailed description of how these effects are accounted for.

We separately calculate $\tau_{\rm sync}$ for both stars 
\citep{1981A&A....99..126H,2002MNRAS.329..897H,2015ApJS..220...15P}: 
\begin{equation}\label{eq:tides_synch}
    \frac{1}{\tau_{\rm sync}} = 3 \left( \frac{k}{T} \right) q^2  \frac{MR^2}{I} \left( \frac{R}{a} \right)^6.
\end{equation}
Here $M$, $R$ and $I$ are the mass, radius and  moment of inertia of the star for which we calculate the tides, $q=M_{\rm comp}/M$ is the binary mass ratio, and $a$ is the orbital separation. $k$ is a dimensionless apsidal motion constant characterizing the central condensation of the star, and $T$ is the characteristic timescale for the orbital evolution due to tides. 
As Eq.~\eqref{eq:tides_synch} shows, $\tau_{\rm sync}$ is strongly dependent on the ratio of the stellar radius to the binary orbital separation. 
In practice, the quantity $k/T$ also varies significantly, depending on whether tidal dissipation occurs principally within convective regions (due to turbulent friction) or radiative regions (due to dynamical tides interacting with stellar oscillations).
As stars may have both convective and radiative regions during their lives,
at every timestep taken by \mesa we separately calculate the dynamical and equilibrium tidal timescales, layer by layer, and apply the shorter of the two.

For radiative regions in a star, we calculate $k/T$ based on the dynamical tidal timescale from
\citet[][see also \citealt{1981A&A....99..126H}]{1977A&A....57..383Z}, where
\begin{equation}\label{eq:kt_rad}
    \left(\frac{k}{T}\right)_\mathrm{rad} = 
\sqrt{\frac{GMR^2}{a^5}} (1+q)^{5/6} E_2, 
\end{equation}
with $E_2$ as the second order tidal coefficient and $G$ the gravitational constant.\footnote{This equation is equivalent to Eq.~(42) of \citet{2002MNRAS.329..897H}, apart from the typo correction of a square root, found in \citet{2007ApJ...667.1170S}.} 
For the calculation of $E_2$, we adopt the latest prescriptions from \citet{2018A&A...616A..28Q}, who investigated the dependence of the parameter on the convective radius $R_{\rm conv}$ for various metallicities and evolutionary stages, finding  
\begin{equation}\label{eq:tides_rad_E2}
    E_2 = 
    \begin{cases}
    10^{-0.42} \left(R_{\rm conv}/R\right)^{7.5} & \text{for hydrogen-rich stars}\\
    10^{-0.93} \left(R_{\rm conv}/R\right)^{6.7} & \text{for stripped-helium stars.}\\
    \end{cases}
\end{equation}

For the equilibrium tidal timescale, we calculate the synchronization timescale for each convective region in the stellar envelope using Eq.~\eqref{eq:tides_synch} and following \citet[][cf.\ Eq.~30]{2002MNRAS.329..897H}:
\begin{equation}
    \left(\frac{k}{T}\right)_\mathrm{conv} = \frac{2}{21}\frac{f_{\rm conv}}{\tau_{\rm conv}} \frac{M_{\rm conv.reg.}}{M},
    \label{eq:kt_conv}
\end{equation}
and use the shortest timescale among them. In Eq.~(\ref{eq:kt_conv}), $M_{\rm conv.reg.}$ is the mass of the convective region, $f_{\rm conv}$ is a non-dimensional numerical factor less than unity that takes into account slow convective shells that cannot contribute to the tidal viscosity within an orbital timescale, and $\tau_{\rm conv}$ is the convective timescale, which we take from 
Eq.~(31) of \citet[based on \citealt{1996ApJ...470.1187R}]{2002MNRAS.329..897H}, adapted to also accomodate convective regions that are below the surface:
\begin{eqnarray}\label{eq:conv_timescale_tides}
    \tau_{\rm conv} &=& 0.431 \left[ \frac{M_{\rm conv.reg.}}{3L} \frac{R_{\rm t,conv.reg.}+R_{\rm b,conv.reg.}}{2} \right. \nonumber \\
    & & \qquad \times \left. (R_{\rm t,conv.reg.}-R_{\rm b,conv.reg.}) \vphantom{\frac{M_{\rm conv.reg.}}{3L}} \right]^{1/3}.
\end{eqnarray}

In the equation above,  $R_{\rm t,conv.reg.}$, $R_{\rm b,conv.reg.}$, and $L$ are the radii of the top and bottom boundaries of the region, and the stellar luminosity respectively, in solar units. 
We use the surface luminosity in all calculations, as it is approximately constant throughout the envelope. 
Typically, the shortest equilibrium tidal timescale corresponds to the outermost convective region. 
In order to avoid tides being dominated by potential artificial convective shells that may appear during the numerical calculation of a star's evolution, we only take into account regions that consist of at least $10$ consecutive shells in our models.

\begin{figure}\center
\includegraphics[width=0.48\textwidth]{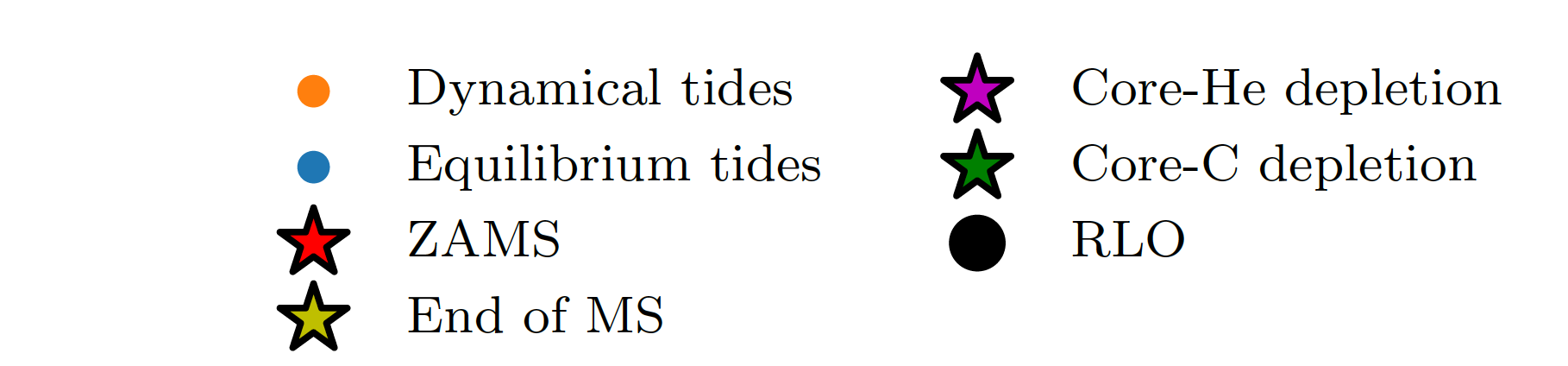}\\
\includegraphics[width=0.48\textwidth]{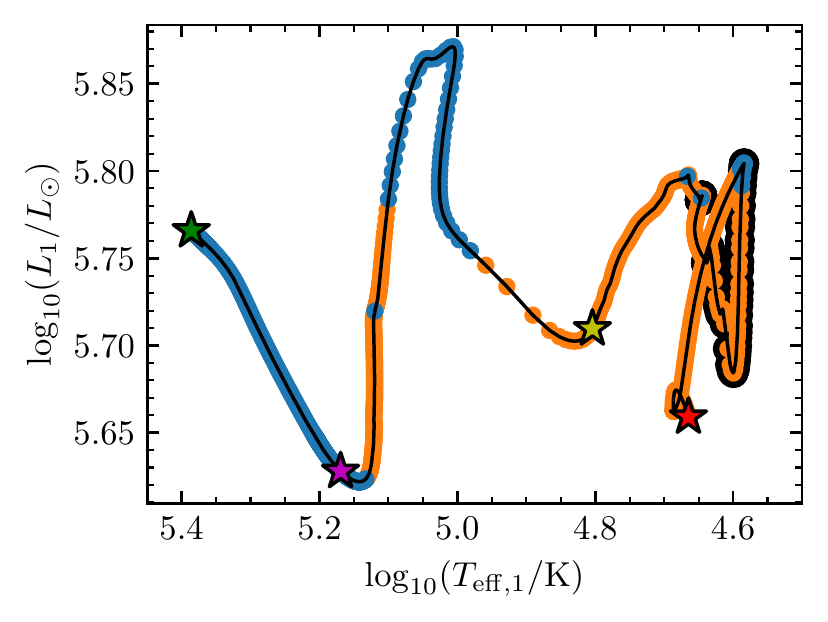}\\
\includegraphics[width=0.48\textwidth]{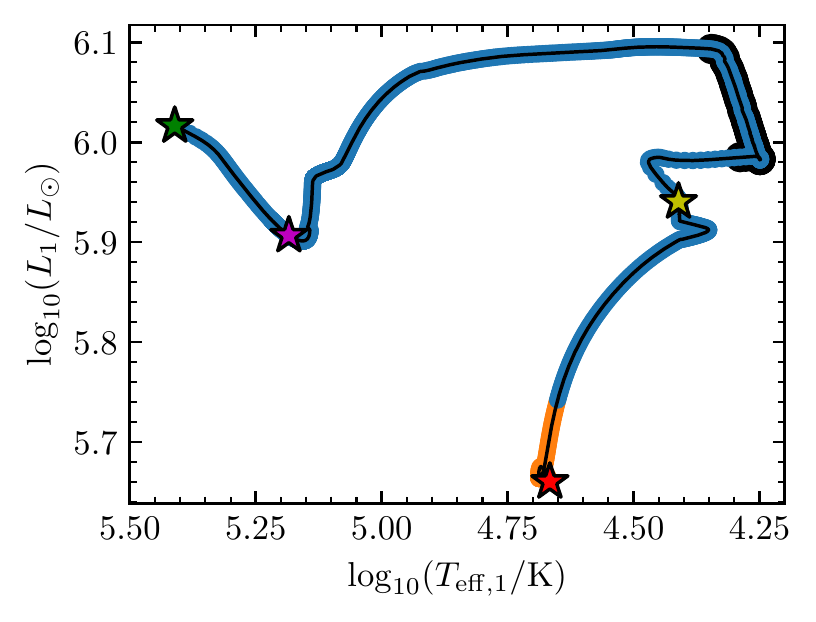}
\caption{Evolution of the donor star in two example close binary systems of initially $M_{1,\rm{initial}}=56.46\, \Msun$ and $M_{2,\rm{initial}}=28.23\, \Msun$, for two different initial orbital periods: $3.16$ days (top) and $31.62$ days (bottom). 
The colors show which tidal timescale is shortest and dominates in the tidal process: equilibrium (blue) or dynamical (orange). 
Star symbols depict the main evolutionary points.     
}
\label{fig:tides}
\end{figure}

In Figure~\ref{fig:tides}, we show two example evolutionary tracks in a Hertzsprung--Russell Diagram of the primary star in an interacting binary of $M_{1,\rm{initial}}=56.46\, \Msun$ and $M_{2,\rm{initial}}=28.23 \,\Msun$, for two different initial orbital periods ($3.16$ days, top panel; $31.62$ days, bottom panel). 
We show which term of the tidal timescale dominates the tidal forces: the dynamical tidal timescale with orange, from Eq.~\eqref{eq:tides_synch} and Eq.~\eqref{eq:kt_rad}, assuming the whole star is radiative, or the equilibrium tidal timescale with blue, from Eq.~\eqref{eq:tides_synch} and Eq.~\eqref{eq:kt_conv}, according to the most important convective region of the star.  
We see that in the beginning of the evolution, the dynamical tidal timescale dominates, as expected for the MS of massive stars that have a radiative envelope.  
The $3.16$ day period system (top) initiates early RLO and does not form convective layers massive enough \textbf{for} equilibrium tides to dominate, until after the end of its MS. 
For the wider binary, even during the MS, the equilibrium tidal timescale tends to become comparable to the dynamical tidal timescale, due to convective regions that appear close to the surface of the star. 
These regions include a small part of the total mass of the star (as low as $10^{-5}\Msun$), but have a significant radial thickness. 
The equilibrium tidal timescale dominates in all the remaining parts of the evolution, apart from the He core burning phase of the stripped primary in the 3 days period system.

\subsection{Mass-transfer}
\label{sec:MT}

Over the course of a binary's evolution, the outermost layers of one of the binary's stellar components may be removed, due to the gravitational pull of its companion. 
As a consequence of either the binary's orbit decay or the expansion of a  star's envelope, this transfer of mass is dictated by the geometry of the Roche potential, namely the gravitational potential constructed in the co-rotating reference frame of the binary system.

\subsubsection{Mass Loss Rates from a Star Overflowing its Roche Lobe}

To calculate mass-loss rates (due to mass transfer only) from main-sequence (MS) stars that overfill their Roche lobes, we use the {\tt contact} scheme within \mesa{}. 
This prescription is a numerical approximation; for stars overfilling their Roche lobes at the beginning of each timestep, mass is removed such that by the end of the timestep, the star remains confined to within its Roche lobe. 
For MS stars, this approximation is consistent with more accurate methods, as MS stars are compact, with relatively small pressure scale heights. 
We choose this prescription, as it allows us to evolve binary systems in which both stars overfill their Roche lobes simultaneously \citep[][]{2016A&A...588A..50M}.

As stars evolve off the MS, however, they tend to expand, forming less dense envelopes as they become giant stars. 
The large pressure scale heights of giant stars cause the {\tt contact} scheme to become inaccurate, and a prescription is required that can more accurately treat these stars' extended envelopes. 
For stars with a central H abundance less than $10^{-6}$, we switch to the {\tt Kolb} scheme \citep{1990A&A...236..385K}.\footnote{The current version of \mesa{} does not allow the reversing of the donor star in this scheme. Occasionally, an once accreting star evolves off the MS, expands, and itself overfills its Roche lobe. In these cases mass transfer is not calculated, and the system predominantly leads to L$_2$ overflow. } 
This prescription allows the star to expand beyond its Roche lobe, and  self-consistently calculates the rate that mass can flow through the inner Lagrangian point based on the local fluid conditions.

\subsubsection{Mass Accretion onto a Non-Degenerate Companion}
\label{sec:MT_non_degenerate}

The evolution of a binary during a mass transfer phase depends not only on the mass-losing star but also on the mass-gaining star. 
Based on the nature of the accretor, the process of accretion is treated differently. 

For binaries with a non-degenerate accretor (those in our grid of two H-rich stars), initially all the mass lost by the donor through RLO is accepted by the accretor. We assume that material being accreted carries the specific angular momentum according to \citet[Appendix A.3.3]{2013ApJ...764..166D}. 
This prescription allows for the distinct treatment of accretion via direct impact of the incoming stream on the stellar surface, or, in case of accretion onto a more compact star, the formation of a Keplerian disk around the accretor. 
The accreted angular momentum spins up the accretor, and mass accretion is restricted when the accretor reaches critical rotation. Mass falling within a critically rotating accretor's gravitational potential will be ejected from the binary with the specific angular momentum of the accretor \citep{2015ApJS..220...15P} in the form of rotationally-enhanced stellar winds, following Eq.~\eqref{ml}. 
At the same time that accretion is spinning up the outer layers of a non-degenerate star, internal mixing processes transport the surface angular momentum toward deeper layers, slowing the star's rotation rate.

\subsubsection{Accretion onto a Degenerate Companion}
\label{sec:MT_degenerate}

Mass transfer onto a degenerate star proceeds similarly as that onto a non-degenerate star, with a few notable exceptions. 
The primary exception is that mass transfer is capped at the Eddington limited rate. 
For sub-Eddington mass-transfer rates onto a CO, mass transfer is assumed to be  conservative. 
However, for super-Eddington rates, the excess matter is lost from the vicinity of the accretor as an isotropic wind (i.e., with the specific angular momentum of the accretor).

We calculate the Eddington-limited rate using standard formulae \citep{2002apa..book.....F}. We first calculate the Eddington luminosity $L_{\rm Edd}$ for an accretor with mass $M_{\rm acc}$:
\begin{equation}\label{Ledd}
L_{\rm Edd} = \frac{4\pi G M_{\rm acc} c }{\kappa},
\end{equation}
where $\kappa$ is the opacity of the incoming material and $c$ is the speed of light. 
For a fully ionized gas, Thompson scattering dominates the opacity, so 
$\kappa = 0.2 (1+X)\,\mathrm{cm}^2\,\mathrm{g}^{-1}$, where $X$ is the hydrogen abundance of the donor. 
By setting $L_{\rm Edd}$ equal to the radiation released by accreted matter as it falls into a CO's potential well ($L_{\rm acc} = \eta \dot{M} c^2$), we can recover the Eddington-limited accretion rate,
\begin{equation}
     \dot{M}_{\rm Edd} = \frac{4\pi G M_{\rm acc}}{\kappa c \eta}.
\end{equation}
The dimensionless constant $\eta$ sets how efficiently the rest mass energy of the incoming matter is converted to outgoing radiation,
\begin{equation}
    \eta = \frac{G M_{\rm acc}}{R_{\rm acc} c^2}.
\end{equation}
For BHs, $R_{\rm acc}$ is set by the spin-dependent innermost stable circular orbit, while for NSs, we use a constant $R_{\rm acc}$ of $12.5$\,km \citep{2018PhRvL.120z1103M, 2019ApJ...887L..24M, 2019ApJ...887L..21R, 2020PhRvD.101l3007L, 2020ApJ...892L...3A, 2021A&A...650A.139K, 2021ApJ...921...63B, 2021ApJ...918L..29R}.
Our grids containing a CO components, are currently focused on NS and BH accretors, for which we simulate a range of masses.  
The type of a CO is determined solely on its mass, with COs having gravitational mass less than $2.5\, \Msun$ being classified as NSs, while those with mass greater than $2.5 \,\Msun$ as BHs.
Within our code, the only difference between these types of accretors is the corresponding $\eta$; otherwise accretion proceeds identically regardless of the type of CO accretor.

As these COs accrete material, they ought to accrete angular momentum. 
In our current version of the grids, we ignore any corresponding increase in the spin rate of NSs. 
For BHs, on the other hand, we self-consistently incorporate the increase in spin frequency as well as its effect on $\eta$ through the radius of innermost stable circular orbit
\citep[ISCO;][]{2003MNRAS.341..385P},
\begin{equation}
    \eta = 1 - \sqrt{1-\bigg(\frac{M_{\rm acc}}{3 M^i_{\rm acc}}\bigg)^2},
\end{equation}
where $M^i_{\rm acc}$ is the initial mass of the accreting BH and $M_{\rm acc}$ its current one. In this equation it is also implicitly assumed that the birth spin of the BH is $\sim$0, as it has been suggested by several studies \citep{2015ApJ...800...17F, 2018A&A...616A..28Q, 2019ApJ...881L...1F}. 
The corresponding increase in the BH's non-dimensional spin rate, $a_{\rm spin}$, can be calculated following
\citet{1974ApJ...191..507T} and \citet{1999MNRAS.305..654K},
\begin{equation}
    a_{\rm spin} = \left(\frac{2}{3}\right)^{1/2} \frac{M^i_{\rm acc}}{M_{\rm acc}}\left\{4 - \left[18\left(\frac{M^i_{\rm acc}}{M_{\rm acc}}\right)^2 - 2\right]^{1/2}  \right\}.
\end{equation}
These equations both assume that the spin-up occurs due to angular momentum accretion from a disk that is truncated at the ISCO.

We do not explicitly stop our simulations if a NS accretes enough mass to cross our 2.5\,\Msun{} threshold thereby collapsing into a BH, but we do switch the $\eta$ instantaneously.

\subsubsection{Onset of CE}\label{sec:binary_physics_instability}

Stars with radiative envelopes entering RLO respond to mass loss by shrinking \citep{1987ApJ...318..794H}; mass transfer then reaches a natural equilibrium set by the strength of some driving force (nuclear evolution, tides, thermal expansion or some other effect), and how quickly the orbital separation, and thus the Roche lobe radius, change due to mass transfer through the inner Lagrangian point.
However, in certain circumstances mass transfer increases in a runaway process, either because stars expand due to mass loss (e.g., stars with deep convective envelopes) or because the binary's orbit shrinks faster than a donor star's radius
\citep[e.g.,][]{1972AcA....22...73P}. 
These phases of binary evolution are notoriously difficult to model as they are intrinsically three-dimensional processes, and they span many orders of magnitude in spatial and temporal scales \citep{2013A&ARv..21...59I}.
We therefore stop our \mesa{} models when binaries enter dynamically unstable mass transfer; we provide a description of how we address this phase in Section~\ref{sec:common_envelope}. 
Here we focus on the conditions we use to identify when a binary enters dynamically unstable mass transfer.

First, we assume that a dynamically unstable RLO phase is initiated whenever mass-transfer rate exceeds $0.1\, \Msun$~yr$^{-1}$. 
It is expected that binaries reaching this limit will only further increase their mass-transfer rates, as this corresponds to a dynamical limit on the mass-loss rate for giant stars (with dynamical timescales of years). 
As a check, we carried out a calibration test where we followed the evolution of a test binary to mass transfer rates even larger than $0.1\, \Msun$~yr$^{-1}$. In every test we ran, we found that the mass-transfer rate increases to arbitrarily high rates, confirming the validity of our limit.
Assigning a limit to the mass-transfer rate also avoids numerical issues caused by the effort of stellar models to converge with such extreme mass loss.

As a second condition, we assume dynamically unstable RLO occurs when the stellar radius of the expanding star extends beyond the gravitational equipotential surface, passing through the second Lagrangian point (L$_2$). 
In such cases the lost matter from the L$_2$ point carries substantial angular momentum, rapidly shrinking the orbit and leading to a runaway process in which the two stars spiral in and trigger a CE  \citep{2011A&A...528A.114T, 2014ApJ...786...39N}. 
We use the prescription from \citet[][cf.\ Eq.~15--19]{2020A&A...642A.174M} to define the spherical-equivalent radius corresponding to L$_2$. 
This condition cannot occur for MS donors, since the {\tt contact} scheme for RLO forces a star's radius to be contained within its Roche lobe. 
For cases where two MS stars overfill both their Roche lobes, in an over-contact binary, we alternatively use the prescription from \citet[][cf.\ Eq.~2]{2016A&A...588A..50M} for the L$_2$ radius, which considers that both stars can contribute to the overflow of the L$_2$ volume together.

\begin{deluxetable*}{ccccccccccccc}
\tabletypesize{\footnotesize}
\tablecolumns{13}
\tablewidth{0pt}
\setlength{\tabcolsep}{3pt}
\tablecaption{ Summary of the five detailed single- and binary-star model grids. \label{tab:grid_properties}}
\tablehead{
\multicolumn{2}{c}{Initial state}& & \multicolumn{8}{c}{Parameters' range and resolution} & &\\
\cline{1-2}\cline{4-11}
  \colhead{Star 1} & \colhead{Star 2}  & & \colhead{$M_1\,[M_\odot]$} & \colhead{$\Delta\log_{10}M_1$} & \colhead{$M_2\,[M_\odot]$}& \colhead{$\Delta\log_{10}M_2$} & \colhead{$q$} & \colhead{$\Delta q$} & \colhead{$P_{\rm orb}\,[{\rm day}]$} & \colhead{$\Delta\log_{10}P_{\rm orb}$} & \colhead{$N$\,\tablenotemark{a}} & \colhead{Failures\,\tablenotemark{b}}
 }
\startdata
ZAMS & - && 0.5--300 & 0.014 & - & - & - & - & - & - & 200 & 1.5\%\\
ZAHeMS\tablenotemark{c} & - && 0.5--80 & 0.055 & - & - & - & - & - & - & 40 & 0\%\\
ZAMS & ZAMS && \change{7}{6.2}--120 & 0.025 & - & - & 0.05--0.95 & 0.05 & 0.72--6105 & 0.07 & \change{56000}{58240} & \change{0.9}{1.5}\%\\
Evolved, H-rich\,\tablenotemark{d} & CO  && 0.5--120 & 0.06 & 1--35.88 & 0.074 & - & - & 1.26-3162 & 0.13 & 25200 & \change{0.8}{0.9}\%\\
ZAHeMS & CO && 0.5--80 & 0.055 & 1--35.88 & 0.074  & - & - & 0.02--1117.2 & 0.09 & 39480 & \change{4.7}{4.8}\%\\
\enddata
\tablenotetext{a}{Total number of models in this grid.}
\tablenotetext{b}{Percentage of models that stopped due to numerical-convergence errors before reaching one of our stopping conditions. 
These rates describe the finalized grids, after a series of re-runs have occurred; see Section~\ref{sec:rerun} for details.}
\tablenotetext{c}{Zero-age He Main Sequence stars.}
\tablenotetext{d}{Although this grid is initialized with H-rich stars at ZAMS, we ignore the portion of each simulated binary's evolution prior to the onset of RLO. The initial state of Star 1 in this grid are therefore somewhat evolved. }
\end{deluxetable*}

For CO accretors, we set a third condition for unstable mass transfer based on the photon trapping radius \citep{1979MNRAS.187..237B, 1999ApJ...519L.169K}. 
Inside that radius photons are advected inward along with accreted matter onto the accretor, while outside that radius, photons diffuse away. 
For stable mass accretion, the photon trapping radius occurs close to the accretor; however, as the accretion rate increases, the photon trapping radius expands.
Once the photon trapping radius reaches the Roche-lobe radius of the accretor it is assumed to lead to a CE phase.
Since the radius of the photon trapping envelope $R_{\rm trap}$ depends on  the Eddington limit of the accretor $\dot{M}_{\rm Edd}$ and the mass-transfer rate from the donor $\dot{M}_{\rm donor}$, we limit the latter assuming an instability condition when \citep{1979MNRAS.187..237B}: 

\begin{equation}
    \dot{M}_{\rm donor} \geq {\dot{M}_{\rm Edd}} \frac{2R_{\rm L,accr}}{R_{\rm acc}}.
\end{equation}

As a final condition, occasionally two stars will both overfill their Roche lobe while one of those stars has evolved off the MS. Since the {\tt contact} scheme in \mesa{} can only evolve binaries in which both stars are on the MS, we assume these binaries automatically enter a CE.

As a test we compared the first three dynamical instability conditions separately to investigate their effect and found that the limiting mass accretion rates are all similar: as soon as a binary reaches any one of them, the other two are close to their limits as well.
Therefore, for a particular binary in our \mesa\ simulation, the binary is considered to enter a CE if any one of them occurs.

\section{Grids of detailed single- and binary-star evolution models \label{sec:grids}}
\label{sec:single_binary_evolution}

While in Section~\ref{sec:stellar_physics} and Section~\ref{sec:binary_physics} we describe the physics we adopt in our simulations, here we provide numerical details about how we produce each of our five \mesa{} grids. 
This includes our procedure for producing initial stellar models for each of our grids (Section~\ref{sec:zams}), our termination conditions common across all of our grids (Section~\ref{sec:terminated}), and a description for each of our five grids of binary simulations (Sections~\ref{sec:HMS}--\ref{sec:CO-HeMS}). 
We summarize the basic properties of each grid in Table~\ref{tab:grid_properties}.

\subsection{Zero Age Main Sequence Models}
\label{sec:zams}

We create our own library of ZAMS models for both H-rich and He-rich stars. 
For the creation of the H-rich ZAMS models we use the \mesa\ revision 11701 template \texttt{create\_zams}.
The process begins with creating a fully-convective star with no nuclear fusion taking place and adopting our protosolar abundances (Section~\ref{sec:microphysics}). 
This model is then evolved with our adopted nuclear reaction network until the H-burning luminosity exceeds 99\% of the total luminosity. 

The He-rich ZAMS (ZAHeMS) models are created in three steps. 
First, we create a pre-MS He star with 100\% $^{4}$He in the same way that we create a H-rich pre-MS star. 
In the second step we adjust the initial metallicity.  In the third step we evolve the model until the He-burning luminosity exceeds 99\% of the total luminosity. 

The two sets of ZAMS and ZAHeMS models are used as a starting point for the five grids of single- and binary-star models.

\subsection{Termination conditions}
\label{sec:terminated}

We set conditions for the termination of our evolutionary models based on both single-star properties and binary-star properties. 
If any one of these conditions are met by an individual simulation, it is terminated at that timestep. 
Our termination conditions are:
\begin{itemize}
    \item A star's age exceeds the age of the Universe (13.8\,Gyr), a condition that is typically only met for the lowest-mass stars we simulate ($M_{\rm init} \lesssim 0.8\,\mathrm{M}_\odot$). 
    In our single-star grids, for numerical purposes, we allow stars to evolve beyond this condition, then truncate their evolution afterwards at the end of the MS, which may extend beyond the age of the Universe.
    \item A star becomes a WD, a condition we quantify by checking if the central degeneracy parameter $\Gamma_c$ (Coulomb coupling parameter) exceeds 10 \citep{2016ApJ...823..102C}.
    \item A star reaches the end of core C-burning, a condition triggered when the fractional abundances of both C and He decrease below $10^{-2}$ and $10^{-6}$, respectively, at the star's center.
    \item A binary enters a CE phase, as described in Section~\ref{sec:common_envelope}. 
    \item A star reaches the thermally-pulsating asymptotic giant branch (TP-AGB) phase and then reaches a point of failure (numerical non-convergence) during thermal pulsations. Because these are not uncommon and we consider the nascent WD to be well-formed be within the AGB star, we consider this evolution to be successful.
\end{itemize}

Our simulations may occasionally end prematurely before any of the aforementioned conditions are reached. 
This may happen because the minimum timestep limit within \mesa{} ($10^{-6}$\,s) is reached  or any individual simulation reaches our maximum runtime on our computing cluster (set to 48 hours). 
We provide details describing how we approach such runs in Section~\ref{sec:rerun}, but these failures are rare, occurring a few percent or less in each grid.

All single- and binary-star models in \posydon{} have a final, internal structutre profile  written to correspond precisely to the last evolutionary phase milestone.  
These profiles are discussed further in Section~\ref{sec:ppq}, Section~\ref{sec:common_envelope}, and Section~\ref{sec:core-collapse}.

\subsection{H-rich, single-star grid}
\label{sec:HMS}
Our first grid of single-star evolutionary models contains a series of non-rotating H-stars with our adopted protosolar composition of $Y=0.2703$ and $Z=0.0142$. 
The grid consists of 200 masses, ranging from $M_\mathrm{init}=0.5\,\Msun$ to $M_\mathrm{init}=300\,\Msun$ with a logarithmic spacing of $\Delta \log_{10} (M_\mathrm{init}/\Msun) = 0.014$\,dex. 
For each star, models were initialized using the procedure described in Section~\ref{sec:zams}, and evolved until one of the termination conditions provided in Section~\ref{sec:terminated} occurs.

\begin{figure}
    \centering
    \includegraphics[width=0.48\textwidth]{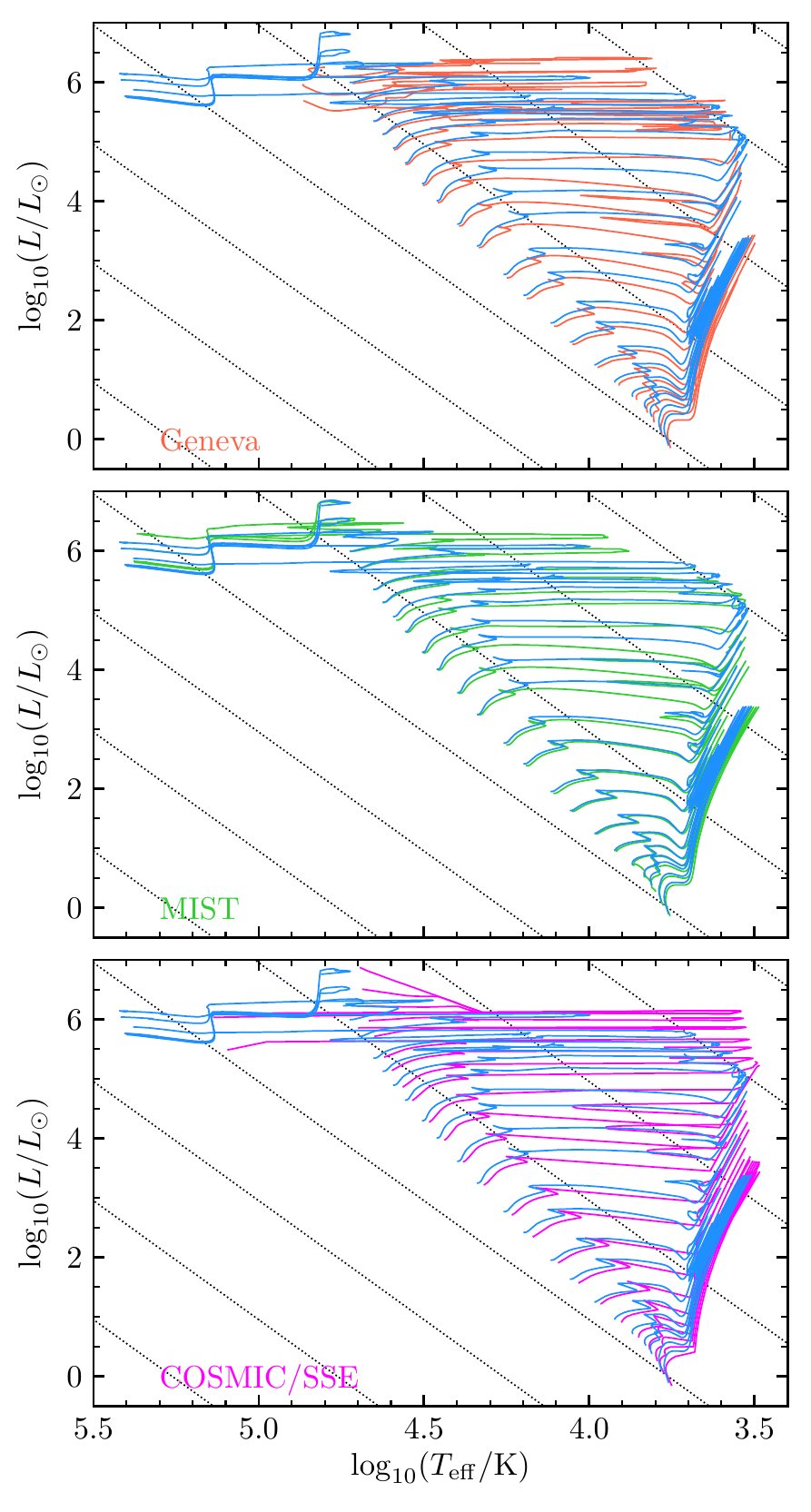}
    \caption{Comparing \posydon{} H-rich ZAMS evolutionary tracks (blue in all panels) with non-rotating {\tt Geneva} \citep[upper;][]{2012A&A...537A.146E}, {\tt MIST} \citep[middle;][]{2016ApJ...823..102C}, and \texttt{SSE} with default wind mass loss prescription from {\tt COSMIC} version 3.4.0 \citep[lower;][]{1998MNRAS.298..525P,2000MNRAS.315..543H,2020ApJ...898...71B} tracks in the Hertzsprung--Russell diagram. 
    The mass range shown is 1--300\,$\Msun$ in all cases. The masses shown are the same as the {\tt Geneva} grid of models between 1 and 120 $\Msun$ with the addition of 175 and 300 $\Msun$ for {\tt POSYDON}, {\tt MIST}, and {\tt COSMIC/SSE}. The dotted, gray lines indicate constant radius at powers of 10 in $R_{\odot}$. } \label{fig:HRD}
\end{figure}

To test their validity, we compare \posydon{} evolutionary tracks to the widely used stellar evolution tracks from the {\tt Geneva} \citep[upper panel;][]{2012A&A...537A.146E}, {\tt MIST} library \citep[center panel;][]{2016ApJ...823..102C}, and {\tt BSE} as implemented in {\tt COSMIC} \citep[lower panel;][]{1998MNRAS.298..525P,2000MNRAS.315..543H,2020ApJ...898...71B} groups in Figure~\ref{fig:HRD}. 
In all cases we show non-rotating models with initial masses between 1\,$\Msun$ and 300\,$\Msun$. 
The pre-MS evolution is omitted from the {\tt MIST} evolutionary tracks and the TP-AGB and post-AGB phases are omitted for clarity. 
All sets of tracks show a similar location for the ZAMS; the subsequent evolution along the MS and through He-burning phases differs due to the way each set of models treats mixing across convective boundaries. 
The clearest differences between the \posydon{} and other models are in the location of the hook feature near the MS turnoff for higher masses, a result of the different adopted core overshoot treatments, and the positions of later phases, a result of the different wind mass-loss treatments among the different groups. 
The {\tt COSMIC} tracks extend to larger radii and cooler effective temperatures, which may place them in the regime of LBVs; however, none of the other sets of evolutionary tracks enter this regime. For a more in depth  comparison see, e.g.,   \citet{2020MNRAS.497.4549A,2021arXiv211202800A}.

\begin{figure}
    \centering
    \includegraphics[width=0.48\textwidth]{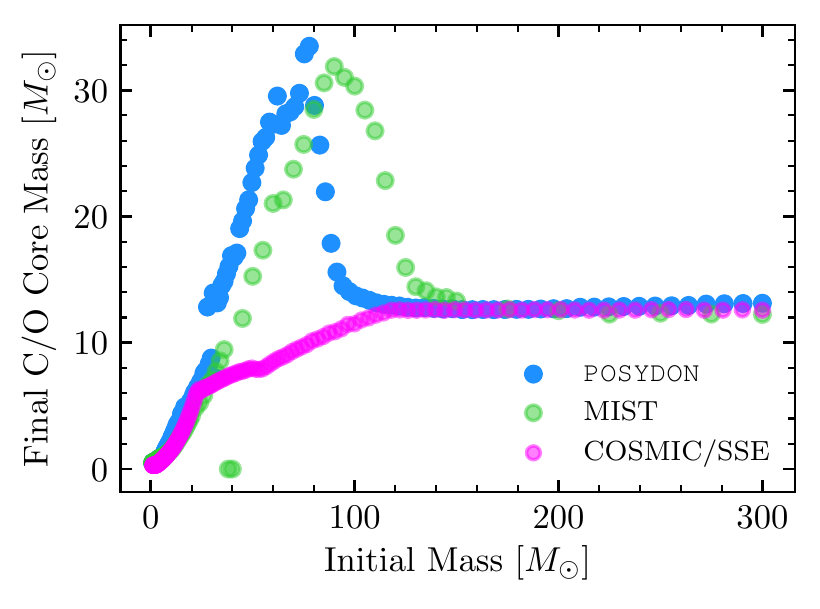}
    \caption{
    Comparison of the final C/O core mass in {\tt SSE} as implemented by {\tt COSMIC} (magenta), {\tt MIST} (green), and \posydon{} (blue). Differences between {\tt MIST} and \posydon{} are due to the larger core overshoot parameter adopted by \posydon{}. Disagreement with the {\tt SSE} models is expected as these models are based on simulations that were only computed for initial masses up to 50\,\Msun; more massive stars are an extrapolation. 
    }.
    \label{fig:core_masses}
\end{figure}

Figure~\ref{fig:core_masses} compares the final C/O core mass between \posydon{}, {\tt MIST}, and {\tt SSE} as implemented by {\tt COSMIC}. 
{\tt SSE} models are calculated until central C-burning, while {\tt MIST} and \posydon{} are calculated through central C-exhaustion for those stars with sufficient mass to ignite carbon or to the WD cooling sequence for lower masses. Differences between the core masses of {\tt MIST} and {\tt POSYDON} are generally due to the different overshooting parameter (we adopt $f_{\rm ov} = 0.0415$ for stars with masses above 8 $\Msun$, compared with $f_{\rm ov} = 0.016$ adopted by {\tt MIST}). The {\tt COSMIC}/{\tt SSE} models exhibit different behavior at larger masses as these prescriptions are based on stellar models that only go up to 50 $\Msun$; larger masses than this are an extrapolation.

\begin{figure}
    \centering
    \includegraphics[width=0.48\textwidth]{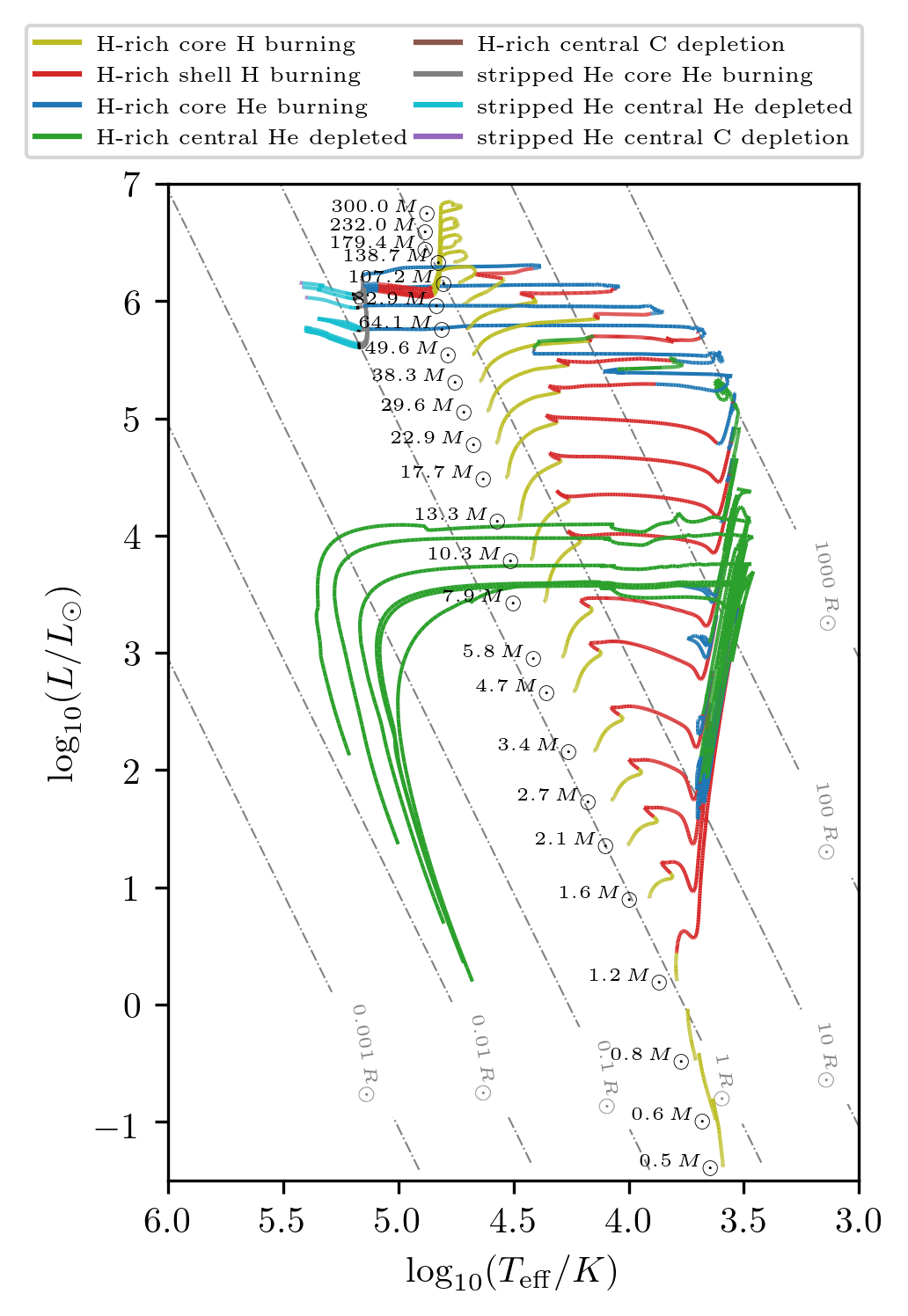}
    \caption{Hertzsprung--Russell diagram of a subsample of \posydon{} single-stellar models where the different \posydon{} stellar states are indicated according to the legend.
    }
    \label{fig:HR_stellar_states}
\end{figure}

As a final comparison, Figure~\ref{fig:HR_stellar_states} shows a Hertzsprung--Russell diagram of a subsample of \posydon{} single stellar model where we indicate the different evolutionary \posydon{} stellar states (Figure~\ref{fig:starstates}) across a range of stellar masses.

\subsection{He-rich, single-star grid}
\label{sec:HeMS}

Our second grid of single-star evolutionary models consists of non-rotating He-rich stars with $Y_{\rm init}=1-Z_{\rm init}$ and our adopted protosolar $Z_{\rm init}=0.0142$. 
This grid consists of 40 masses ranging from $M_{\rm init}=0.5\,\Msun$ to $M_{\rm init}=80\,\Msun$ with a logarithmic spacing of $\Delta \log_{10} (M_{\rm init}/\Msun) = 0.055$\,dex. 
For these masses, stellar evolution models were computed starting from ZAHeMS models (Section~\ref{sec:zams}) and evolved until one of the termination conditions described in Section~\ref{sec:terminated} occurs.
For all but the lowest-mass cases, the core C depletion condition is the relevant one; models with initial masses below $1.1\,\Msun$ do not ignite C-burning in the core, and therefore terminate as He-core WDs.

\begin{figure}
    \centering
    \includegraphics[width=0.48\textwidth]{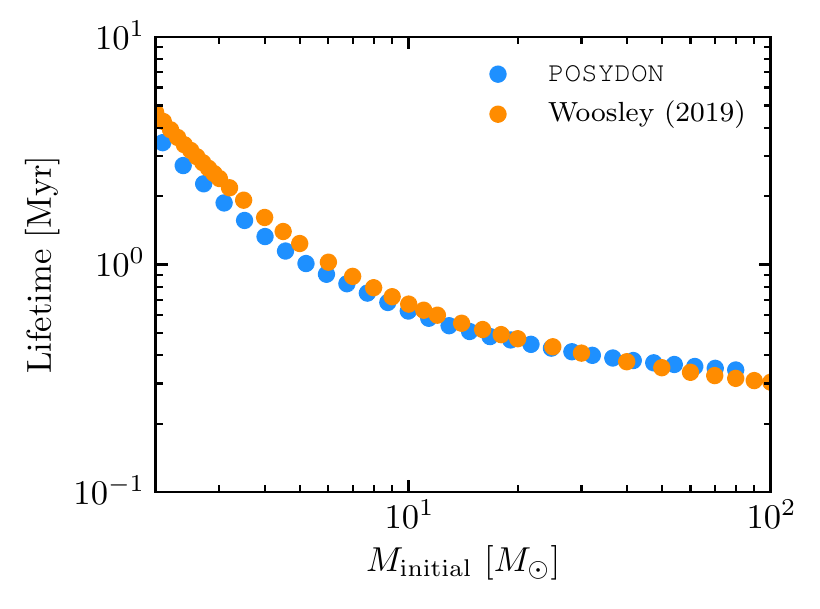}
    \caption{Lifetimes of the \posydon{} and \citet{2019ApJ...878...49W} single He-star evolution models match to within $\simeq$0.1\,dex. }
    \label{fig:He_star_lifetime}
\end{figure}

As a test of our \posydon{} He-star models, we compare their lifetimes to those of the He-star models of \citet{2019ApJ...878...49W} in Figure~\ref{fig:He_star_lifetime}. 
Only the overlapping range of initial masses is shown here; the \citet{2019ApJ...878...49W} grid includes models with masses from 1.8--120\,$\Msun$. 
These models match to within $\simeq 0.1$\,dex in log lifetime across the entire range of initial masses.

\begin{figure}
    \centering
    \includegraphics[width=0.48\textwidth]{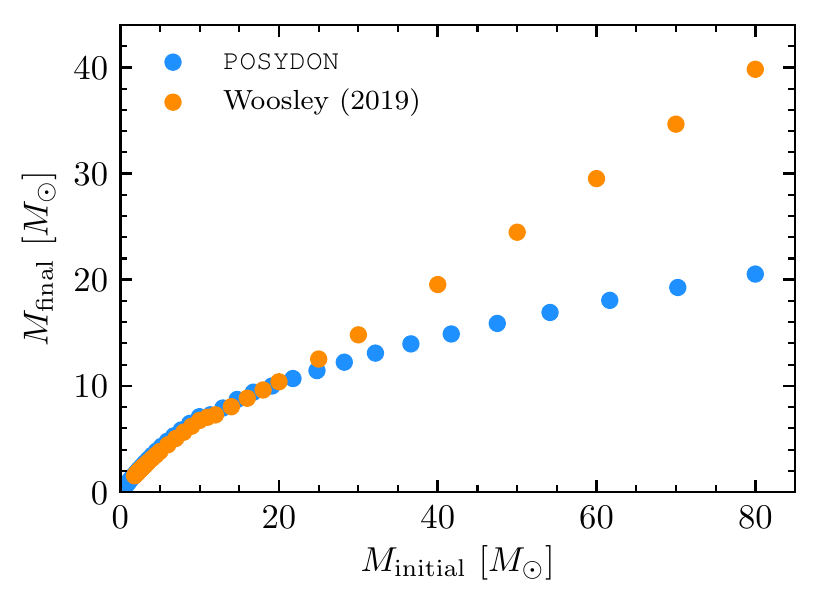}
    \caption{Final masses of the \posydon{} and \citet{2019ApJ...878...49W} single He-star evolution models. Differences at $M_{\rm initial} \gtrsim 20\,\Msun$ are due to the stronger wind mass loss prescription adopted by \posydon{}.}
    \label{fig:He_star_mass}
\end{figure}

As a second test, we compare the final masses between the same two model grids in Figure~\ref{fig:He_star_mass}. 
Although the lifetimes are similar, the final masses show a significant difference, particularly at higher initial He-star masses. 
\citet{2019ApJ...878...49W} notes that the change in slope of the initial-final mass relation around $M_{\rm final}$ of 11\,$\Msun$ is due to the mass-loss prescription adopted for exposed CO cores; at larger initial He-star masses, the entire He-star mass is burned to heavier elements. 
For all the single- and binary-star model grids in \posydon{}, we adopt the mass-loss prescription from \citet{nugis2000} for He-rich stars. The latter predicts on average stronger wind mass loss than the prescription from \citet{2017MNRAS.470.3970Y} adopted by \citet{2019ApJ...878...49W}, leading to the substantially different final masses between the two prescriptions at $M_{\rm initial} \gtrsim 20\,\Msun$.

\begin{figure}
    \centering
    \includegraphics[width=0.48\textwidth]{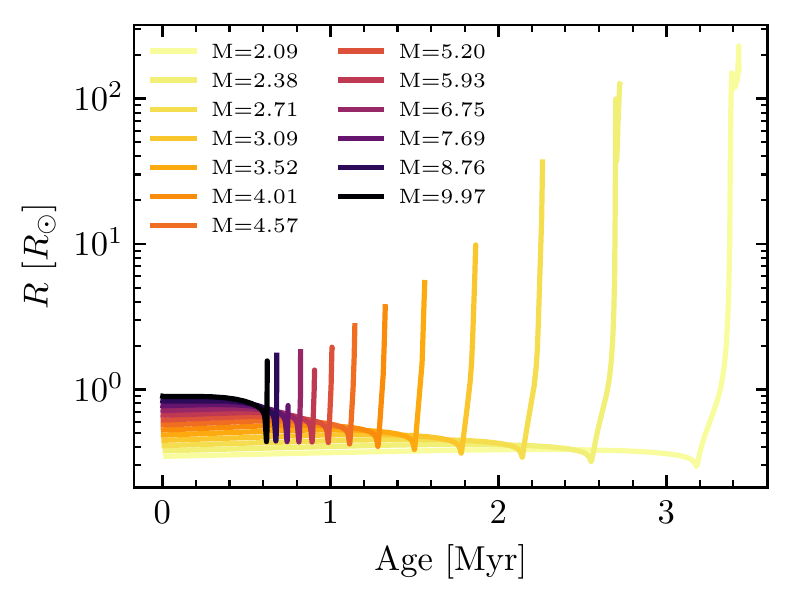}
    \caption{Evolution of radius for He stars with masses between 2 and 10 $M_{\odot}$. Less massive He-stars evolve slower, but expand farther when they become giant stars. This trend implies that more-massive He-stars in binary systems will undergo RLO for a relatively narrow range of orbital periods, a behavior exhibited by our binary star simulations and seen in Figure~\ref{fig:CO-HeMS_MESA_grid_TF12}.}
    \label{fig:He_radius}
\end{figure}

Finally, we show the radius evolution of the He star tracks for the mass interval of 2--10\,\Msun\ in Figure~\ref{fig:He_radius}. He-rich stars exhibit a peculiar feature where less-massive stars expand farther on the giant branch than their more massive counterparts, in agreement with results by \citet{1986A&A...167...61H}. When in a binary system, this implies that there is a relatively narrow range of orbital periods in which massive He stars will undergo RLO. Less massive He stars expand to hundreds of \Rsun, leading to a wide range of orbital periods in which these stars can interact with a putative companion. This behavior is realized in our binary star grids, and its effects are seen explicitly in Figure~\ref{fig:CO-HeMS_MESA_grid_TF12}.

\subsection{Binaries consisting of two hydrogen-rich main-sequence stars \label{sec:HMS-HMS}}
\begin{figure*}[t]\center
\includegraphics{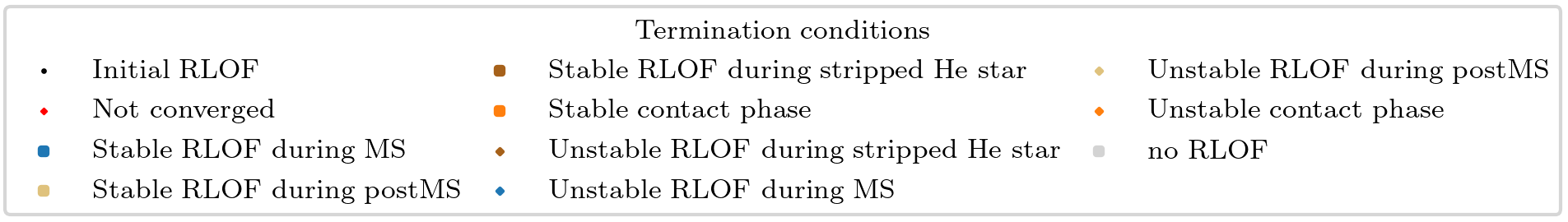}\\
\includegraphics{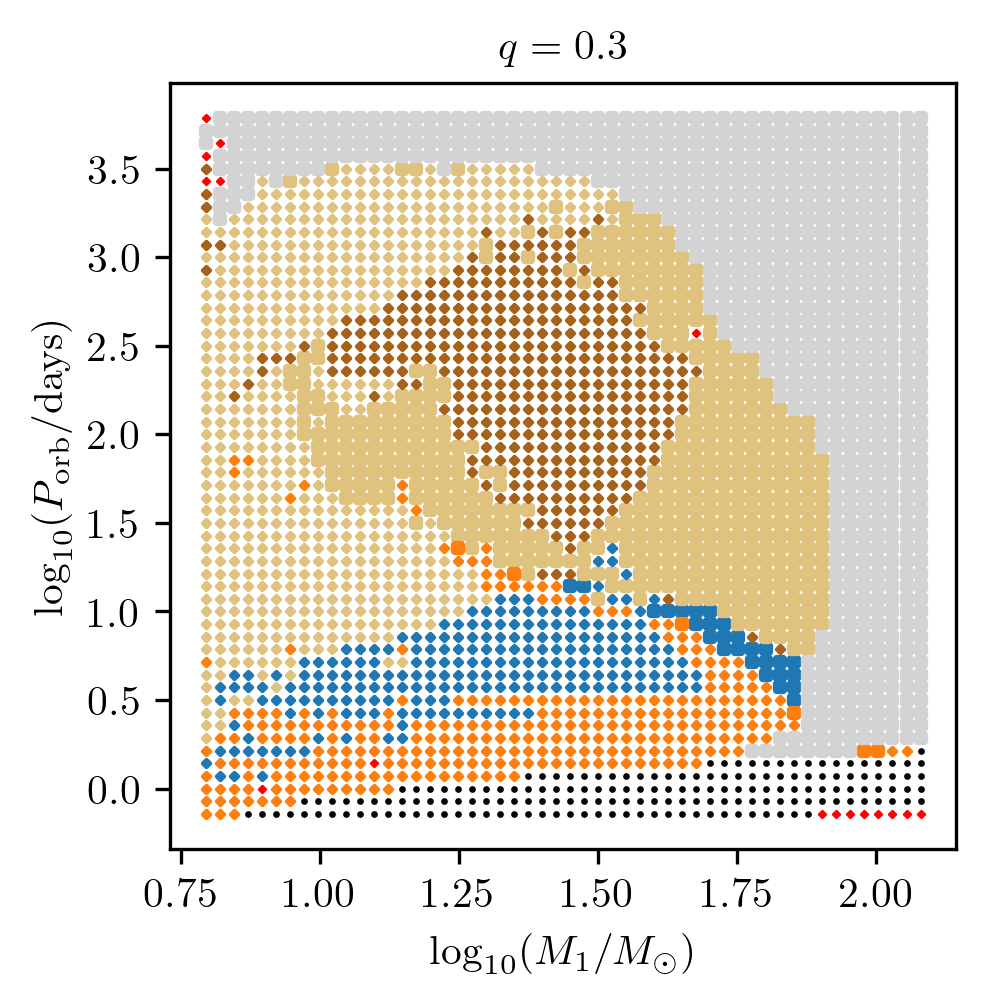}
\includegraphics{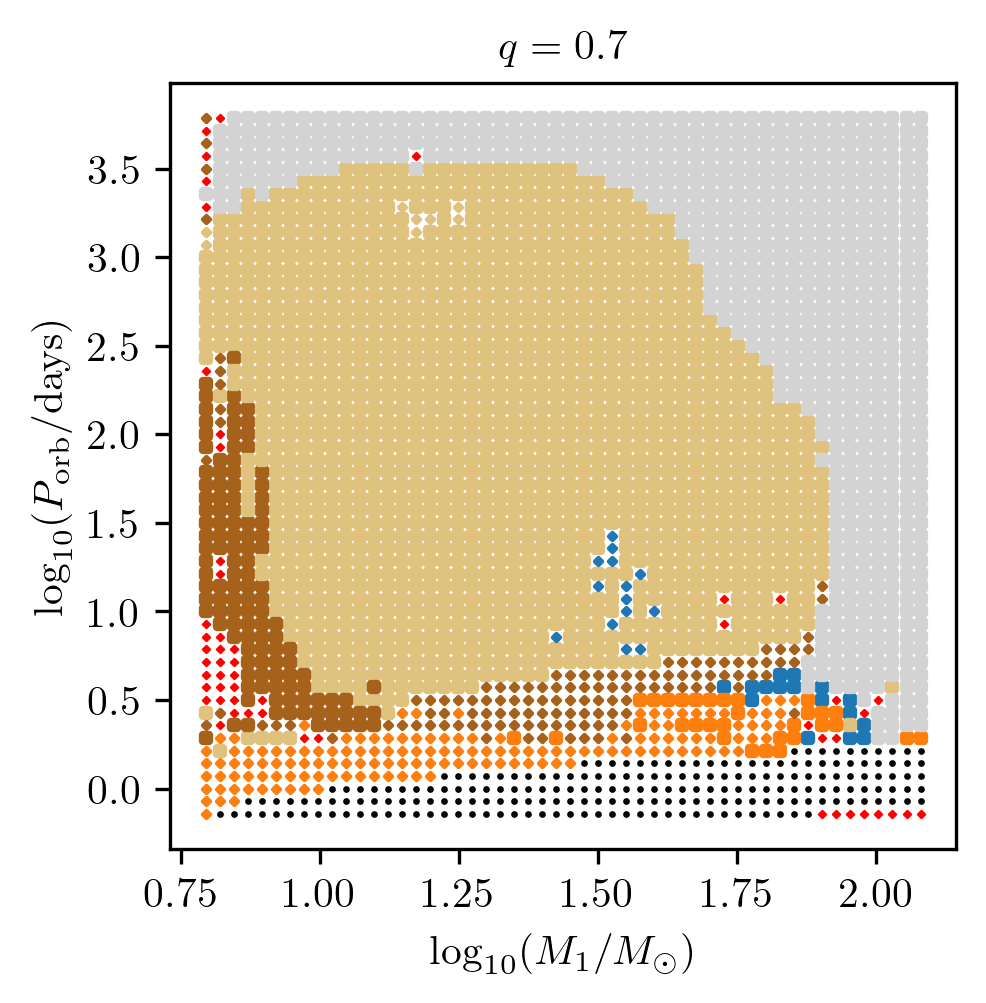}
\caption{ View of two grid slices, for two different values of initial binary mass ratio ($q=0.3$ on the left, $q=0.7$ on the right),  from our grid of binary-star models consisting of two H-rich stars, initially at ZAMS. 
The different symbols summarize the evolution of each of the models. 
We distinguish between models that experienced stable or no mass transfer (squares), reaching the end of the life of one of the stars, and the ones that stopped during mass transfer due to one of our conditions for dynamical instability (diamonds). 
Different colors distinguish the evolutionary phase of the donor star during the latest episode of mass transfer (or no RLO at all for grey). 
Small black dots at low initial periods depict systems that were in initial RLO at birth and red diamonds represent the models that stopped prematurely for numerical reasons. }
\label{fig:HMS-HMS_MESA_grid_TF12}
\end{figure*}

\begin{figure*}[t]\center
\includegraphics{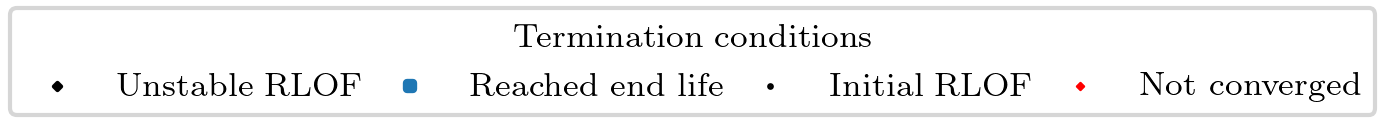}
\\
\includegraphics{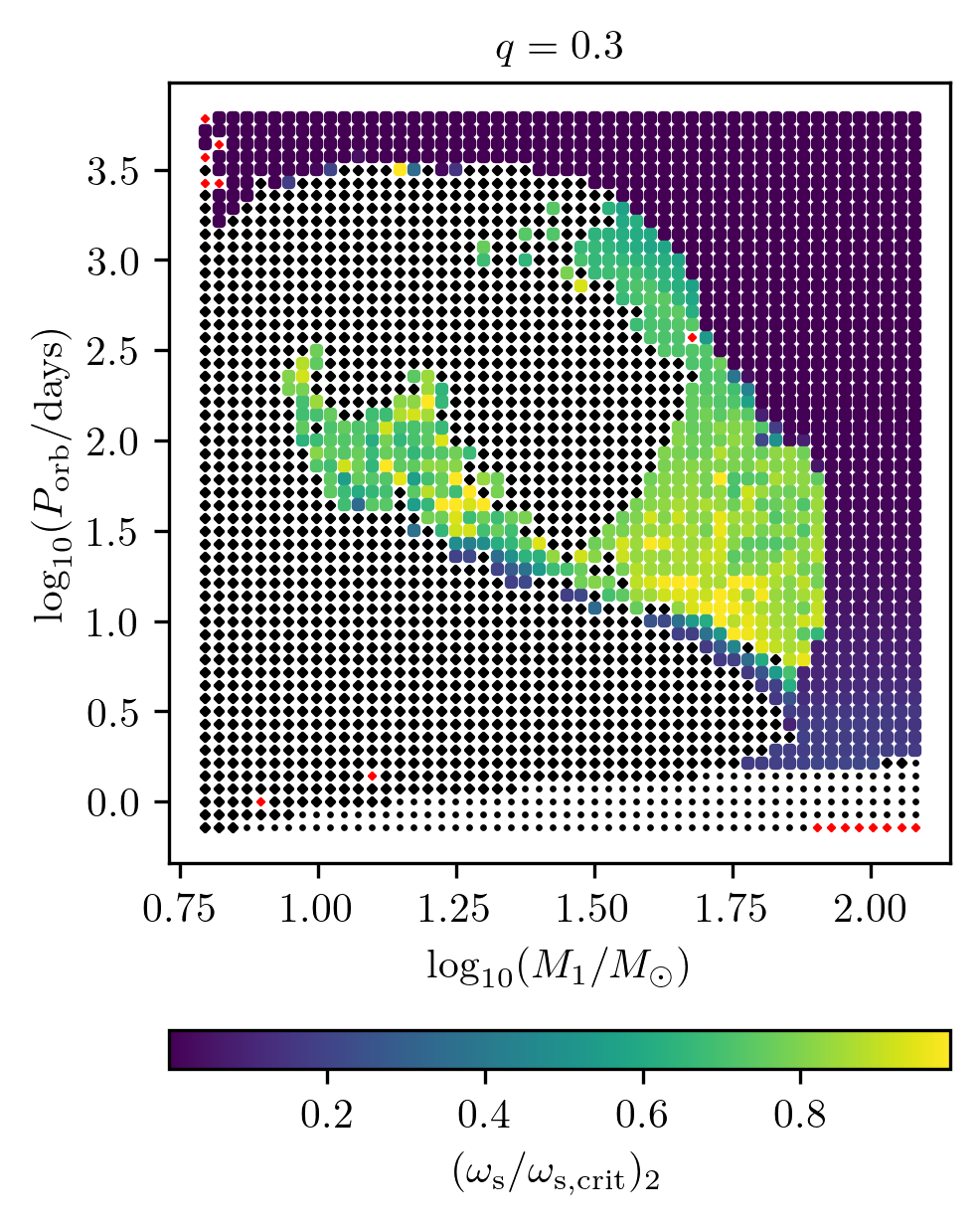}
\includegraphics{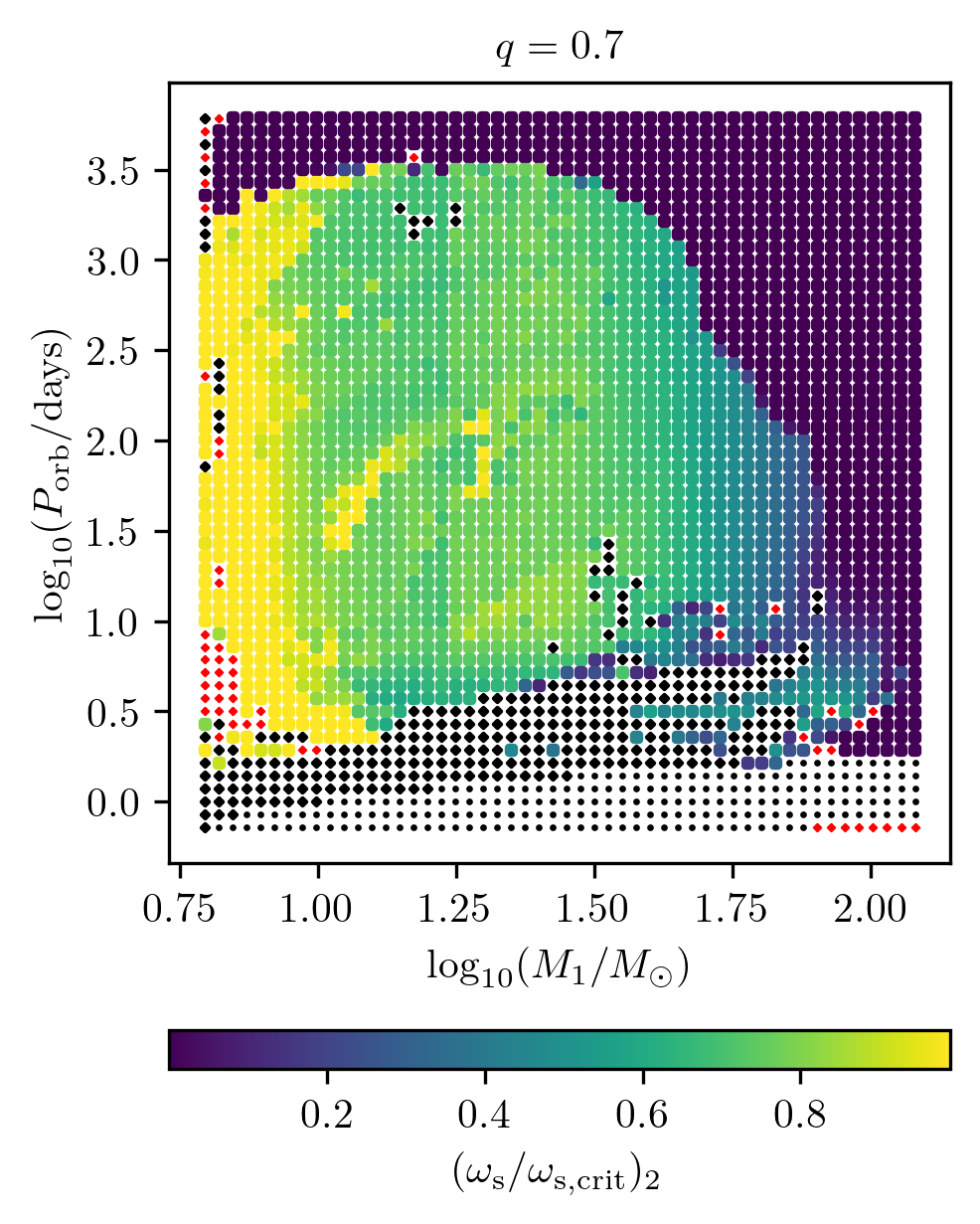}
\caption{For the same grid slices shown in Figure~\ref{fig:HMS-HMS_MESA_grid_TF12}, and only for systems where one of the two stars  reached the end of its life,  we depict the final ratio of the angular velocity of the secondary star (the initially less massive) divided by its critical rotation rate, $(\omega_{\rm s}/\omega_{\rm s,crit})_2$. 
In most cases where mass transfer occurred, the secondary star accreted mass and spun up, remaining highly spinning until the end of the life of the initially more massive star.  
}
\label{fig:HMS-HMS_MESA_grid_TF1}
\end{figure*}

For modeling the evolution of two ZAMS stars in a binary system, we run a grid of 58,240 separate binary evolution models, varying the initial mass of the primary star $M_1$, the initial binary mass ratio $q=M_2/M_1$ (where $M_2$ is the mass of the companion star), and orbital period $P_{\rm orb}$. 
We consider $52$ values of initial primary masses, ranging from $M_{1}=6.23\,\Msun$ to $M_{1}=120\,\Msun$ with a logarithmic spacing of $\Delta \log_{10}( M_{1}/\Msun) = 0.025$\,dex, and $20$ values of initial binary mass ratios, ranging from $q=0.05$ to $q=1$ with a spacing of $\Delta q = 0.05$. 
Finally, we cover $56$ values of initial orbital period, ranging from $P_{\rm orb}=0.7$\,days to $P_{\rm orb}=6105$\,days with a logarithmic spacing of $\Delta \log_{10} (P_{\rm orb}/{\rm days}) = 0.07$\,dex, in order to explore all binary configurations ranging from close systems in initial RLO to wide systems that never exchange any mass.

We simulate binaries by first separately initializing two H-rich, single stars at ZAMS following the procedure defined in Section~\ref{sec:zams}. 
We then place those stars in a binary with a second relaxation step, where we force their  their rotation periods to be synchronized with the orbital period, implicitly assuming that the synchronization has happened during the pre-MS phase. The latter might not be true for wide binaries, but our assumption induces negligible rotation to the stellar components of those systems and does not affect their further evolution.
As long as both stars in the binary are under-filling their Roche lobes after this relaxation step, we start to evolve the binary. 
Evolution continues until one of the  termination conditions described in Section~\ref{sec:terminated} occurs.

In Figure~\ref{fig:HMS-HMS_MESA_grid_TF12} we provide two two-dimensional slices of this grid, where we show our simulation outcomes as a function of $M_1$ and $P_{\rm orb}$ for fixed $q$ values.
In the left panel we show one example of a mass ratio $q=0.3$, and on the right a more-equal mass $q=0.7$ slice. 
Each point in the panels represents a separate simulation from our grid. 
Diamond markers represent models that terminated in a CE, while square markers represent models that terminated when one of the stars completed its evolution (e.g., reached core C exhaustion).
These are systems that experienced either only stable mass-transfer episodes or no mass transfer at all, so their evolution can be continuously modeled.   
At the bottom of each panel we see systems that are born filling their Roche lobes (black dots). 
These systems are assumed to merge, and therefore never produce a viable binary. Finally, a small fraction of systems never complete their evolution, producing binary stellar models that at some point fail to converge (red diamonds).

Separately, the color of each marker indicates that particular binary's mass transfer history. 
Systems with sufficiently close initial $P_{\rm orb}$ tend to lead to contact phases (orange) where both stars fill their Roche lobes simultaneously. 
Most, but not all, of these system end up entering a CE phase. 
Sufficiently widely separated (or very massive) systems never fill their Roche lobes, and therefore never interact (gray markers). For intermediate orbital periods, the colors differentiate the evolutionary state of the donor when the latest mass transfer phase was initiated, ranging from MS (blue) to post-MS (tan), to stripped He-MS (brown). Stable mass transfer causes the donor star to be almost completely stripped of its H-rich envelope. In the latter case (brown) the low-mass stripped donors initiate a second mass transfer phase (Case BB mass-transfer) when they re-expand \citep{1981A&A....96..142D, 2020A&A...637A...6L}.

\begin{figure*}[t]\center
\includegraphics{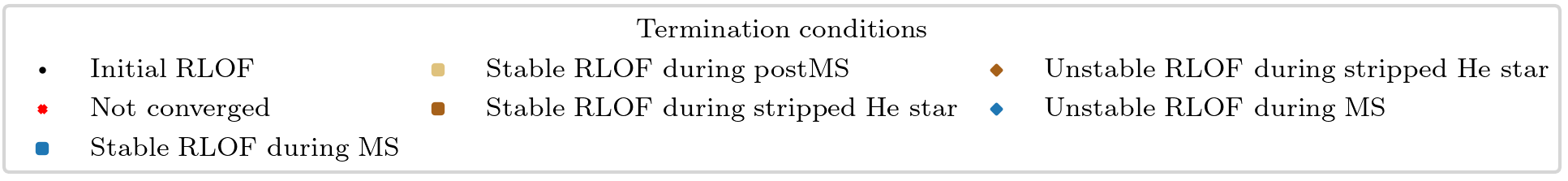}\\
\includegraphics{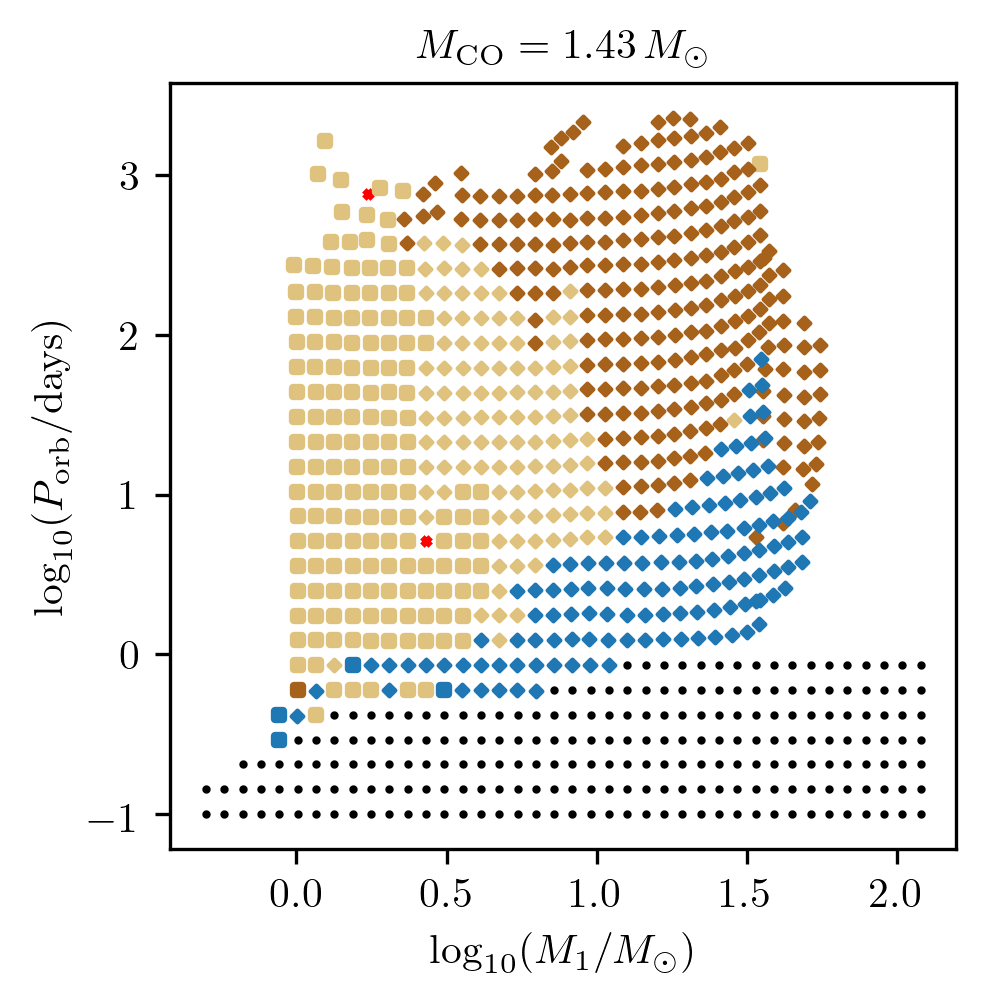}
\includegraphics{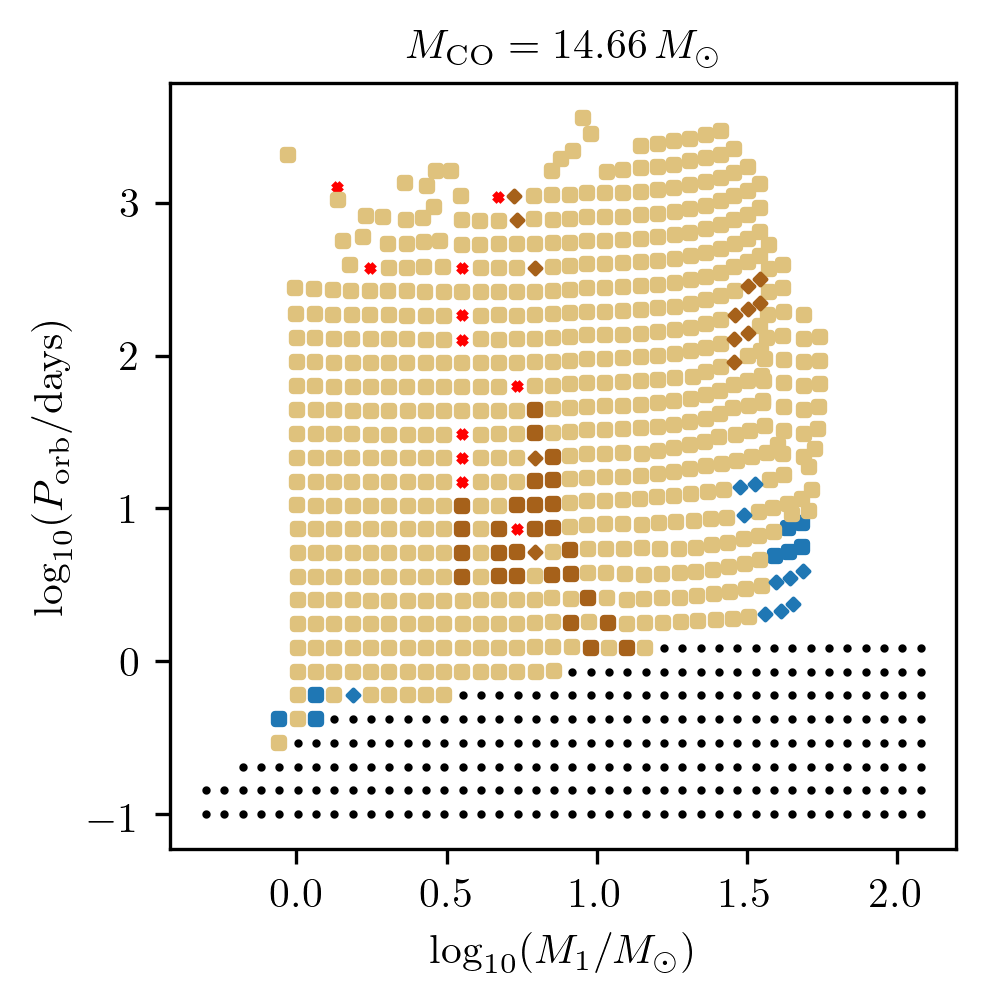}
\caption{View of two slices, for two different values of initial CO masses ($M_{\rm CO}=1.43\,\rm M_{\odot}$ on the left, $M_{\rm CO}=14.66\,\rm M_{\odot}$ on the right),  from our grid of binary-star models consisting of H-rich star and a CO at the onset of RLO. 
The different symbols summarize the evolution of each of the models, as in Figure~\ref{fig:HMS-HMS_MESA_grid_TF12}. 
Binaries that never initiated mass transfer are not shown here.}

\label{fig:CO-HMS_MESA_grid_TF12}
\end{figure*}

Comparison between the two panels shows that the mass ratio leads to a stark difference in the mass-transfer outcomes. 
Whereas nearly all systems with an initial $q=0.7$ result in stable mass transfer, the opposite is true for our $q=0.3$ systems.
At the same time, some features between the two mass ratios are similar: \emph{(i)} The boundary between interacting and non-interacting systems seems to be insensitive to $q$ (and therefore the secondary's mass). 
At the largest orbital periods, stars do not expand far enough to overfill their Roche lobes. 
At the largest masses, stars have extremely strong winds that widen their orbits, simultaneously stripping the primary of its H-rich envelope, and these stars never expand enough to fill their Roche lobes. 
\emph{(ii)} Systems with initial $P_{\rm orb}\lesssim5$ days tend to result in dynamically unstable mass transfer. 
\emph{(iii)} There is a large region of binaries with initial primary mass $\simeq$40--50 \Msun\ that stably overfill their Roche lobes as post-MS stars.
These stars achieve their mass transfer stability mainly due to their strong stellar winds, which increases the mass ratio and the orbit of the system until the moment of overflow.

We model, and keep track of, the properties of both stars in the binary system throughout their evolution, as well as their detailed internal structure at the end of the models. 
In Figure~\ref{fig:HMS-HMS_MESA_grid_TF1} we show, for the same two mass-ratio slices as in Figure~\ref{fig:HMS-HMS_MESA_grid_TF12}, the final rotational rate of the secondary (the initially less massive) stars for systems that avoid dynamically unstable mass transfer. 
Each marker's color is set by how close each star's rotation rate is to its critical rate.
Highly rotating secondary stars have all experienced substantial mass and angular-momentum accretion during their evolution. 
Many of them have reached critical rotation, $(\omega_{\rm s}/\omega_{\rm s,crit})_2=1$, early during mass transfer, at which point further mass accretion becomes non-conservative (c.f.\ Section~\ref{sec:MT_non_degenerate}). 
The right-hand panel shows that the companion's rotation rate is closely linked with $M_1$, as companion stars with lower mass primary stars also have lower masses and therefore do not lose as much angular momentum through their own stellar winds. 
This behavior is independent of the assumed initial rotation of the stars.

We find a small subset of initially very close systems in the bottom right corner (log$_{10} (M_1/\Msun)> 1.75$ and log$_{10} (P_{\rm orb}/{\rm days}) \simeq 0.5$) that retain a significant rotational rate even though they avoid mass transfer. 
In binaries with such tight orbits, tidal forces between the stars are sufficiently strong to keep them fast rotating, despite their strong winds.

\subsection{Binaries consisting of a compact object and a hydrogen-rich star, at the onset of Roche-lobe overflow \label{sec:CO-HMS_RLO}}

\begin{figure*}[t]\center
\includegraphics{figures/TF1_legend.png}\\
\includegraphics{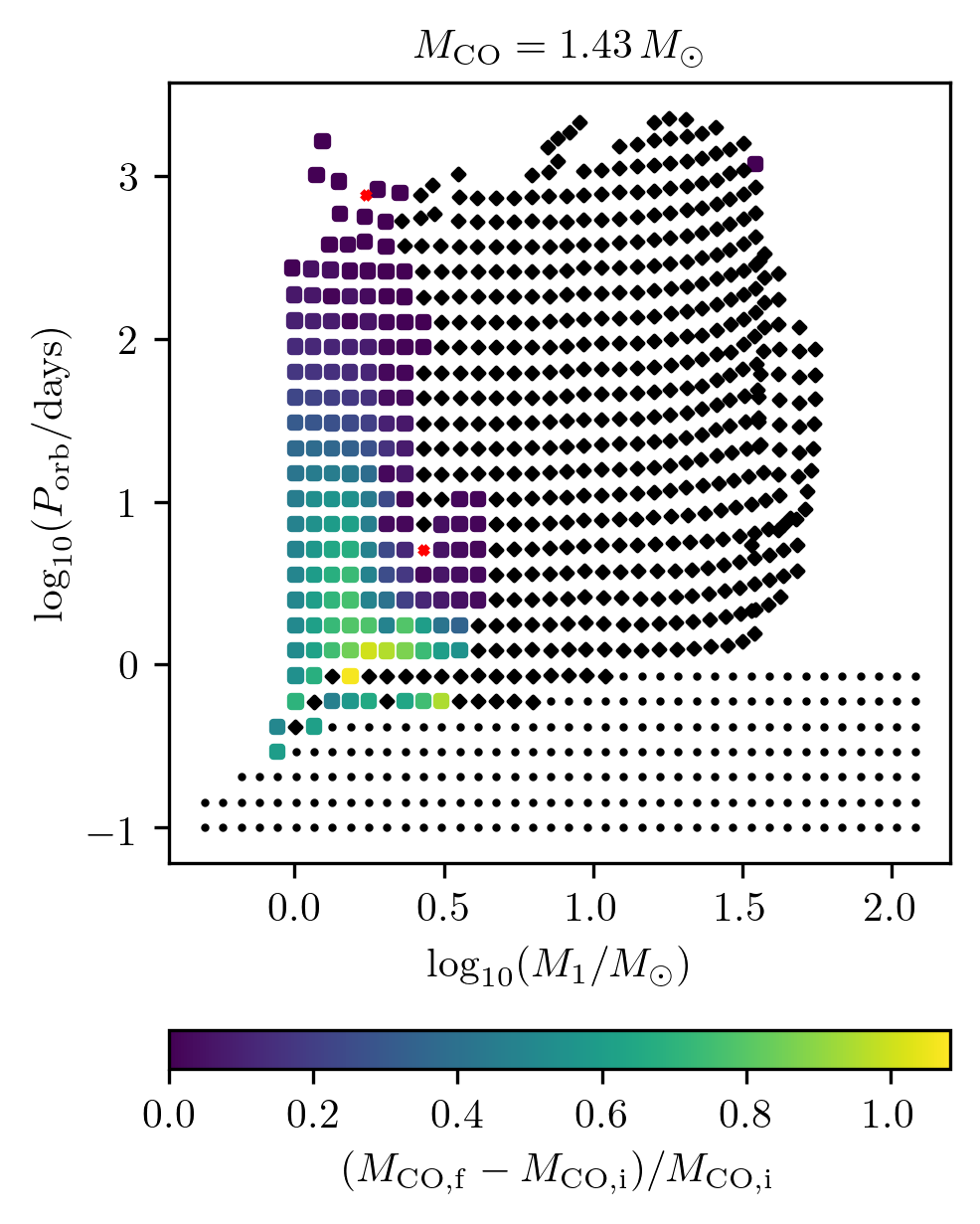}
\includegraphics{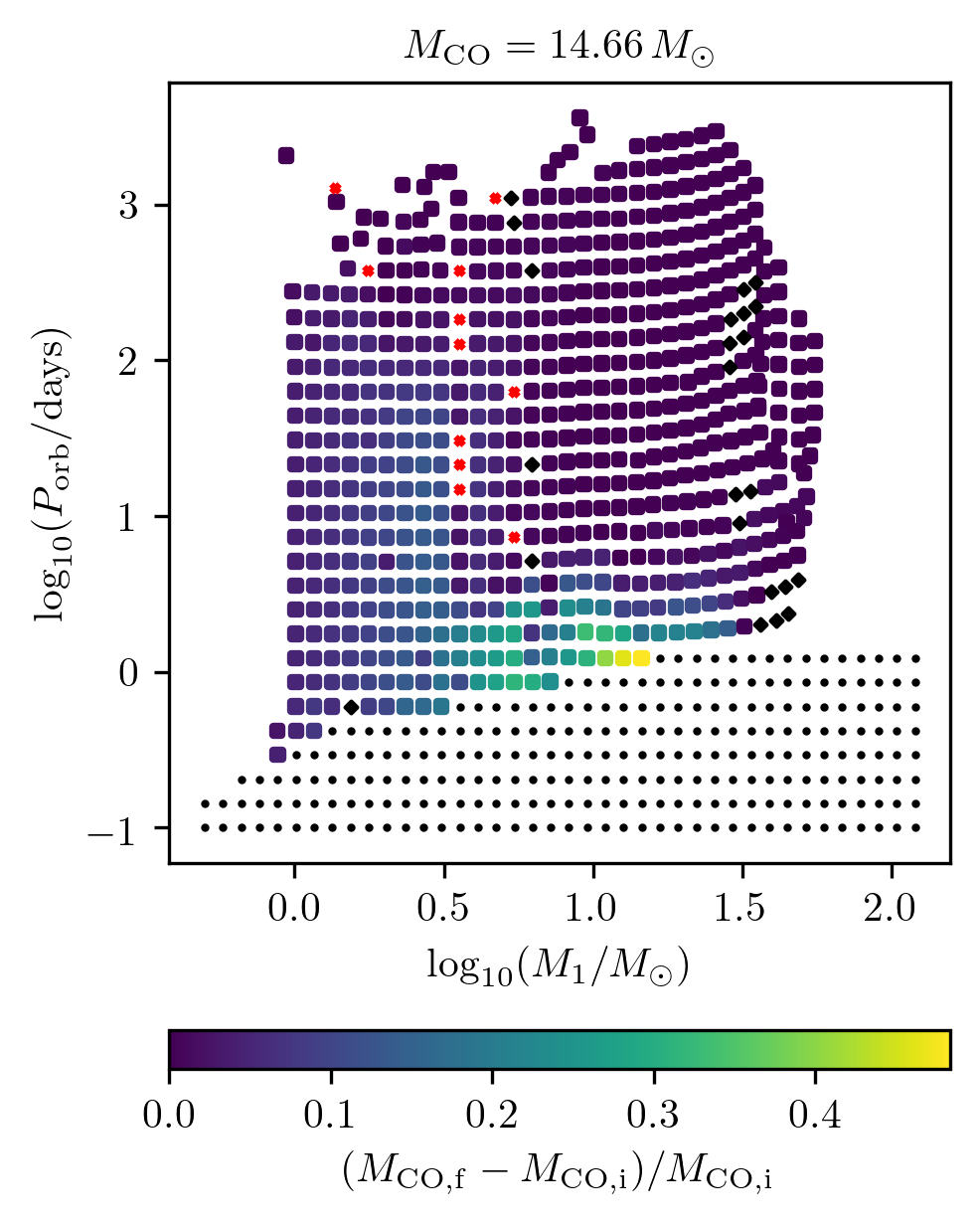}
\caption{Relative increase in the mass of the CO $(M_{\rm CO,f}-M_{\rm CO,i})/M_{\rm CO,i}$ due to accretion for systems where the non-degenerate star reached the end of its life. 
The grid slices are the same as shown in Figure~\ref{fig:CO-HMS_MESA_grid_TF12}. 
Although accretion is Eddington-limited, COs in binaries with pre-RLO mass ratios in the range $q\sim 1$--$2$ (defining $q=M_{\rm CO,i}/M_{\rm 1,i}$) and short initial periods, which will experience long-duration mass-transfer phases, manage to accrete a significant amount of mass.}

\label{fig:CO-HMS_MESA_grid_TF1_relative_mass}
\end{figure*}

\begin{figure*}[t]\center
\includegraphics{figures/TF1_legend.png}\\
\includegraphics{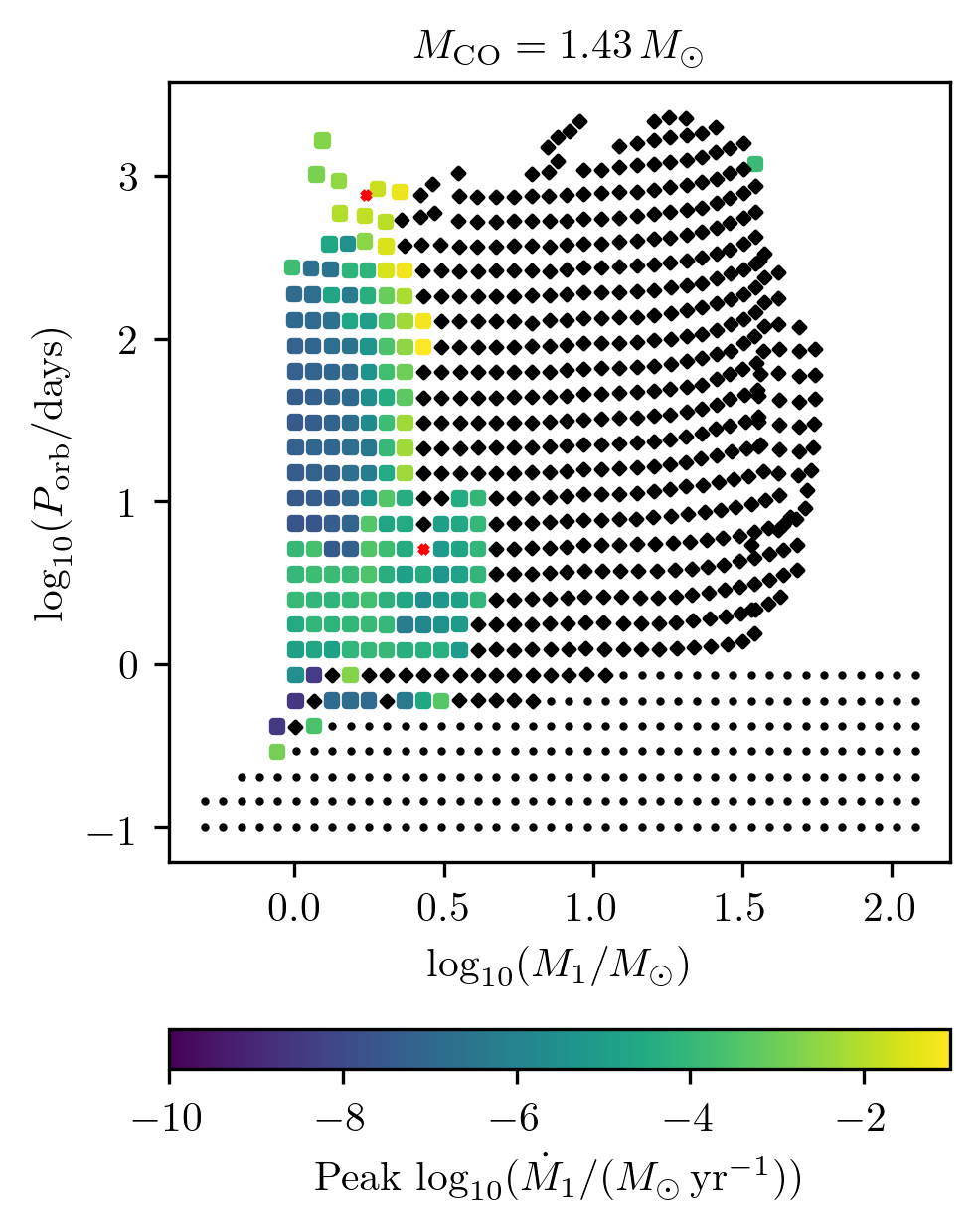}
\includegraphics{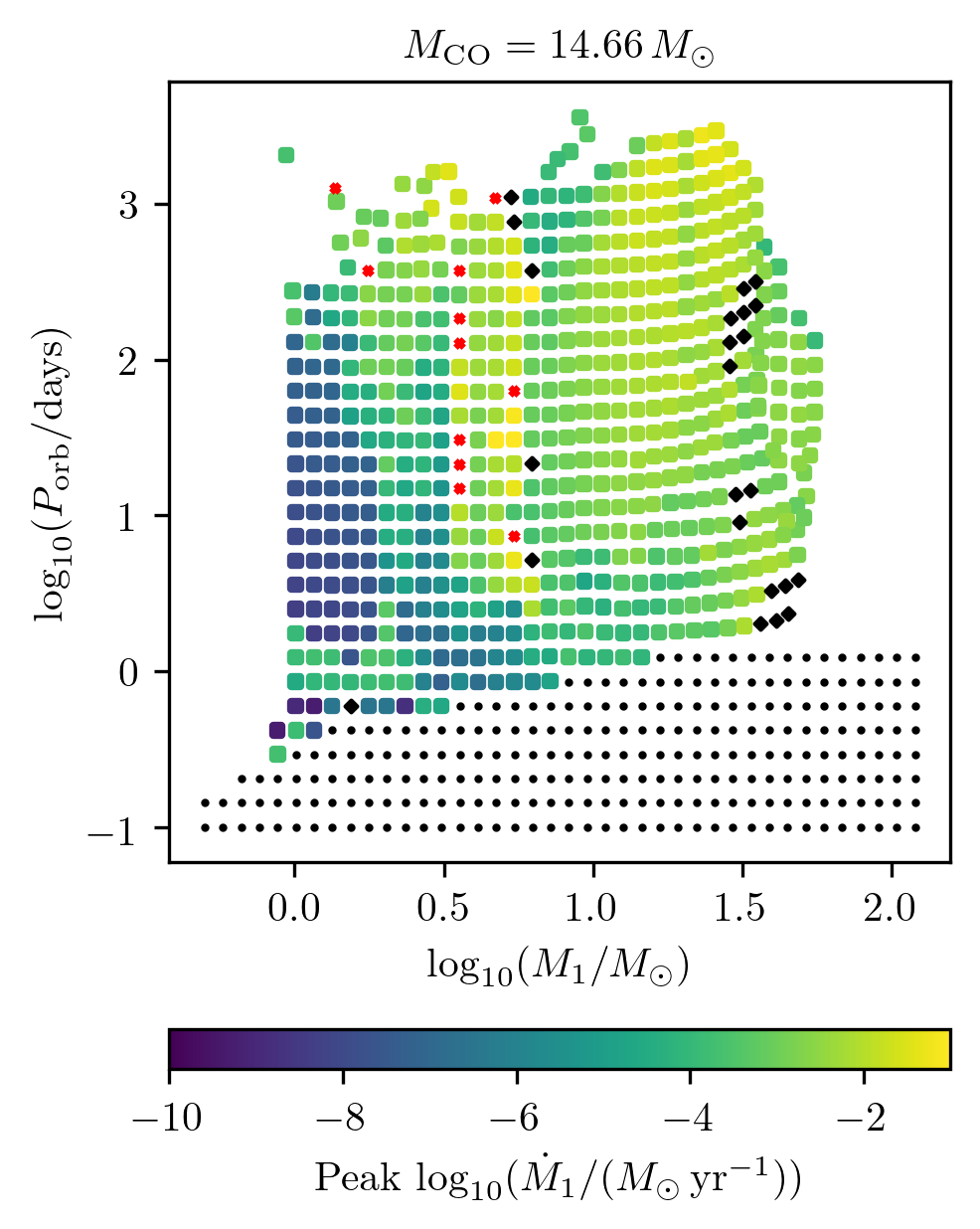}
\caption{Same as Figure~\ref{fig:CO-HMS_MESA_grid_TF1_relative_mass}, but now the color of the symbols depict the maximum mass-transfer rate that occurred in the evolution of each binary. 
A significant part of the parameter space leads to highly super-Eddington mass-transfer rates, albeit in short-lived phases for most cases, and thus to the potential formation of ultra-luminous X-ray sources. This peak mass-transfer rate refers to the rate the donor star is losing mass through RLO; accretion onto the accretor is limited to the Eddington rate.}
\label{fig:CO-HMS_MESA_grid_TF1_max_mt}
\end{figure*}

Our second grid of binary star simulations consists of a H-rich star in a binary with a CO at the onset of RLO. 
This grid consists of 25,200 binary evolution models, where we vary the initial mass of the primary star $M_1$, the initial mass of the CO $M_{\rm CO}$, and the orbital period $P_{\rm orb}$. 
We consider $40$ values of initial primary masses, ranging from $M_{1}=0.5\,\Msun$ to $M_{1}=120\,\Msun$ with a logarithmic spacing of $\Delta \log_{10} (M_{1}/\Msun) = 0.06$\,dex, and $21$ values of initial CO masses, ranging from $M_{\rm CO}=1\,\Msun$ to $M_{\rm CO}=35.88\,\Msun$ with a logarithmic spacing of $\Delta \log_{10} (M_{\rm CO}/\Msun) = 0.074$\,dex. 
Finally, we cover $30$ values of initial orbital period, ranging from $P_{\rm orb}=1.26$\,days to $P_{\rm orb}=3162$\,days with a logarithmic spacing of $\Delta \log_{10} (P_{\rm orb}/{\rm days}) = 0.13$\,dex. 
Our choice of CO mass range covers massive WD, NS, and BH accretors.

Our procedure in constructing this grid is different from what was described in Section~\ref{sec:HMS-HMS}. 
We start each of the simulations with binaries composed of a ZAMS H-rich star and a CO, which in the \mesa{} code is approximated by a point mass. 
Initially, and until each of the binary models reach the onset of RLO, we neglect orbital angular-momentum loss mechanisms, such as tides, magnetic breaking and gravitational radiation, while we artificially enforce the synchronization of the non-degenerate star with the orbit at all times. 
We do, however, allow for wind mass-loss from the non-degenerate star, which also results to  a widening of the orbit. 
Once the onset of RLO is reached, we include the effects of all orbital angular-momentum loss mechanisms and discard the prior evolution of the system, treating the onset of RLO as the effective starting point of our models. 
Furthermore, from that point onward, we do not artificially enforce the synchronization of the non-degenerate star's spin rotation with the orbit, but we instead follow the tidal synchronization process self-consistently, following the prescriptions described in Section~\ref{sec:Tides}. 
Finally, binaries that never reach the onset of RLO are not considered further; these detached binaries are modeled separately as described in Section~\ref{sec:detached}. 
There, we also provide a full explanation of how we use this binary-star grid, composed of a H-rich star and a CO at the onset of RLO, within a larger infrastructure to completely evolve binaries from ZAMS to double COs.

Figure~\ref{fig:CO-HMS_MESA_grid_TF12} shows two slices of the grid with different CO masses, $M_{\rm CO}=1.43$\,M$_{\odot}$ to represent a NS accretor and $M_{\rm CO}=14.66$\,M$_{\odot}$ to represent a more-massive BH accretor. 
The symbols depicted in Fig.~\ref{fig:CO-HMS_MESA_grid_TF12} have the same meaning as in Figure~\ref{fig:HMS-HMS_MESA_grid_TF12}. 
Although our true initial binary parameters are regularly spaced, $M_1$ and $P_{\rm orb}$ on the axes shown in Figure~\ref{fig:CO-HMS_MESA_grid_TF12} are the binary's quantities at the onset of RLO, the effective starting point of the models; therefore, the grid does not appear to be regularly spaced (strong winds exhibited by massive stars tend to expand binary orbits prior to mass transfer). 
We do not show those binaries that never interact (even though we ran these simulations). As already seen in the binary-star model grid composed of two H-rich stars  (Figure~\ref{fig:HMS-HMS_MESA_grid_TF12}), binaries too widely separated will never overfill their Roche lobes, and binaries with massive H-rich stars have winds too strong to expand into giant phases. 
In this grid, Figure~\ref{fig:CO-HMS_MESA_grid_TF12} shows an additional region of white space at low mass ($M\lesssim1$\,\Msun) that occurs because these stars remain on the MS for the entirety of the simulation, never expanding to fill their Roche lobes within the age of the Universe.

Examining the stability of the mass-transfer phase, Figure~\ref{fig:CO-HMS_MESA_grid_TF12} shows that nearly every donor star accreting onto a $14.66$\,M$_{\odot}$ BH does so stably, whereas only the lower mass accretors ($M\lesssim$4.5\,\Msun) do so for NS accretors. 
This difference is because the stability of a mass transfer in a binary primarily depends on the mass ratio, with a higher accretor mass allowing for higher donor masses. 
Our findings, at least for the case of NSs, are consistent with recent results from \cite{2020A&A...642A.174M}, who use the same criteria to define the onset of L$_2$ overflow leading to dynamical instability as done in our work.

\begin{figure*}[t]\center
\includegraphics{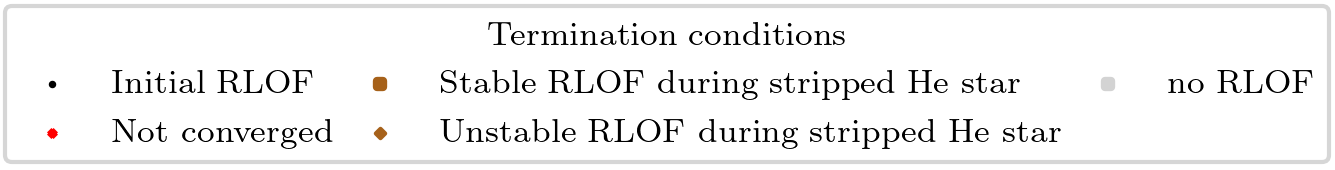}\\
\includegraphics{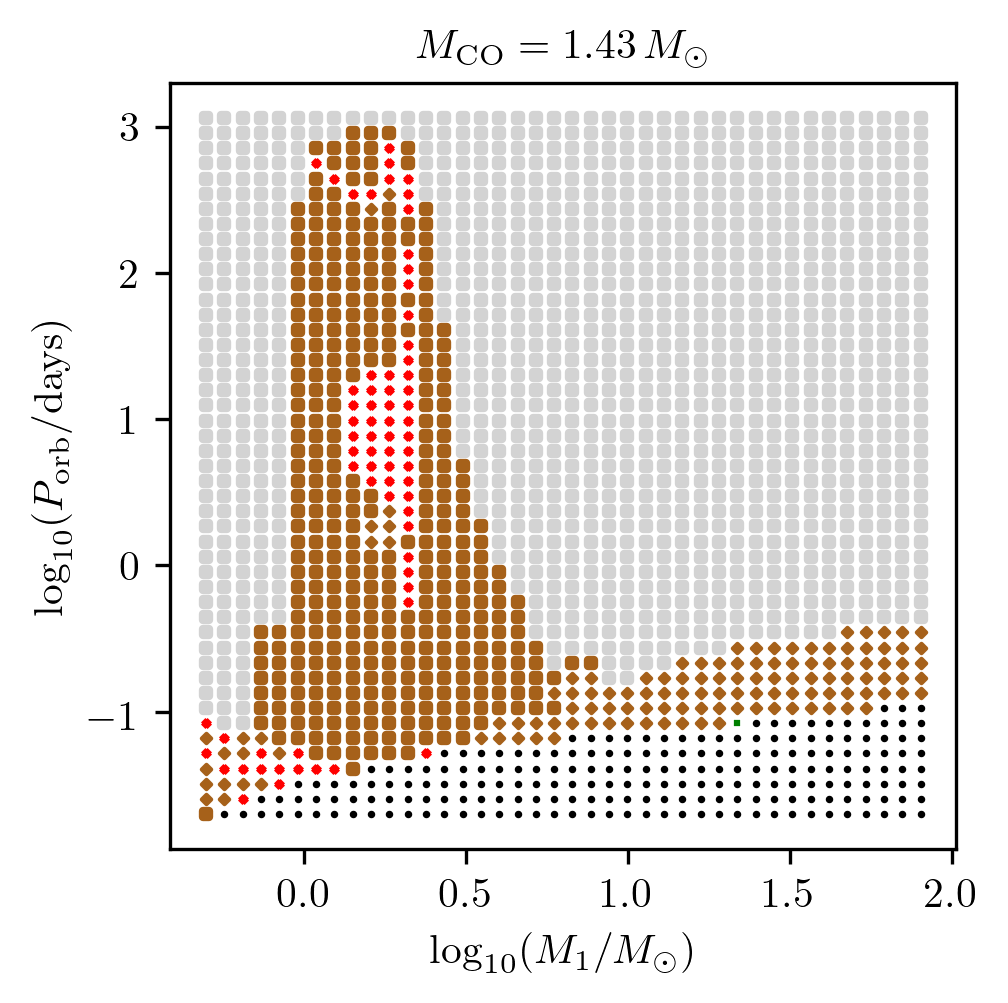}
\includegraphics{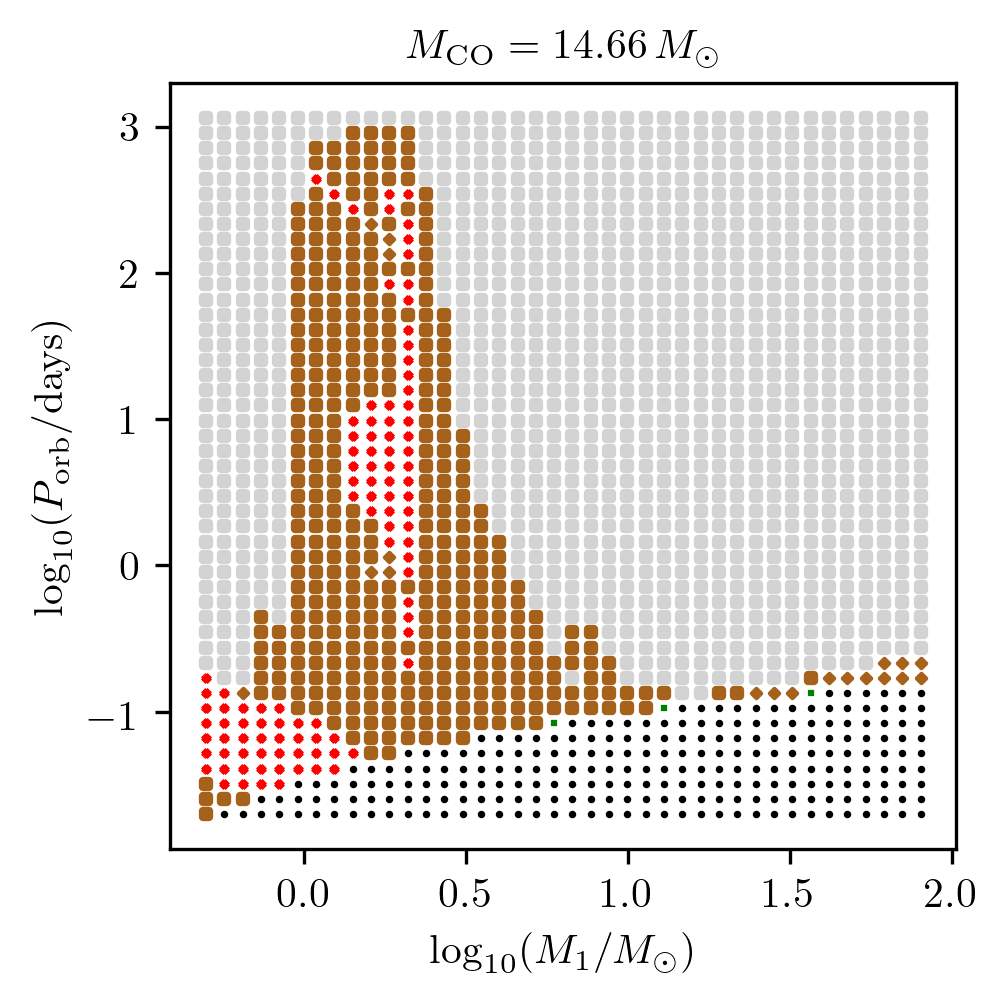}
\caption{
View of two grid slices for two different values of initial CO masses ($M_{\rm CO}=1.43\,\rm M_{\odot}$ on the left, $M_{\rm CO}=14.66\,\rm M_{\odot}$ on the right), from our grid of binary-star models consisting of He-rich stars and CO companions. 
The different symbols summarize the evolution of each of the models, as in Figure~\ref{fig:HMS-HMS_MESA_grid_TF12}. 
}
\label{fig:CO-HeMS_MESA_grid_TF12}
\end{figure*}

Figure~\ref{fig:CO-HMS_MESA_grid_TF1_relative_mass} shows the relative changes in the accretor masses in the same two slices in $M_{\rm CO}$ as Figure~\ref{fig:CO-HeMS_MESA_grid_TF12}.  
High amounts of accretion mainly depends on two factors: a sufficiently high-mass accretion rate and a long-lasting RLO phase. 
In both panels, this happens for binaries with short periods $\simeq1$~day, and  pre-RLO mass ratios in the range $q\sim 1$--$2$ (defining $q=M_{\rm CO,i}/M_{\rm 1,i}$). 
Despite our assumption of Eddington-limited accretion, for these binaries, stable accretion occurs for over a long time, and in both cases the binaries transition to low-mass X-ray binaries. 
These findings are in agreement with earlier works by \citet{2003MNRAS.341..385P,2015ApJ...800...17F,2020A&A...642A.174M}.

The high mass-transfer rates achieved by most initial binary configurations are explicitly shown in Figure~\ref{fig:CO-HMS_MESA_grid_TF1_max_mt}, where each marker's color corresponds to the peak mass-transfer rate for each binary. These rates refer to the  mass being lost by the donor star due to RLO; accretion onto the accretor is still Eddington-limited.
In both panels, super-Eddington mass-transfer rates occur in most binaries with higher peak mass-transfer rates encounters in binaries with higher periods and larger donor star masses. 
However, since the larger orbital separation of these binaries implies the donors in these systems would be more evolved at RLO onset, compared with initially shorter-period binaries, these mass transfer phases tend to be short-lived. 
Therefore, binaries with short orbital periods (but not so short that they overfill their Roche lobes initially) will lead to the most accretion onto a CO.

\subsection{Binaries consisting of a compact object and a He-rich star \label{sec:CO-HeMS}}

\begin{figure*}[t]\center
\includegraphics{figures/TF1_legend.png}\\
\includegraphics{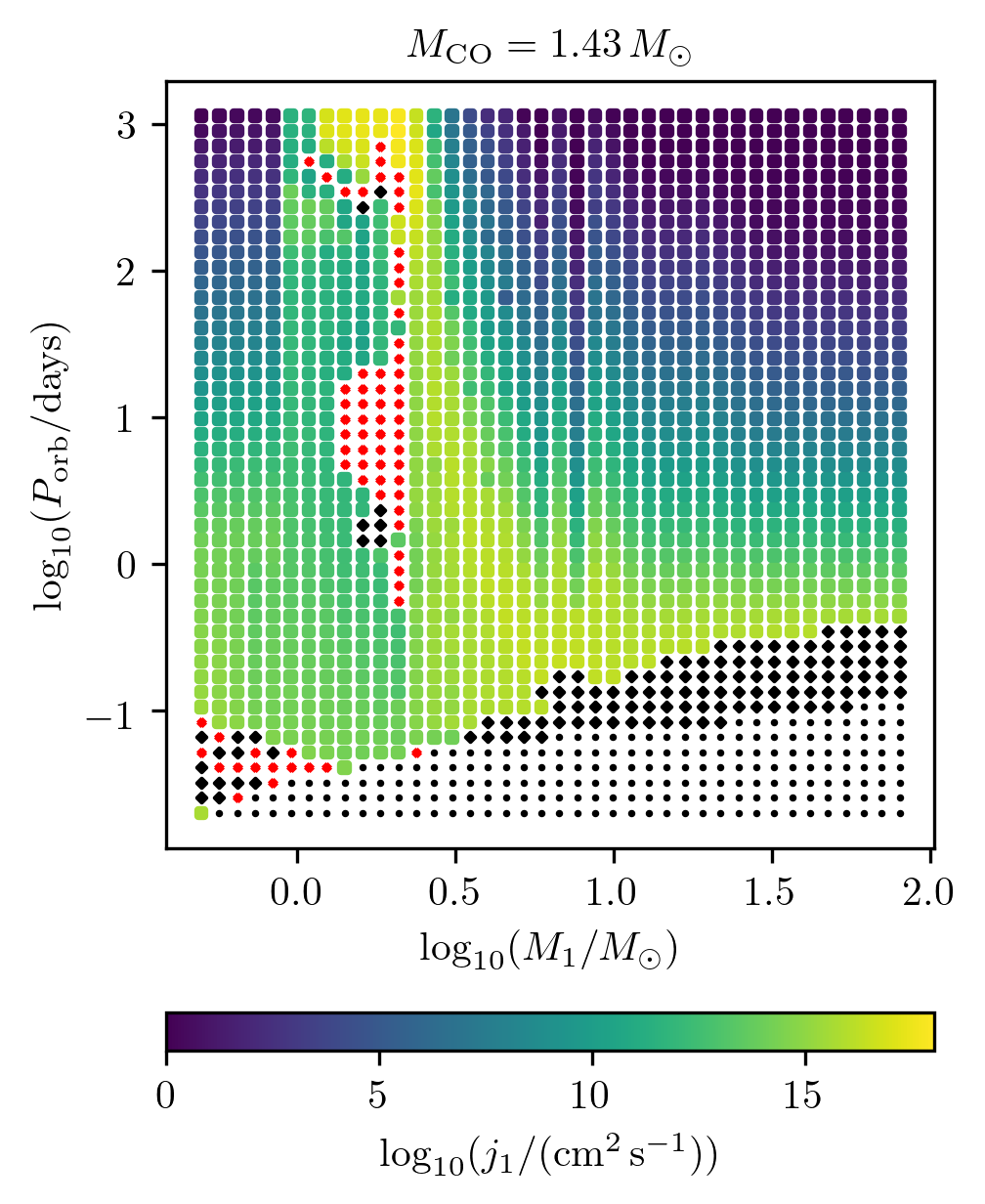}
\includegraphics{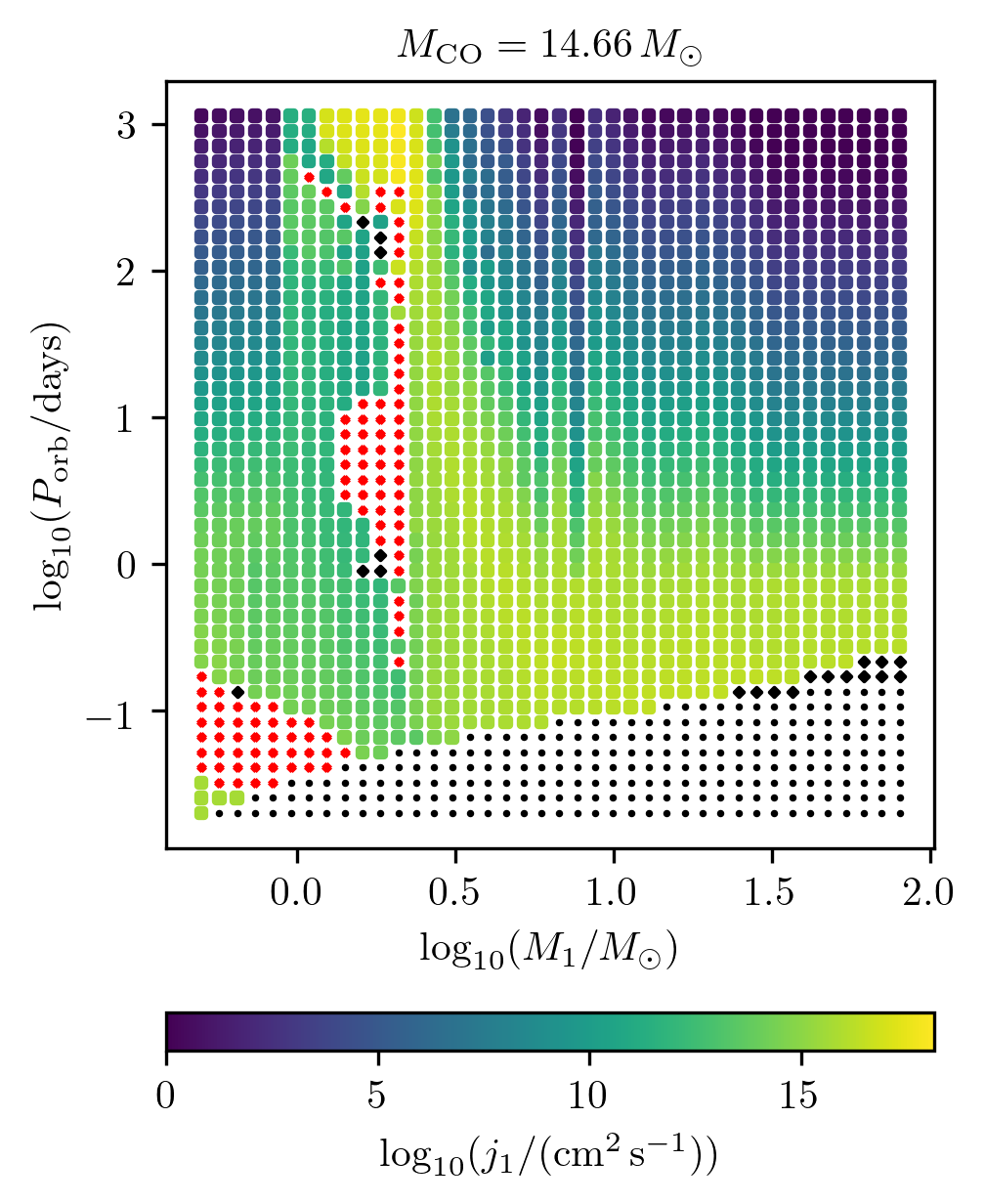}
\caption{The final specific angular momentum $j_1 = J_1/M_1$ where $J_1$ is the He-star AM and $M_1$ its mass, at carbon depletion for our grid of He-stars with CO companions. We only show $j_1$ for systems where the non-degenerate star reached the end of its life. 
The grid slices are the same as shown in Figure~\ref{fig:CO-HeMS_MESA_grid_TF12}.}
\label{fig:CO-HeMS_MESA_grid_TF1}
\end{figure*}

Our final grid of detailed binary-star simulations consists of 39,480 models of He-rich stars with CO companions, where we vary the initial mass of the primary star $M_1$, the initial mass of the CO $M_{\rm CO}$, and the orbital period $P_{\rm orb}$. 
We consider $40$ values of initial primary masses, ranging from $M_{1}=0.5\,\Msun$ to $M_{1}=80\,\Msun$ with a logarithmic spacing of $\Delta \log_{10} (M_{1}/\Msun) = 0.055$\,dex, and 21 values of initial CO masses, ranging from $M_{\rm CO}=1\,\Msun$ to $M_{\rm CO}=35.88\,\Msun$ with a logarithmic spacing of $\Delta \log_{10} (M_{\rm CO}/\Msun) = 0.074$\,dex. 
Finally, we cover $47$ values of initial orbital period, ranging from $P_{\rm orb}=0.02$\,days to $P_{\rm orb}=1117.2$\,days with a logarithmic spacing of $\Delta \log_{10} (P_{\rm orb}/{\rm days}) = 0.09$\,dex. 
Our procedure for generating these binaries closely follows the process described in Section~\ref{sec:HMS-HMS} for the grid of binary-star modes composed of two H-rich stars. Here, we replace the initial primary-star models with He-rich stars at ZAHeMS, while the companion COs are modeled as point masses.

Figure~\ref{fig:CO-HeMS_MESA_grid_TF12} shows an example of two slices of this grid, one corresponding to a NS companion (with $M_\mathrm{CO}\simeq1.43\,\Msun$) and one corresponding to a BH (with a $M_\mathrm{CO}\simeq14.66\,\Msun$). 
Marker shapes and color scheme follow the same convention as in Figure~\ref{fig:CO-HMS_MESA_grid_TF12}, but since these simulations are initialized with He-stars, the symbol key is simplified in Figure~\ref{fig:CO-HeMS_MESA_grid_TF12}.

When comparing the two panels, the most apparent difference occurs at large $M_{1}$ and short orbital period: whereas accreting NSs enter unstable mass transfer (these systems typically end up merging in a CE, cf.\ Section~\ref{sec:populations}), 
the corresponding accreting BHs typically either overfill their Roche lobes at ZAMS or avoid mass transfer altogether. 
In contrast, we find that independently of the CO mass, systems with low He-star masses ($M_\mathrm{1} \leq 3\,\Msun$) mass transfer up to wide orbital periods ($P_{\rm orb} < 10^3 \, \mathrm{days}$). 
This occurs because low-mass He-stars expand their He-rich envelope much farther during their later He-shell and C-burning phases (Figure~\ref{fig:He_radius}).

Both slices of the grid present two islands of failed simulations, one with $M_\mathrm{1} \simeq 1.8\,\Msun$ and $P_{\rm orb}$ of the order of days and another island with $M_\mathrm{1} \lesssim 1\,\Msun$ and $P_{\rm orb}$  of the order of hours. 
\mesa{} has difficulty modeling the envelope's structure as it expands to large radii in the first island, whereas the second, short-$P_{\rm orb}$ island is due to \mesa{} having difficulty following a star's evolution into a He WD after it has been spun up due to tides and mass-transfer. 
Combined, failed runs account for $\simeq 5$\% of the models in this grid.
In practice we find these failed runs do not bias our population synthesis results of merging NSs and BHs as these portions of the parameter space predominantly lead to the formation of WDs. 

In Figure~\ref{fig:CO-HeMS_MESA_grid_TF1} we show the same two grid slices, but now the marker color corresponds to the specific angular momentum of the He-star $j_{1}$, at the end of the simulation.
\mesa{} allows us to track this quantity, as it self-consistently models the interplay between tides (which spin up the star), stellar winds (which spin down the star and widen the binary), mass transfer (which alters the orbital period), and internal angular momentum transport. 
Comparing Figure~\ref{fig:CO-HeMS_MESA_grid_TF12} and Figure~\ref{fig:CO-HeMS_MESA_grid_TF1}, we find that the He-stars with the highest specific angular momenta are those with either short $P_{\rm orb}$ or stable mass transfer.

The binary-star grid, composed of a He-rich star and a CO companion, presented in this section closely agree with those of \citet{2018A&A...616A..28Q} and \citet{2020A&A...635A..97B,2021A&A...647A.153B}. 
In contrast to these previous works, the present grid further expands the parameter space coverage to lower He-star masses and to larger orbital periods.

\section{Grid Post-Processing \label{sec:postprocessing}}

Each single- or binary-star evolution simulation produces a series of data files which must be parsed, analyzed, and collated before we can use them within \posydon{}. 
Our process includes: (1) re-running any failed simulations; (2) adding post-processed quantities to our data grids; (3) a post-processing procedure used exclusively on our single, H-rich and He-rich star grids, which allows for an efficient interpolation among tracks of different masses; (4) the downsampling of our grids to reduce data size; (5) classifying each model within our grids based on the different resulting stellar and binary types, and (6) fitting classifiers and interpolators over the stellar and binary parameters in each grid. 
We describe the first 4 steps next, while the steps of classification and interpolation are discussed in Section~\ref{sec:machine_learning}.

\subsection{Re-running Failed Models} \label{sec:rerun}

\begin{figure}[t]\center
\includegraphics[width=\columnwidth]{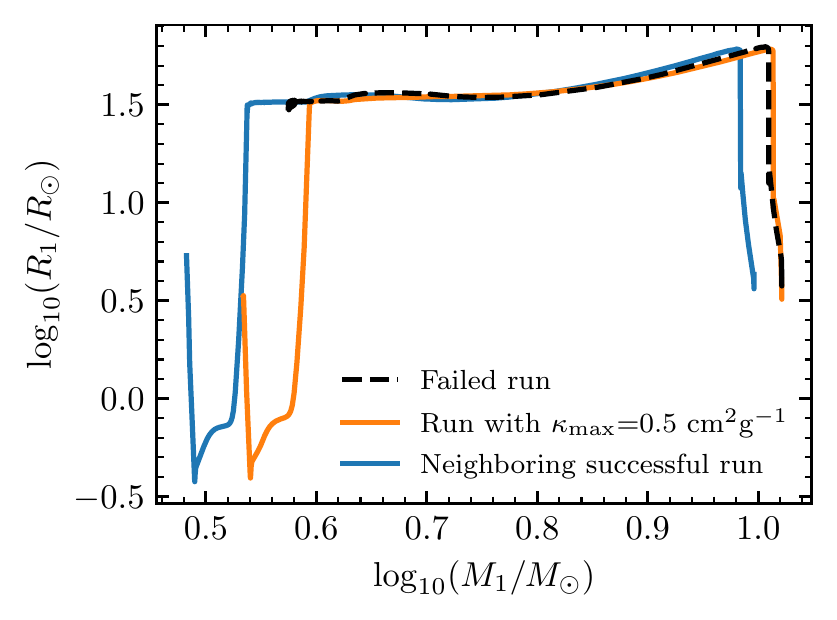}
\includegraphics[width=\columnwidth]{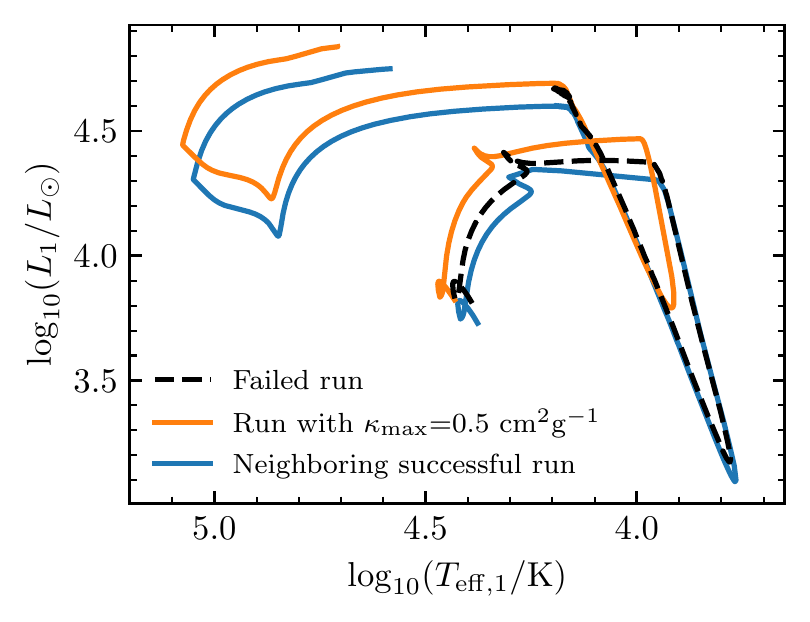}
\caption{Typical example of the evolution of a binary-star model that failed to reach the end of the simulation due to over-resolved stellar oscillations that eventually lead \mesa{} to convergence problems. 
We show the evolution of the primary's radius, as a function of its mass (top panel) and its track in the Hertzsprung--Russell diagram (bottom panel). 
The binary initially consists of a 10.50\,M$_\odot$ and a 5.25\,M$_\odot$ H-rich ZAMS stars at an orbital period of 43.94\,days. 
Comparison between our original, failed simulation (dashed, black line) and a successful simulation where we artificially limit the radiative opacity to $0.5\,\mathrm{cm}^2/\mathrm{g}$ (orange line) shows that our approximation is typically accurate to within 0.1 dex for all stellar parameters. We additionally show the evolution of an adjacent simulation in our binary grid (blue line; same mass ratio and orbital period but a primary mass of 9.91\,$\Msun$), which shows that any inaccuracy induced by our opacity approximation is a similar magnitude to the differences between neighboring simulations.}
\label{fig:opacity_fix}
\end{figure}

After having computed our grids of single- and binary-star models, we first identify those runs that did not reach our desired end point (cf.\ Section~\ref{sec:terminated}). 
This can happen for a variety of reasons, many of which we have not yet been able to eliminate. 
For example, one source of problematic runs appears to deal with stellar oscillations; in certain cases, \mesa{} tries to resolve short-timescale evolution driven by the $\kappa$-mechanism, which dramatically shortens the size of successive \mesa{} steps. 
We address this problem by re-running our failed binary simulations with a maximum radiative opacity ($\kappa_{\mathrm{max}}$) set to $0.5\,\mathrm{cm}^2\,\mathrm{g^{-1}}$.
This approximation reduces the failure rate of each binary grid from $\simeq10.9\%$, $\simeq8.0\%$, and $\simeq11.8\%$, for the binary-star grids composed of two H-rich stars, a H-rich star with a CO companion at the onset of RLO, and a He-rich star with a CO companion, respectively, to $\simeq$0.9\%, $\simeq$1.5\%, and $\simeq$4.8\%. 
The differences in the resulting evolutionary tracks with and without the opacity limit are generally small when compared to differences in tracks of adjacent points in our initial parameter space and compared to our interpolation accuracy (Section~\ref{sec:interpolation_accuracy}).

Figure~\ref{fig:opacity_fix} shows a typical example of a binary-star model, initially composed of two H-rich ZAMS stars with masses 10.50\,M$_\odot$ and 5.25\,M$_\odot$ and an orbital period of 43.94\,days. This binary initially failed to reach the end of the simulation (dashed, black line; \mesa{} exceeded its minimum timestep limit), but did so successfully when re-run with an upper limit to the radiative opacity (orange line). 
The top panel shows that the stellar radius evolves similarly between the two simulations as the donor star loses mass. For the radius and effective temperature (bottom panel), the two properties most affected by an opacity limit, differences between the two tracks are typically less than 0.1 dex.

For further comparison we show an adjacent binary-star model in the same grid with stars of the same mass ratio and orbital period but slightly less massive primary star of $M_{1}=9.9\,\Msun$, that successfully reached the end of the simulation without the need to limit the opacity. Although the neighboring simulation is better able to match our failed run when the luminosity dips to low values, comparison between all three tracks in Figure~\ref{fig:opacity_fix} suggests that any inaccuracies accrued by our opacity limit are of a similar magnitude to any differences between adjacent simulations in our model grids.

\begin{table*}
  \centering
  \caption{A list of the different stellar types we adopt in \posydon{}.
  }
  \label{tab:starstates}
  \begin{tabular}{l lll}
  \hline\hline
  Compact & \multicolumn{3}{c}{Non-degenerate star states} \\
    \hline
    \verb|WD| & \verb|H-rich_Core_H_burning| &
                \verb|H-rich_Core_He_burning| &
                \verb|H-rich_Shell_H_burning| 
    \\
    \verb|NS| & \verb|H-rich_Central_He_depleted| &
                \verb|H-rich_Central_C_depletion| &
                \verb|H-rich_non_burning| 
    \\
    \verb|BH| & \verb|stripped_He_Core_H_burning| &
                \verb|stripped_He_Core_He_burning| &
                \verb|stripped_He_Shell_H_burning| 
    \\
              & \verb|stripped_He_Central_He_depleted| &
                \verb|stripped_He_Central_C_depletion| &
                \verb|stripped_He_non_burning|\\
  \hline
  \end{tabular}
\end{table*}

\begin{figure*}
    \centering
    \includegraphics[width=0.99\textwidth]{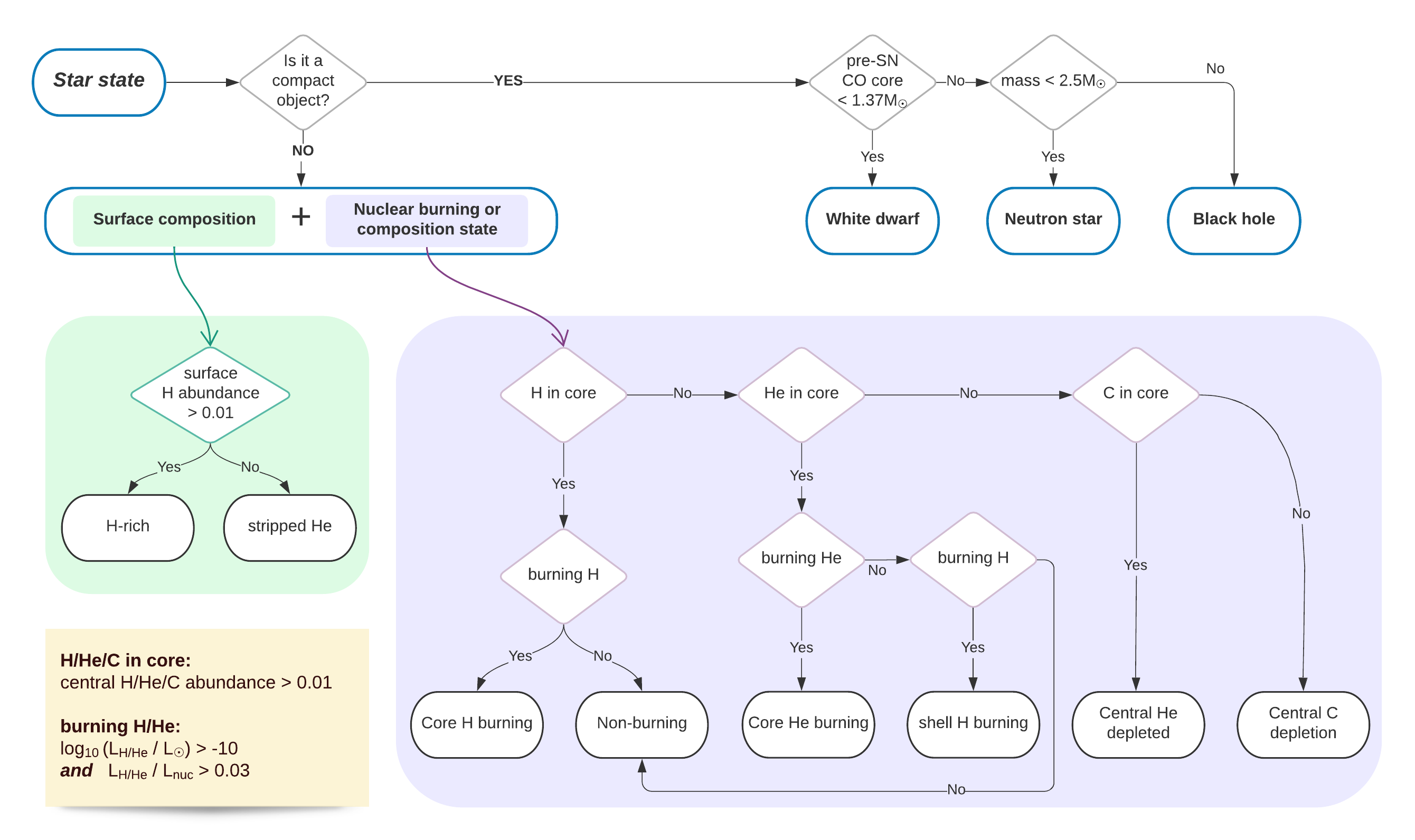}
    \caption{The state of compact objects is determined during the core-collapse step: if the mass of the pre-SN C/O core is less than $1.37\,M_\odot$, then we form a WD. Otherwise a SN will occur, and the state of the CO is determined based on its mass. 
    For non-degenerate stars, the state is a combination of a surface-composition state, and a nuclear burning state. 
    The former is decided solely by the presence or absence of hydrogen at the surface, whereas the burning state depends on whether a species (H, He or C) has been depleted from the core, and the burning of which species still contributes to the nuclear luminosity ($L_{\mathrm{nuc}}$).
    }
    \label{fig:starstates}
\end{figure*}

\subsection{Post-Processed Quantities}\label{sec:ppq}

Our single- and binary-star \mesa{} simulations result in two types of files: history files which contain the time evolution of the binary's and its component stars' properties and profile files which define each non-degenerate star's structure. 
In this first version of \posydon{}, we save the profiles of stars only at the end of the simulations. 
Combined with the \mesa{} terminal output, we have all the information necessary to analyze each simulation.

As a first step, we analyze the final binary properties and terminal output to broadly determine how and why each binary simulation ended. 
We first identify the small subset of binaries in which, despite the process described in Section~\ref{sec:rerun}, the \mesa{} simulation failed to converge; these are ignored throughout the remainder of this work.
For the successful binary-star simulations, we have four separate conditions (e.g., Figure~\ref{fig:CO-HeMS_MESA_grid_TF1}): (1) binaries that were already in RLO when initialized at ZA(He)MS; (2) binaries in which one star reached the end of its lifetime (i.e., one of the first three termination conditions described in Section~\ref{sec:terminated}) and went through stable mass transfer; (3) binaries which avoided mass transfer and the two component stars essentially evolved in isolation; and (4) binaries that entered unstable mass transfer (described in Section~\ref{sec:common_envelope}). 

As a second step, we separately analyze each star and assign it a stellar type at the end of each simulation. For COs, this is a straightforward task, as the type of CO is dependent only on its mass.
Non-degenerate stars present a more difficult object to typecast. 
Most pBPS codes rely upon the $k$-type stellar classification introduced in \citet[][based on \citealt{1997MNRAS.291..732T}]{2000MNRAS.315..543H}  
In \posydon{}, we use instead a two-term classification system, dependent upon both what part of the star (if any) is undergoing nuclear burning and the envelope's composition. Table~\ref{tab:starstates} provides a list of the possible stellar type combinations, while in Figure~\ref{fig:starstates} we show the algorithm for determining them.

As a third step, we assign a designation to the resulting binary configuration in each of our simulations. 
These include {\tt detached} for binaries in which both stars are confined within their respective Roche lobes, {\tt RLO1} or {\tt RLO2} for binaries in which the primary or secondary star, respectively, is overfilling its Roche lobe; {\tt contact} for binaries in which both stars are overfilling their respective Roche lobes; {\tt not\_converged} for systems where the binary-star simulations ran into numerical convergence problems, and {\tt initial\_MT}  for binaries initialized in RLO. 
These designations of stars and binary states are used throughout \posydon{} and are updated by each evolutionary step. Therefore evolutionary phases modeled with on-the-fly calculations (see Section \ref{sec:other_physics}) also affect the star and binary states, which result in two additional possible designations: {\tt merger} for those binaries that have merged, and {\tt disrupted} for those binaries that have become unbound due to some process.

As a fourth step, we analyze the mass-transfer history of the binaries we simulate, identifying the donor star's state when RLO initiated, and whether or not that mass transfer phase was stable or unstable (e.g., {\tt caseA\_from\_star1}). Note that we use the canonical definitions for Case A, Case B, and Case C mass transfer \citep[for a review, see e.g.,][]{1991ApJS...76...55I}. Specifically these labels (which can be identified by the differing symbol types and colors shown in, e.g., Figure \ref{fig:CO-HMS_MESA_grid_TF12}) identify whether the donor star was on the MS, on the post-MS, or a stripped He star. 
In cases where systems evolve through multiple phases of mass transfer, all phases are included in the label (e.g., {\tt caseA/B\_from\_star1} if a Case A mass-transfer phase is followed by Case B one).

As a final step, we calculate a number of post-processed quantities, ranging from 
parameters related to specific core-collapse mechanisms (Section~\ref{sec:core_collapse}) to different CE prescriptions (Section~\ref{sec:common_envelope}). These are typically parameters that require integrals over all or part of a star's structure, which for efficiency we pre-compute. 
All post-processed quantities are summarized in Table~\ref{table:post_processed}.

\subsection{Resampling of single-star grids using Equivalent Evolutionary Phases}\label{sec:EEPs}

For our single-star grids, we perform an additional post-processing step, which resamples the history output of the \mesa{} code in a way that facilitates the interpolation of entire evolutionary tracks. 
This is necessary for the computations described in Section~\ref{sec:detached}. 
Using the method from \citet{2016ApJS..222....8D}, we assign \emph{equivalent evolutionary phases} (EEPs) throughout the evolution of a star. 
This method designates \emph{primary} EEPs to major structural changes to a star (e.g., He ignition), and regularly spaced \emph{secondary} EEPs in between. 
Primary EEPs are extracted directly from the computed stellar tracks. 
We then interpolate between the time-steps to identify every quantity of a star that we track at each secondary EEP. 
By applying this method to each of our single star tracks (both H-rich and He-rich) we can more easily interpolate within our single-star grids to find the quantities  (e.g., radius, core mass) characterizing a star of any mass, and at any point throughout its evolution. 

The methodology described above is unfortunately not directly applicable to binary-star evolutionary tracks. For this interpolation method to work, the defined EEPs must be strictly ordered a priori. However, binary interactions, can happen at any point during the lifetime of a binary, often more than once, changing the order of EEPs in a non-tractable way. The interpolation of entire binary-star evolutionary tracks will be addressed in future releases of \posydon{}.

\subsection{Downsampling of binary-star grids}
\label{sec:downsampling}

The evolutionary timesteps taken by \mesa{} are typically small, producing high-resolution binary histories and final profiles of the individual stars. 
To reduce the memory footprint of the data, and decrease computation times when modeling binary populations, we downsample the binary tracks (i.e., keep a subset of the steps).

For each individual run in a binary-star grid, we obtain from the \mesa{} simulation the evolution of the binary and individual stars' parameters, as well as the post-processed quantities described in Section~\ref{sec:ppq}. 
For a total number of parameters $m$, the state of the binary is encoded in the $m$-dimensional vector $\bh$.
The evolution of the binary is a multivariate time-series given by $\bh_i = \bh(t_i)$, with $i = 1, \cdots, N$, where $N$ is the number of steps, each corresponding to an age $t_i$. Before the downsampling, the independent variable (age), and non-physical parameters (e.g., model number in \mesa{}) are excluded from $\bh$, while all other parameters are rescaled linearly from $0$ to $1$.

\begin{figure}
    \centering
    \includegraphics[width=0.95\columnwidth]{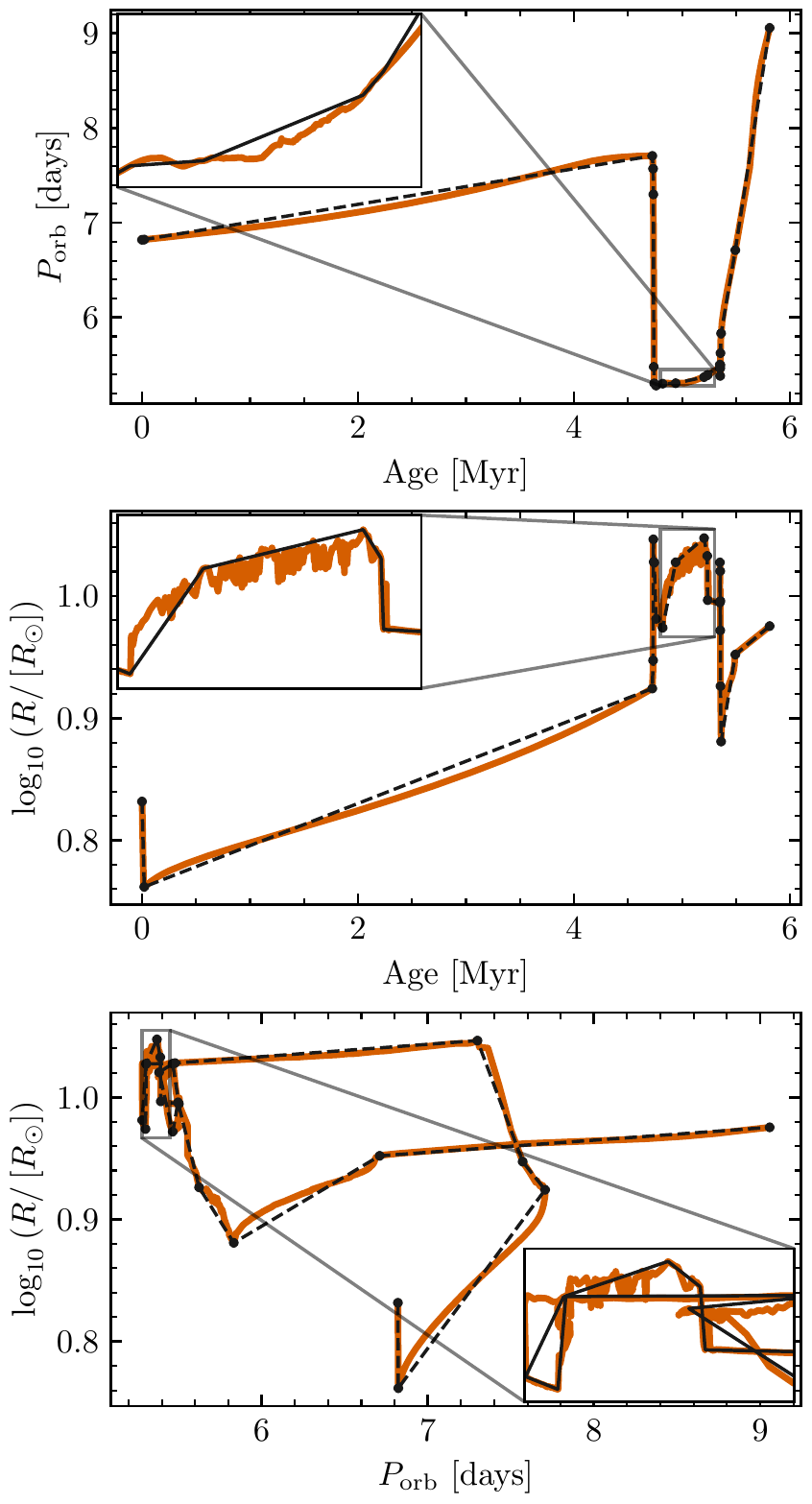}
    \caption{The evolution of a $37.6\,M_\odot$ star with a $22.6\,M_\odot$ companion with an initial $P_{\rm orb}$ of $6.82\,\rm days$. 
    We compare the complete track provided by \mesa{} (orange) comprised of 3412 steps to our downsampled track (black dots) containing on 122 steps for the binary's orbital period (top panel) and each stars' radius (bottom two panels). 
    In this particular case, the compression ratio is ${\sim}155$, but the ratio varies from star to star and depends on which parameters are accounted for by the algorithm. 
    The downsampling algorithm captures even the rapid variations seen in the two stars' radii between 4.8\,Myr and 5.3\,Myr (shown in the insets).}
    \label{fig:downsampling}
\end{figure}

\begin{figure}
    \centering
    \includegraphics[width=0.99\columnwidth]{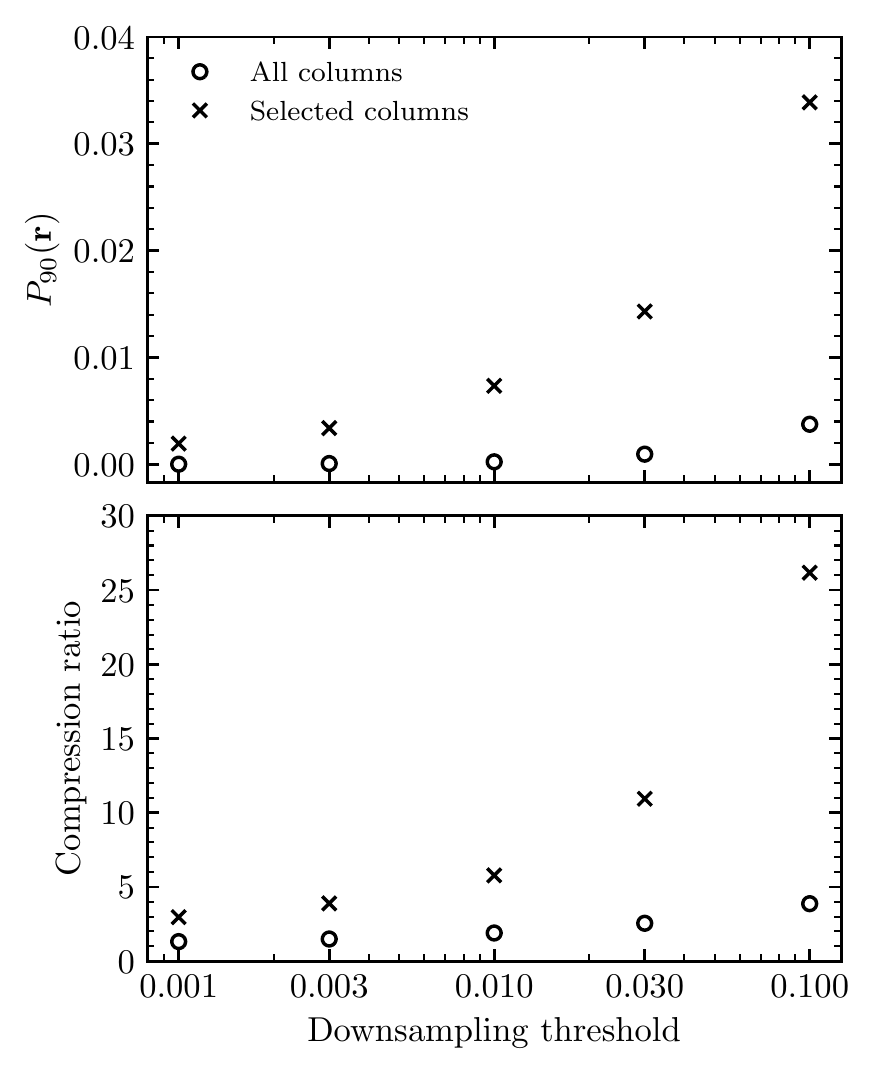}
    \caption{The accuracy ($P_{90}(\mathbf{r})$; top panel) and compression ratio (bottom panel) of our downsampling algorithm as applied to the HMS--HMS binary grid, as a function of the downsampling threshold. 
    We compare the performance of the algorithm when it is applied to all output columns in the data (circle markers) or only a selected list of columns ($\times$; see excluded parameters annotated with an asterisk in Table~\ref{table:MESA_singlestar_attributes} and Table~\ref{table:binary_properties}). 
    As the downsampling threshold $\epsilon$ increases from $10^{-3}$ to $10^{-1}$, the compression ratio dramatically improves, but at the cost of the accuracy ($P_{90\%}$ represents the average of the 90-th percentile of the interpolation relative errors across all runs and either all or selected parameters). 
    For our grids we use only selected columns with $\epsilon=10^{-1}$, which gives us a compression ratio of $\simeq$26, while limiting any errors that could be accrued from this process to within a few per cent.}
        \label{fig:downsampling_report}
\end{figure}

The downsampling algorithm selects a subset of the original steps, so that if interpolated at the original timesteps $t_i$ the interpolation absolute scaled error is below a chosen threshold $\epsilon$:
\begin{equation}
    e_i = \| \bh_i - \hat{\bh}_i \| < \epsilon,
    \label{eq:downsamplingcriterion}
\end{equation}
where $\hat{\bh}_i$ is the interpolated point using the age as the independent variable. 
We use linear interpolation,
\begin{equation}
    \hat{\bh}_i = \bh_j + \frac{t_i - t_j}{t_{j+1} - t_j} \left(\bh_{j+1}-\bh_{j}\right),
\end{equation}
where $j$ and $j{+}1$ are adjacent steps of the downsampled time-series so that $t_j<t_i<t_{j+1}$.

The search for the steps in the downsampled data is performed as follows.
Initially, we include only the first ($\bh_1$) and last ($\bh_N$) points. Then we search for the intermediate point ($\bh_j$ where $1<j<N$) with maximum interpolation error. 
If this error is below the threshold ($e_j < \epsilon$), then the algorithm has finished; otherwise it includes this point and continues the search in the two parts of the time-series before ($\bh_1$ to $\bh_j$) and after ($\bh_j$ to $\bh_N$) the intermediate point. 
The process continues until all original points are well approximated by the interpolation of the selected subset of steps, i.e., $e_i < \epsilon, \forall i$.

Additionally, we apply this downsampling method to the stellar profile data, following the exact procedure outlined above with the mass coordinate as the independent variable. A shell is kept not only when the interpolation error exceeds the predefined threshold, but also when the adjacent shells that are kept have difference in mass is larger then 0.5\% of the total stellar mass.

To demonstrate the validity of our method, Figure~\ref{fig:downsampling} shows an example of the downsampling of a track using the same interpolation error threshold $\epsilon=0.1$ we use for our grids. However, here we apply it to only two parameters (orbital period and radius of secondary star) for visualization purposes; when the algorithm operates in a higher-dimensional space, it retains a large fraction of the initial points to capture the overall shape, making it hard to inspect its performance through two-dimensional plots. The downsampled version of the track is able to follow even the rapid oscillations occurring during the late-stage evolution of this particular binary.

The choice of $\epsilon$ is a balance between data compression and interpolation accuracy, a trade-off we demonstrate explicitly for our model grid composed of two H-rich stars in Figure~\ref{fig:downsampling_report}.
For our grids in \posydon{}, we set the downsampling threshold to $0.1$ and enforce it only for a list of 22 columns from the simulation output (see excluded parameters annotated with an asterisk in Table~\ref{table:MESA_singlestar_attributes},  Table~\ref{table:binary_properties}, and Table~\ref{table:profile_columns}). 
This results to a compression factor of ${\sim}26$ with respect to original simulation data, but still sufficiently high accuracy with respect to the original grid. 
The final size of the three binary-star grids, after downsampling, is $\sim 9.3$\,GB.

\section{Our Classification and Interpolation Approach \label{sec:machine_learning}}
Even after our various stages of post-processing, we cannot use the grids of binary-star simulations within \posydon{} as is for modeling populations (except if we follow a nearest-neighbor matching approach).
While our binary-star simulations have only been run for a select, finite combination of initial masses and orbital periods, BPS requires us to have the capability to evolve a binary anywhere within the domain of interest.
We solve this problem in two steps. 
First, we apply a classification method to each of our grids to identify regions that undergo qualitatively different classes of evolution. 
Then we separately apply an interpolation method to each class to calculate stellar and binary properties. 
We describe the details of those methods in Section~\ref{sec:classification} and Section~\ref{sec:interp_interp}, respectively.

\subsection{Transformation and Rescaling of Grid Data}
\label{sec:ml_preprocessing}
For classification and interpolation purposes, we can interpret each of our binary grids as a data set that comprises $N$ input binaries, $\{\bx_n\}_{n=1}^N$, along with its corresponding scalar $\{\by_n\}_{n=1}^N$ and class $\{\bz_n\}_{n=1}^N$ targets. These sets can be columnwise stacked into the matrices $\bX \in \real^{3\times N}$, $\bY \in \real^{M\times N}$ (where $M$ is the number of output quantities), and $\bZ$ of dimension ${4\times N}$ (because we classify each run into one of four broadly defined categories; see Section \ref{sec:classification}). 
More specifically, each $\bx_n \in \real^3$ contains the initial masses and orbital period of the $n$-th binary in the grid, whereas $\by_n$ and $\bz_n$ denote the collection of final binary and single-star quantities, and their associated classification, respectively.
The $N$ runs are distributed in a uniform three-dimensional mesh, either on a linear or logarithmic scale following the description provided in Section~\ref{sec:grids}. 
This uniform grid constitutes an initial and naive way of thoroughly covering the parameter space, which is feasible due to the low dimensionality of the data.

\begin{figure}[ht]
    \centering
    \includegraphics[width=\columnwidth]{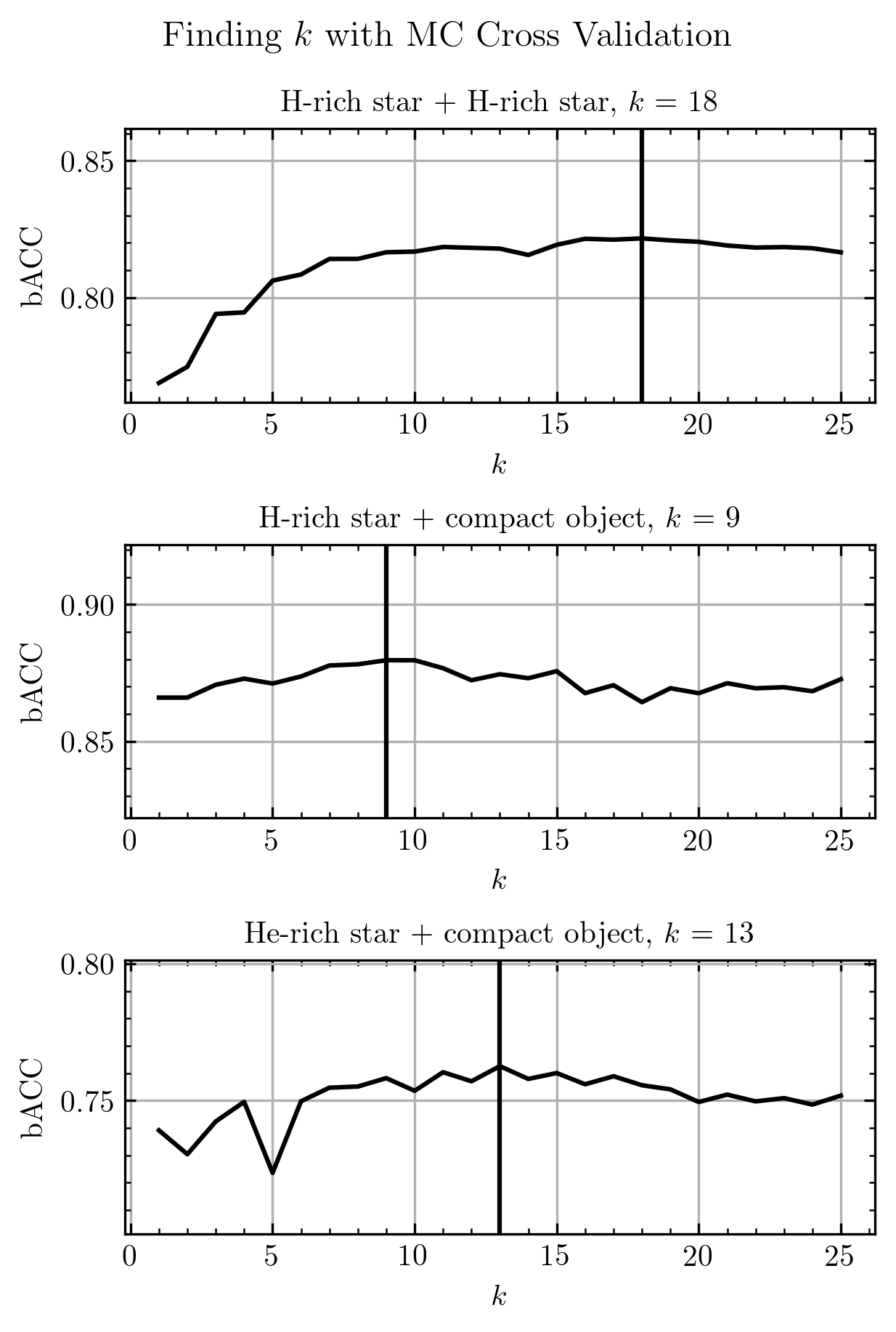}
    \caption{The optimal $k$ in our $k$NN classification scheme for each of our three grids. 
    The bACC is calculated using 10-fold MC cross-validation, and averages the statistical recall of each of our four classes described in Section~\ref{sec:classification}. 
    The highest bACC for each grid is provided at the top of each panel and is indicated by the vertical lines; we use these $k$s when classifying our grids for running populations. Although we use the optimal $k$ for each grid, our results are relatively insensitive to that choice.
    }    
    \label{fig:knn_xval}
\end{figure}

\begin{figure*}[ht]
    \centering
    \includegraphics[width=\textwidth]{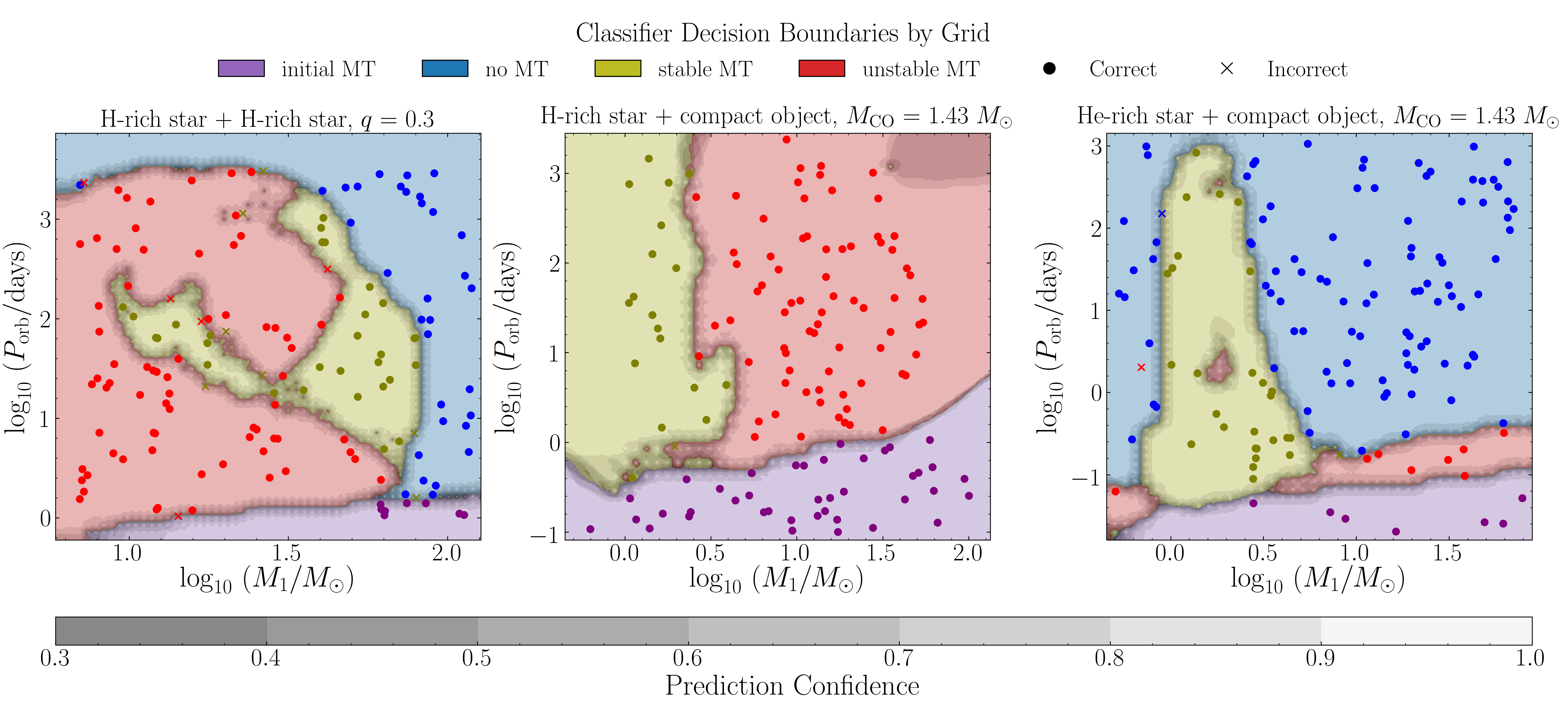}
    \caption{Decision boundaries of the $k$NN classifiers for a single slice in each of the three grids as a function of the primary star's mass on the horizontal axis and the orbital period on the vertical axis (the choice of $q$ or $M_{\rm CO}$ for each slice is indicated in each panel's title). Shaded gray regions overlaid onto class regions represent the confidence of the classifier in that region. Points on top of the decision boundaries represent the validation data, where the edge color of each point shows the ground truth of the given point, and the fill color shows the classifier's prediction. Only in rare circumstances and only near classification boundaries does our classifier make incorrect predictions for our validation set.} \label{fig:classif_dec_boundary}
\end{figure*}

A convenient preprocessing of the data is crucial for both interpolation and classification. 
We apply a series of non-linear and linear transformations to numeric data. Choosing the optimal transformation depends on the task (interpolation or classification), the method used for each task, and whether we are dealing with an input or an output quantity.

First, we consider a non-linear transformation of the data using the logarithm: inputs can be transformed as $\log \bx_i$ and targets as $\log \by_i$ or $\log (-\by_i)$ if $\by_i<0$.
Classification accuracy will improve when our algorithm uses the logarithm of the inputs for data sampled evenly in log-space.
The effect on interpolation is different: e.g., a linear interpolation in the log-space results in a non-linear interpolation on the untransformed space.
This is similar to an approach where a non-linear space is transformed through a kernel to a space in which a linear model allows for modeling behavior appropriately.

We automatically choose whether to apply a logarithmic scaling using a cross-validation scheme. The optimal scaling for both inputs and outputs is chosen using the lowest relative error out of all the feasible scalings that could be applied to the given variable.

As a second step, we apply a min--max scaling to the inputs so that the transformed features $\bx_i^\mathrm{t}$ are confined to the range $[-1,1]$, and we standardize the outputs such that they have zero-mean and unit variance,
\begin{equation}
    \by_i^\mathrm{t} = \frac{\by_i-\overline{\by}_i}{\sigma_{\by_i}}.
\end{equation}
The choice of scaling for the inputs derives from the uniform nature of the input grid data. Although it is possible that we sampled our data in a non-optimal way, in practice we find the best results occur when our data scaling follows our grid sampling.
In the case of the interpolated quantities, standarization produces improved metrics, particularly because it is less sensitive to outliers.

\subsection{Classification of our Grids}
\label{sec:classification}

Accurate classification is a critical aspect of the \posydon{} approach to evolving binary systems. 
Therefore, we separate our binaries into four categories based on their mass-transfer histories. 
The categories are: stable mass transfer, unstable mass transfer, binaries that never interact, and those in RLO at ZAMS (Section~\ref{sec:ppq}). 
In addition to using their mass-transfer history, we could further segregate binaries into more refined classes; however, we find this to be currently unnecessary, and we can accurately interpolate our binaries given these four broad classes.

\begin{figure*}[ht]
    \centering
    \includegraphics[width=\textwidth]{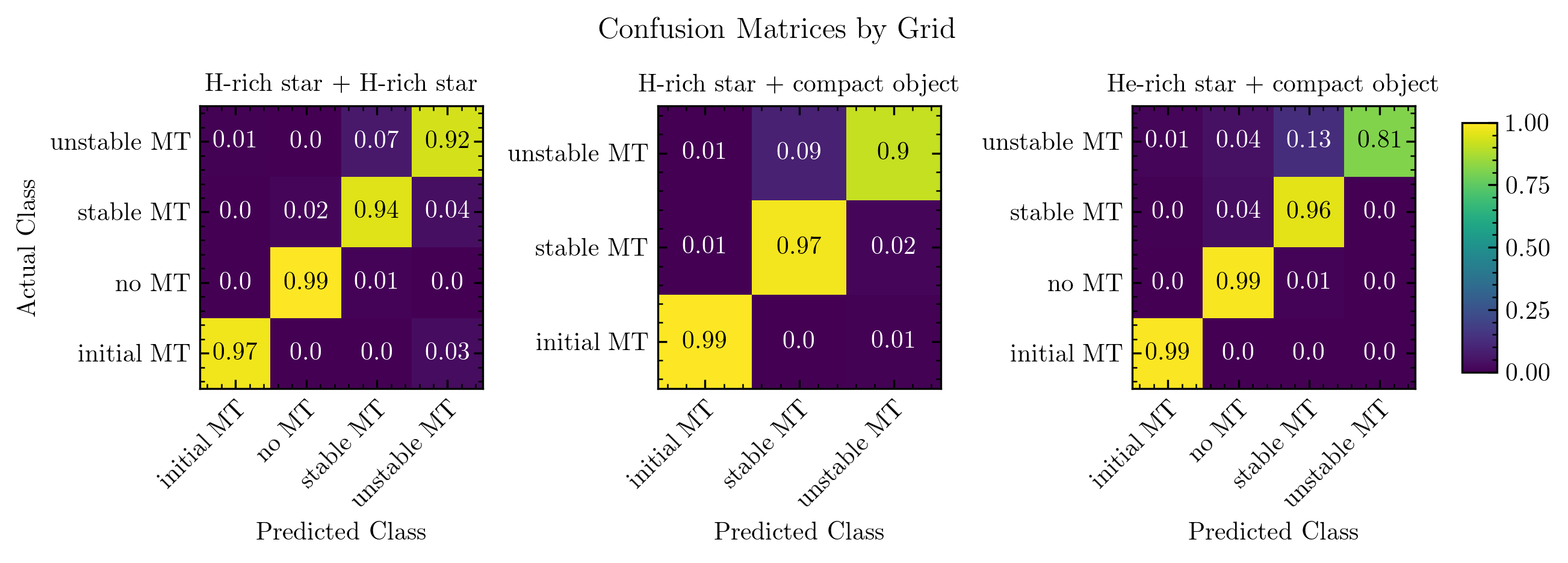}
    \caption{Confusion matrices for each of our three binary grids. 
    Each value at grid cell $c_{ij}$ represents the fraction of binaries that belong to class $j$ (vertical axis) and were classified as class $i$ (horizontal axis). 
    Each row is normalized so that the sum of each row is $1$, and the color of each cell indicates the magnitude of the value in the cell.
    Accuracies are all above 90\%, with the exception of unstable mass transfer for the He-rich star with a CO companion.}
    \label{fig:classif_confusion}
\end{figure*}

For each of our three binary-star grids, we generate a classification object that determines which of the four previously defined outcomes will be the result of a binary with any particular combination of two masses and orbital period. 
In this first version of \posydon\ we use a $k$-nearest neighbors ($k$NN) classifier, a simple and robust classifier that achieves high precision in this task. 
We use the Euclidean distance as distance metric for the transformed input grid
and weight each neighbor in the neighborhood proportionally to their inverse distance. 

We optimize the number of nearest neighbors we use by applying a Monte Carlo (MC) cross-validation scheme and selecting the $k$ that produces a higher balanced accuracy (bACC). 
The bACC metric averages the statistical recall for each class (recall is the number of true positives divided by the combined number of true positives and false negatives) 
to produce a metric that accounts for any imbalances between classes. 

Figure~\ref{fig:knn_xval} shows the average cross-validation performance for our three grids in terms of bACC as a function of the number of neighbors in the $k$NN classifier starting from $k=1$ and highlights the location of the optimum. 
We train our final classifier on our regularly spaced grids using the optimal $k$, listed at the top of each panel in Figure~\ref{fig:knn_xval} and indicated by a vertical black line.

Figure~\ref{fig:classif_dec_boundary} shows our classifier applied to one slice in each of our three grids, with the colors indicating different regions. 
Overlaid grey contours, pronounced near class boundaries, indicate classification uncertainty. 
We ignore the no mass-transfer class for the grid of H-rich stars with CO companions, as this grid only applies to interacting binaries.

To evaluate the accuracy of our classifiers, we use the validation data set associated with each of our three binary grids. 
For each grid this validation set is comprised of binaries randomly sampled with the same range and scale, linear or logarithmic, as its training counterpart. 
Each of the three validation sets contain $3000$ samples which roughly represent  $\sim$5--12\% of the number of binaries in the regular grid.
By applying our classifiers to the same initial values as those of our validation binaries, we can evaluate the accuracy of our classifiers. In Figure~\ref{fig:classif_dec_boundary} our validation data is indicated by points (correctly classified) and crosses (incorrectly classified). It is evident that incorrectly classified validation binaries are very rare.

To evaluate the quantitative accuracy of our classifier, we provide a confusion matrix for each of three grids in Figure~\ref{fig:classif_confusion}.
Diagonal squares indicate the fraction of systems that were correctly classified, while off-diagonal squares indicate the fraction of incorrectly classified systems. 
The matrices are calculated such that each row sums to unity. 
All classes in all grids have an accuracy in excess of 90\%, often much more so, except for unstable mass transfer for binaries with a He-rich star and a CO companion. 
Examination of two slices of this grid in Figure~\ref{fig:CO-HeMS_MESA_grid_TF1} shows that the unstable mass-transfer class comprises a relatively small portion of the overall grid, existing at small orbital periods, small CO masses and large companion masses. 
Reliable classification of small classes is difficult, but improving our classification accuracy will be a focus of future efforts (Section~\ref{sec:future}).

\subsection{Interpolation of our Grids}
\label{sec:interp_interp}

Once classified based on their mass-transfer characteristics we separately interpolate binaries falling into each class for each of our three binary-star simulation grids. 
We only interpolate quantities for three of our binary classes, since those binaries overfilling their Roche lobe at ZA(He)MS 
are dismissed. 

\begin{table*}
    \centering
    \begin{tabular}{p{0.45\textwidth}p{0.50\textwidth}}
        \hline\hline
        Constraint & Relation \\
        \hline
    \multicolumn{2}{c}{\textit{Type I constraints: Equations}}\\ \hline
        Kepler's Third Law &
            $a = \left[G \left(M_1 + M_2\right) P_{\rm orb}^2 / 4 \pi^2 \right]^{{1}/{3}}$
        \\
        Mass-transfer fraction\textsuperscript{(a)} &
            $x = 1 - \dot{M}_{\rm{sys,2}} / \dot{M}_{\rm{tr}} $
        \\
        Stefan-Boltzmann Law &  
            $T_{\rm{eff}} = \left(L / 4\pi R^2 \sigma_{\rm{SB}} \right)^{{1}/{4}}$
        \\
        Sum of nuclear luminosities & 
            $L_{\rm nuc} = L_{\rm H} + L_{\rm He} + L_{\rm Z}$
        \\\hline
    \multicolumn{2}{c}{\textit{Type II constraints: Inequalities}}\\ \hline
        Mass-loss from the system from the vicinity of a star\textsuperscript{(b)} &
            $\dot{M}_{\rm{sys,1}} < \dot{M}_{\rm{tr}}$ and
            $\dot{M}_{\rm{sys,2}} < \dot{M}_{\rm{tr}}$
        \\
         Core masses and radii &
            $M_{\rm{C/O-core}} < M_{\rm{He-core}} < M$ and
            $R_{\rm{C/O-core}} < R_{\rm{He-core}} < R$
        \\
        Envelope masses and core radii\textsuperscript{(c)} &
            $M_{\rm{env}}{<}M$ and
            $R_{\rm{core}}{<}R$
        \\
        Mass, thickness and middle radius of the convective region (for tides) & 
            $M_{\rm{conv.reg.}} < M$ and \par
            $0 < R_{\rm{conv.reg.}} - (1/2) D_{\rm{conv.reg.}} < R_{\rm{conv.reg.}} + (1/2) D_{\rm{conv.reg.}} < R$
        \\
        Remnant baryonic mass &
            $M_{\rm{rembar}} < M$
        \\
        \hline
    \multicolumn{2}{c}{\textit{Type III constraints: constrained sum}}\\ \hline
        Central abundances &
            $X_{\rm{c,H1}} + X_{\rm{c,He4}} + X_{\rm{c,C12}} + X_{\rm{c,N14}} + X_{\rm{c,O16}} + X_{\rm{c,other}} = 1$
        \\
        Surface abundances &
            $X_{\rm{s,H1}} + X_{\rm{s,He4}} + X_{\rm{s,C12}} + X_{\rm{s,N14}} + X_{\rm{s,O16}} + X_{\rm{s,other}} = 1$
        \\\hline\hline
    \end{tabular}
    \caption{Constraints that are ensured for our interpolated quantities. Type I relations are written such as the left-hand side indicates the quantity that is inferred from the rest. 
    Notes:\\
    \textsuperscript{(a)} This constraint is applied after constraint \textsuperscript{(b)} since the mass-loss rate has to be less than the mass-transfer rate. 
    Moreover, $x$ is set to $1$ if no mass is transferred. \\
    \textsuperscript{(c)} There are four pairs of these quantities corresponding to the radii where the $^1$H fraction drops below 1\%, 10\% and 30\%, and finally where the $^4$He fraction drops below 10\% for pure He stars.}
    \label{tab:interp_constraints}
\end{table*}

We use an $N$-dimensional (where $N$ is the number of binaries) linear interpolation: the data is divided into a set of $N$-simplices, tetrahedra in our three-dimensional data, by means of a Delaunay triangulation (which is not unique given the regular structure of our grids). 
The interpolated value for a given point corresponds to the value at the hyperplane that passes through the vertices of the simplex which contains the point. 
The choice of whether to apply a non-linear transformation on $\by_i$, $\log \by_i$ depends directly on that magnitude. 
For each output magnitude we select the optimal scaling via MC cross-validation with $x$ iterations and $p\%$ of test data comparing the average relative error across iterations. The final interpolator is trained using all binaries within a particular class for each grid.

The linear interpolation method is not capable of extrapolation: the value for any point which lies outside the convex hull defined by the constructed Delaunay triangulation will be undetermined. 
Although we are not, in general, interested in interpolating outside the training grid, there will be a small region between the convex hull of the linear interpolation and the decision boundary provided by the classifier where we still want to obtain system properties. For this small sliver of parameter space, we adopt values of the nearest point in parameter space of the same class. This is a problematic region where the probability of belonging to the interpolable class will be low, expressing the uncertainty we have about those binaries with the current resolution of the grids. We are currently exploring a method of tackling this problem by incorporating new simulations along the decision surface, identified using an active-learning scheme \citep{2022arXiv220316683A}.

\subsection{Ensuring Physical Congruity of Interpolated Values}
\label{sec:sanitization}

The linear interpolation method described here treats each feature independently without preserving possible physical correlations. 
However, the interpolated results may produce incongruous quantities within a resultant star. 
For example, the Stefan--Boltzmann law connecting the luminosity, radius, and effective temperature of a star might not hold for an interpolated binary. 
As another example, the He-core mass of a star must always be less than the star's total mass. 
We have carefully identified a number of physical constraints within the quantities that we are interpolating that must be satisfied by any realistic star, each of which we list in Table~\ref{tab:interp_constraints}.

\begin{figure}
    \centering
    \includegraphics[width=\columnwidth]{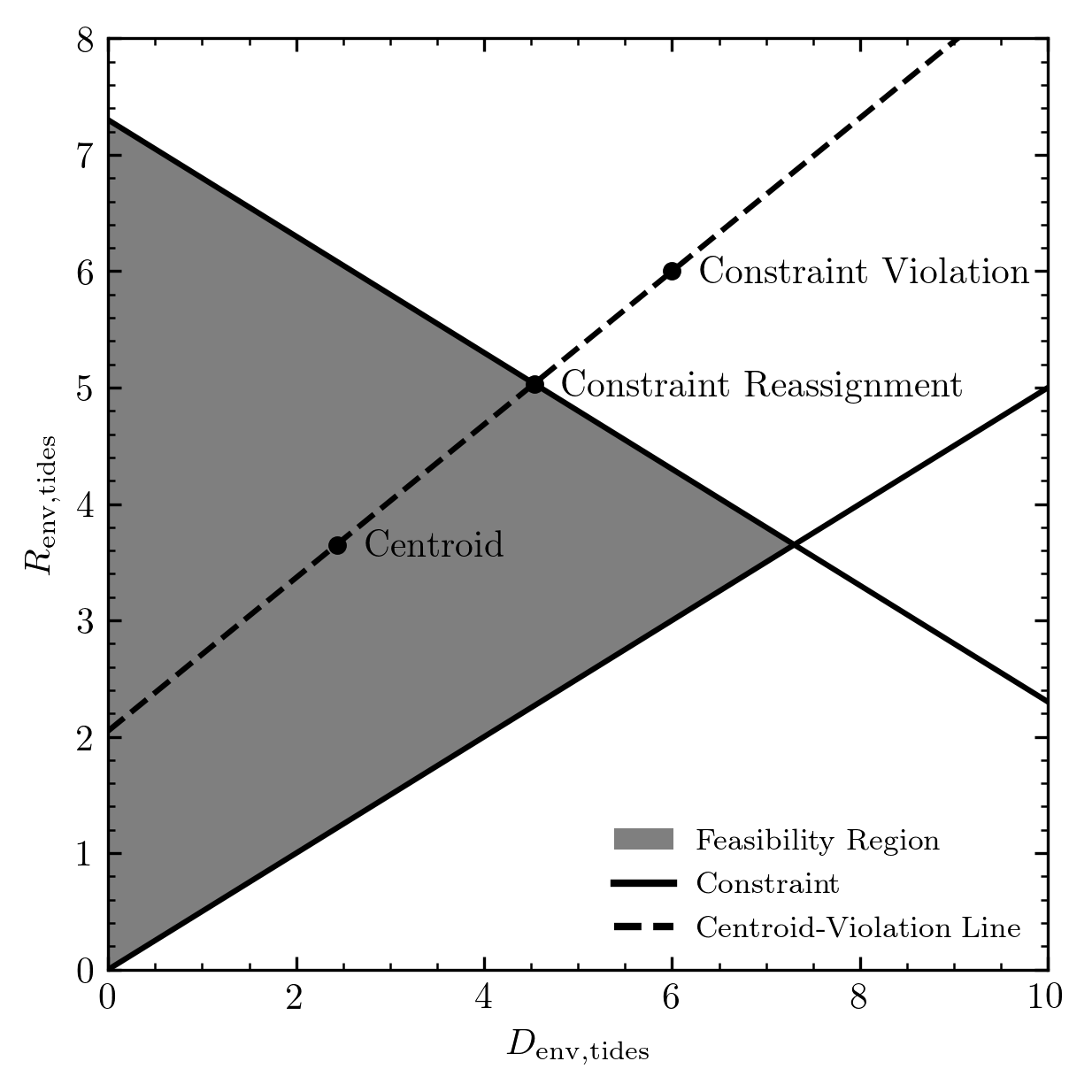}
    \caption{
    The convective envelope radius $\Renv$ and width $\Denv$ need to simultaneously agree with three separate inequalities defined in Table~\ref{tab:interp_constraints}. 
    The feasibility region (dark grey) represents the overlap of all three inequalities decomposed from the constraint. 
    When our interpolation method proposed a point outside the feasibility region, we reassign it to a new point determined from the intersection of the border of the feasibility region and the line drawn from the centroid of the region to the constraint-violating, proposed point. }
    \label{fig:2.7_sanit}
\end{figure}

To address this issue, we process the interpolated quantities for a given binary so that they respect this list of constraints, which fall into one of three different \emph{Types} depending on the basis of the corrective action required. 
When quantities are connected via an equation (Type I), then the interpolated value of one parameter is ignored and inferred by solving the equation on the interpolated values of the remaining parameters. 
The Stefan--Boltzmann equation provides an example of a Type I constraint: we only interpolate each stars' $R$ and $L$, while $T_{\rm eff}$ of the interpolated star is derived. 
In the case of inequalities between quantities (Type II), the quantity that must be less than another is limited by the value of the latter. 
Finally, there are cases where all quantities ought to add to a certain value (Type III). 
For instance, the fractional chemical abundances of a star's core must, by definition, sum to unity. 
We ensure these constraints are satisfied by normalizing our interpolated outputs. 
In one case described in the Table~\ref{tab:interp_constraints} footnotes, a parameter is subject to two separate constraints, in which case we are careful to apply them in the correct order.

In the case of the constraint involving the interpolated quantities \Renv and \Denv, the middle point ($\equiv \left(R_{\rm t,conv.reg}+R_{\rm b,conv.reg}\right)/2$) and the thickness ($\equiv R_{\rm t,conv.reg}-R_{\rm b,conv.reg}$) of the convective region for the computation of tides, respectively, a special treatment is required. 
Both quantities must be positive and less than the star's radius. 
However, constraining them independently as in other Type II constraints does not work as the inner and outer radius of the convective region must both be inside the star: $0 < \Renv - \Denv/2 < \Renv +  \Denv/2 < R$.
We decompose this relationship into three inequalities: $ \Denv/2 \geq 0$, $\Renv \geq \Denv/2$, and $\Renv + D_{\rm{conv.reg.}}/2 \leq R$. 
In the $\Denv$--$\Renv$ plane, the constraints form a feasibility region in the shape of a triangle with the vertices $(0, 0), (0, R),$ and $(R, R/2)$, where $R$ is a fixed values.
If the constraints are violated, then the interpolated values lie outside of the triangle.
The triangle's centroid $(R/3, R/2)$ and the point corresponding to the interpolated values define a line $l$.
The intersection between $l$ and the border of the triangle satisfies the constraint inequality, and is used to assign new values to the parameters.
Figure~\ref{fig:2.7_sanit} provides a pictorial representation of the algorithm.

To assess how often our constraints defined in Table~\ref{tab:interp_constraints} are violated in practice without imposing constraints, we interpolated 3000 binaries for each one of the three grids using the random initial conditions of their validation sets (Section~\ref{sec:interpolation_accuracy}). 
For each binary, we checked all the constraints (two checks per binary system, and $23$ checks per non-degenerate companion star) and counted the violations. 
In the case of Type I constraints (equations), we consider violation a relative error of more than $0.001$ in the inferred quantity. 
Overall, we found 57,035 violations in the 274,560 checks (${\sim}20.8\%$) we performed. 
After applying the algorithm defined here, all violations were corrected.

\subsection{How Accurate are our Interpolation Methods?}
\label{sec:interpolation_accuracy}

\begin{figure*}
    \centering
    \includegraphics[width=\textwidth]{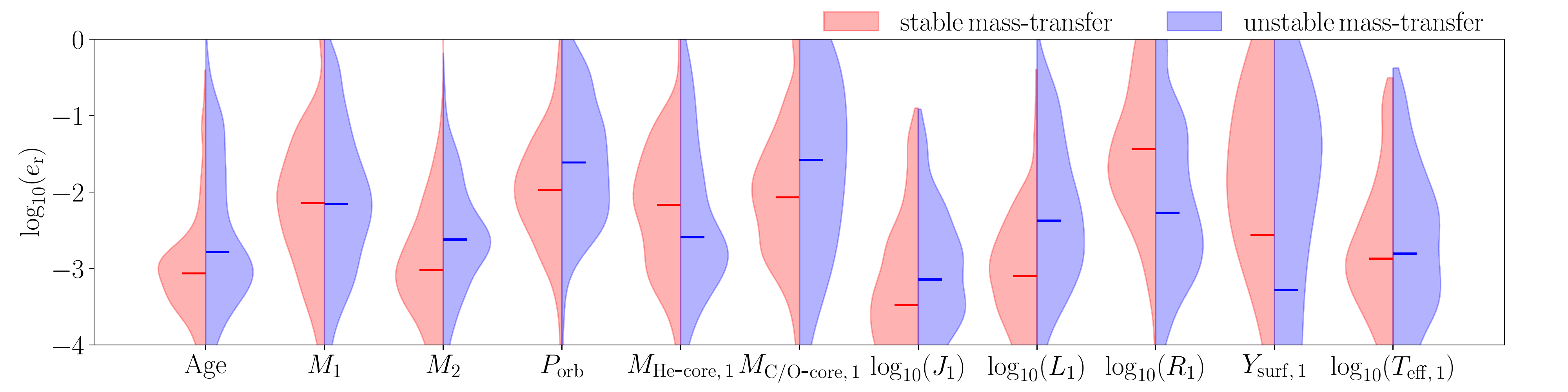}
    \includegraphics[width=\textwidth]{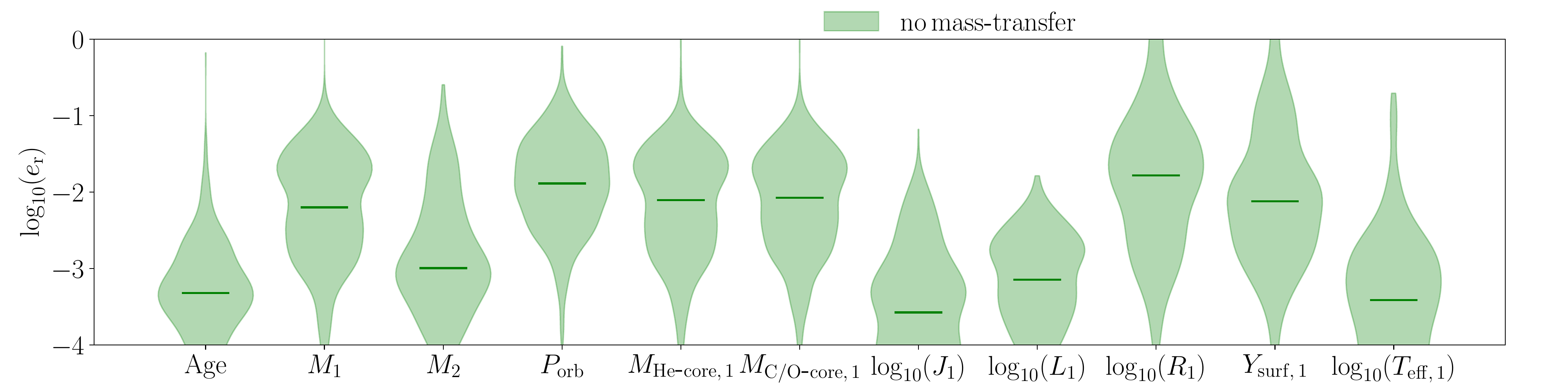}
    \caption{ Interpolation scheme accuracy for ten selected parameters when applied to our grid of two H-rich stars, as calculated using our set of validation binaries. 
    We separate our sample by their different mass-transfer histories to independently evaluate their individual accuracy. Median relative errors ($e_{\rm r}$) indicated by the horizontal lines in each distribution are typically 1\% or lower for the stable mass-transfer and unstable mass-transfer cases (top panel) and the no mass-transfer case (green; bottom panel). 
    Improving this accuracy will be a focus of future work.}
    \label{fig:interp_er_v_HMS-HMS}
\end{figure*}

\begin{figure*}
    \centering
    \includegraphics[width=\textwidth]{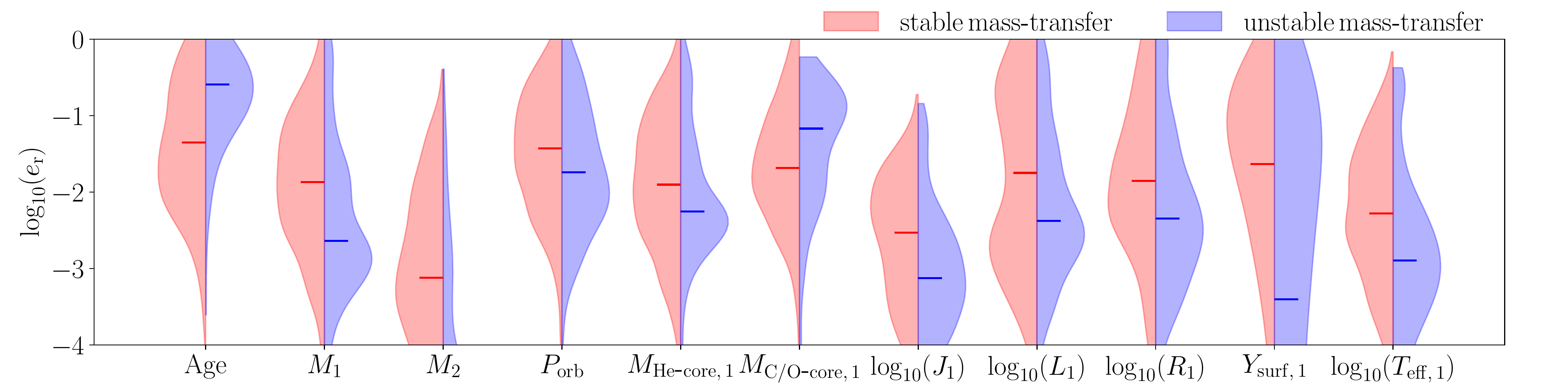}
    \caption{The interpolation accuracy for the same 10 parameters as in Figure~\ref{fig:interp_er_v_HMS-HMS} for our grid of H-rich stars with a CO companion. 
    Since we never use the models from this grid that do not undergo RLO, we do not evaluate the no mass-transfer binaries. 
    In most cases, typical errors are 1\% or better, but several of the distributions have tails extending towards larger $e_{\rm r}$. }
    \label{fig:interp_er_v_CO-HMS_RLO}
\end{figure*}

To assess the performance of the interpolation scheme, we use the same validation data sets that we used to evaluate our classification accuracy, described in Section~\ref{sec:classification}.
Our trained interpolators are applied to the same initial binary parameters as those of the three sets of 3000 binaries comprising our validation sets, one for each of our three binary grids. 
Since these binaries are not used in the training phase, the difference between this set and our interpolated predictions for them provide an ideal comparison from which we can determine the accuracy of our methods.

Figure~\ref{fig:interp_er_v_HMS-HMS} provides the accuracies for eleven selected binary and stellar parameters for our grid of two H-rich stars evolved from ZAMS. 
We have split these samples by their mass-transfer histories so we can separately identify our algorithm's accuracy for the stable mass transfer (red) and unstable mass transfer (blue) cases in the top panel and no mass transfer case (green) in the bottom panel. 
For nearly all parameters and all classes, our median errors are below 1\%. Some parameters such as age and $J_1$ are significantly more accurately interpolated, while others such as $M_{{\rm C/O-core}, 1}$ may be somewhat less accurate. The distributions are quite broad, suggesting that inaccuracies may exist when parameters show sharp variations as a function of input binary parameters. This is particularly apparent for our unstable mass transfer channel, likely a result of the relatively smaller number of simulations that  enter unstable mass transfer and the varying evolutionary stages of the donor stars at the onset of the dynamical instability.  For instance, the relatively large error distribution for $M_{C/O-{\rm core, 1}}$ for our unstable mass-transfer class is likely due to the rapid core growth during the giant phase when donor stars typically enter dynamical instability. Furthermore, despite their large error distributions, some parameters, such as $R_1$, have little impact on the evolution of a binary. Whether the primary star's next evolutionary phase is a core collapse (in the case of the stable mass-transfer scenario) or a CE (in the case of unstable mass transfer) the mass at the outermost part of the star has little impact on the binary's outcome. Nevertheless, we plan to improve these accuracies with future enhancements to our interpolation schemes.

In Figure~\ref{fig:interp_er_v_CO-HMS_RLO} we provide analogous results for our stable mass-transfer and unstable mass-transfer binaries for our grid of H-rich stars with a CO companion. 
Our models tend to show larger variations in accuracy compared with our grid of two H-rich stars. Median errors tend to range from 1\% to 10\% with certain parameters such as age and $M_{{\rm C/O-core}, 1}$ performing noticeably worse. At the same time certain parameters like $M_2$ are very accurately determined, as these parameters vary during the evolution of the binaries in this grid (CO companions in this grid typically accrete little mass). One ought to consider the importance of each parameter when evaluating the accuracy of our models. For instance for stars going through unstable MT $M_{{\rm He}, 1}$ is a much more important parameter than $M_{{\rm C/O}, 1}$ and $Y_{\rm surf}$ has no impact on a binary's future evolution.

\begin{figure*}[t]
    \centering
    \includegraphics[width=\textwidth]{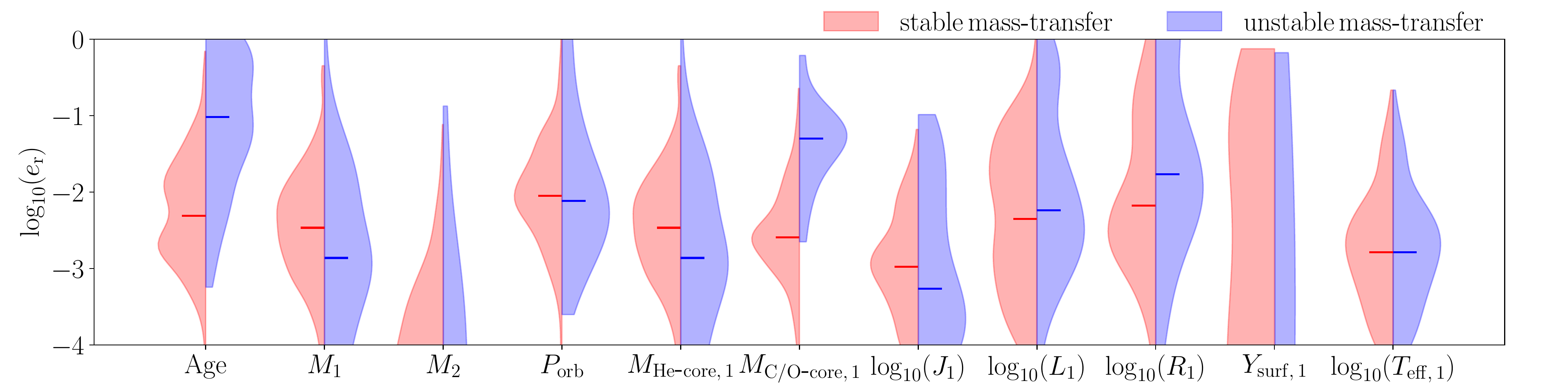}
    \includegraphics[width=\textwidth]{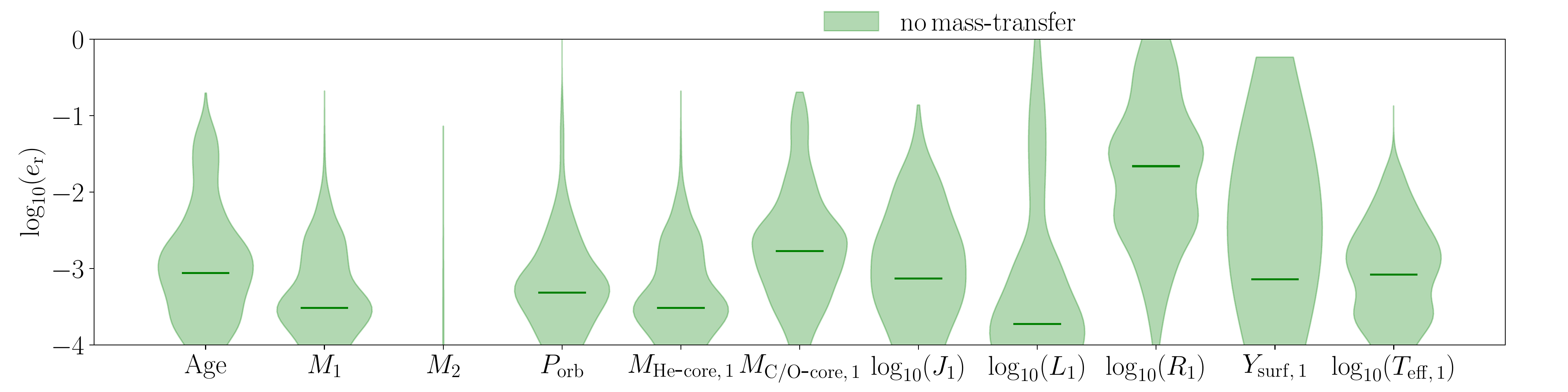}
    \caption{The interpolation accuracy for the same 10 parameters as in Figure~\ref{fig:interp_er_v_HMS-HMS} and Figure~\ref{fig:interp_er_v_CO-HMS_RLO}, but for our grid of He-stars with CO companions. 
    As with our other two grids, this grid has typical median errors below 1\%; however, tails of the distribution extend towards larger $e_{\rm r}$, especially for the stable mass-transfer and unstable mass-transfer cases (top panel).    
    Our no mass-transfer case (bottom panel) is much more accurate than the corresponding binaries in our grid of two H-rich stars in Figure~\ref{fig:interp_er_v_HMS-HMS}. The apparent truncations in the distributions (e.g., $M_{\rm C/O-core, 1}$ and $\log_{10} (J_1)$) are genuine representations of the data.}
    \label{fig:interp_er_v_CO-HeMS}
\end{figure*}

Finally, Figure~\ref{fig:interp_er_v_CO-HeMS} shows the accuracies for our grid of He-rich stars with CO companions. Any edges in the distributions are genuine representations of the underlying data.
Our trained interpolators provide the most accurate predictions of our three grids; median accuracies are typically between 0.1\% and 1\% for the stable mass transfer and unstable mass transfer classes, and somewhat better for the no mass transfer class.

The accuracies provided in Figures~\ref{fig:interp_er_v_HMS-HMS}, \ref{fig:interp_er_v_CO-HMS_RLO}, and \ref{fig:interp_er_v_CO-HeMS} all refer to the data sets and associated interpolation objects provided in v1.0 of \posydon{}. 
One could use the \posydon{} infrastructure to evolve larger numbers of binaries than we have provided along with v1.0, which would improve our interpolation accuracy. A focus on regions where our interpolation methods are least accurate would provide the largest benefit.  
Using a combination of active-learning techniques, more complex machine-learning algorithms, and much more computation time, we expect that future versions of \posydon{} will only exhibit substantially improved classification and interpolation accuracies (Section~\ref{sec:future}).

\subsection{Limitations of our Approach}
\label{sec:ml_limitations}
There are a few limitations of our approach. First, our approach first classifies the binary's type and subsequently performs interpolation. The effect of such a technique is that by performing two optimization problems, the second of which relies on the first, it is possible to propagate error throughout the pipeline. Treating the entire problem as one optimization problem has the potential to reduce error.

Additionally, we transform the grid space by logarithmic transformations before performing linear interpolation, which results in a non-linear interpolation model. Such an approach is similar to using a kernel, where a space is transformed through a kernel function to a space in which a linear model allows for accurate modeling of the behavior of the space \citep{Theodoridis.Konstantinos2009}. A more systematic approach is to consider a kernelized interpolation approach, by applying kernel selection techniques. In the case of a Gaussian process, for example, we may consider a whole family of functions which are specified by a kernel function, to allow for more flexibility on the prior belief of the space \citep{MacKay2003}.

Finally, upon finding an interpolated value, we physically enforce the constraints, as detailed in Section~\ref{sec:sanitization}. However, in principle such a technique, which does not consider the constraints in the optimization objective itself does not guarantee an optimal solution subject to the constraints. 
One way to incorporate the constraints in our model is to add a regularizer term in our loss function to enforce the constraints \citep{astroMLText}, i.e., the loss function balloons when constraints are violated.

\section{Evolutionary Processes Separate from Single- and Binary-Star Model Grids}
\label{sec:other_physics}

\begin{figure*}\center
\includegraphics[width=\textwidth]{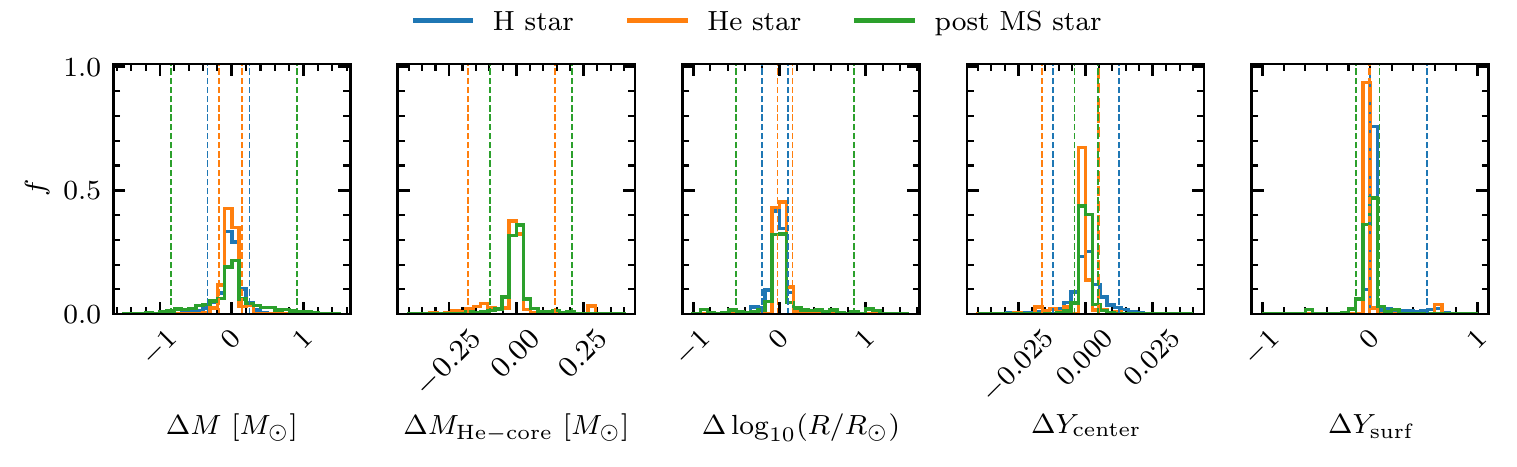}
\caption{Distributions of the difference between the matching point of a single star model in the beginning of the detached step and the values from the previous step, for various physical quantities of a non-degenerate star. 
We show the relative difference in mass $\Delta M$, He-core mass $\Delta M_{\rm He-core}$  and logarithm of the radius $\Delta \log_{10}(R/\Rsun)$, as well as the difference in the He central and surface abundances $Y_{\mathrm{center}}$ and $Y_{\mathrm{surf}}$, respectively. 
Vertical, dashed lines from left to right delineate the 5-th and 95-th percentiles of the distributions. The distributions all suggest that the non-degenerate stars in detached binaries can be accurately matched to a single star model. }
\label{fig:matching}
\end{figure*}

Besides computing, processing, classifying, and interpolating the five separate grids of single- and binary-star models, additional steps are required to follow the complete evolution of  a stellar binary from ZAMS to double CO formation (and potentially its merger). 
These are defined by three separate processes: orbital evolution in eccentric, detached binaries, CE evolution, and stellar core-collapse. 
While the latter two  are standard elements of BPS codes, the need for the former requires some explanation. 
Binaries are intrinsically eccentric after a SN occurs, yet our pre-calculated grids of binary-star models are initiated with circular orbits. 
Including eccentricity as an input to our \mesa{} models would add an additional dimension to our simulation grids, challenging our computational capabilities. Furthermore, self-consistently modeling binary mass transfer along with stellar evolution and tides in eccentric orbits is an active area of study \citep{2007ApJ...667.1170S, 2009ApJ...702.1387S, 2010ApJ...724..546S, 2016ApJ...825...70D, 2016ApJ...825...71D, 2019ApJ...872..119H}, and to date no detailed binary evolution grids have included initially non-circular binaries. 

Nevertheless, in a detached binary, tidal forces cause an eccentric binary to both circularize and synchronize, an effect that must be taken into account, along with other orbital angular-momentum loss processes (e.g., wind mass loss, gravitational radiation and magnetic breaking). To specifically address this, we evolve binaries after the a SN event using a separate process described in Section~\ref{sec:detached}.
We only switch back to using the pre-calculated grid of binary-star models once RLO occurs. 

In the current version of \posydon, binaries that successfully exit from a CE phase initiated by two non-degenerate stars, are also modeled following the process described in Section~\ref{sec:detached} (see also Figure~\ref{fig:structure}). These are binaries consisted of a H-rich and a He-rich, or two He-rich stars in a close circular orbit. We follow their evolution as detached binaries until one of the two stars reaches core-collapse. In a small fraction of post-CE binaries ($\sim 0.3\%$ of the total population) typically consisted of a low-mass ($\lesssim 4\,\Msun$) He-rich and a H-rich MS star, the He-rich star overflows its Roche-lobe as it expands to become a giant He star, and initiates mass transfer onto the H-rich MS star. Since we do not have a grid of detailed binary-star models covering this part of the parameter space, we cannot follow further the evolution of these binaries. We plan to address this in future versions of \posydon. Finally, the evolution of all binaries that are in orbits wider than what is covered by our grids of detailed binary-star models, and thus will never initiate RLO, is also followed as described in Section~\ref{sec:detached}. 

In the subsequent sections, we provide details about how we evolve binaries through an eccentric, detached phase, a CE evolution phase, and core collapse.

\subsection{Evolution of eccentric, detached binaries \label{sec:detached}}

\subsubsection{Matching with a single-star track}\label{sec:matching}

Even though a non-degenerate star in a detached binary is influenced by its companion (for instance, due to tides), we are making the assumption that so long as RLO is not occurring, our single-star, non-rotating models provide reasonable approximations for the evolution of non-degenerate stars in these detached binaries.
We first match the non-degenerate star in the detached binary system with the closest model (searching across different masses at all ages) from our single-star (both  H-rich and He-rich) evolutionary tracks. 
The matching is achieved by minimizing the sum of the squares of key parameters describing the structure of a star. 
These parameters differ depending on the evolutionary phase of the star. 

For the matching process, we distinguish among: (i) MS stars that still have H in their core (with central H mass abundance $X_{\mathrm{center}}>0.01$), (ii) stars that evolved off the MS and retain even a thin H-rich envelope (post-MS, with $X_{\mathrm{center}}<0.01$ and surface H mass abundance $X_{\mathrm{surf}}>0.01$), and (iii) evolved stripped stars that are effectively a H-deficient core ($X_{\mathrm{surf}}<0.01$). 
In case (i) the parameters whose differences are minimized are the total mass of the star, its central H abundance $X_{\mathrm{center}}$ and the radius of the MS star. 
After the first core collapse of the system occurs, most binary companions are in this state, being the initially less massive secondaries that evolved slower than the primary that has formed a CO. 
For case (ii), we replace the central H abundance with the He one ($Y_{\mathrm{center}}$) and the radius with  the mass of the fully developed He core. 
In case (iii), for stripped He-rich stars, we use as minimized parameters the He-core mass of the star (which is equal to its total mass), the radius, and its center He mass abundance. 

We normalize the chosen quantities, such that they have similar weighting in the minimization process. 
The normalization factors are chosen from typical ranges of each parameter for the stars that we focus on: 20\,\Msun for the total mass of MS or post-MS stars; 10\,\Msun{} for He-core masses of stripped He-rich stars, 2.0 for $\log_{10}(R/\Rsun{})$; and $1.0$ for chemical mass-fraction abundances.

We quantify the quality of the matching by calculating the difference of various quantities from the previous step. 
In Figure~\ref{fig:matching}, we see that the difference between the new interpolated total mass in the beginning of the detached step and its previous value, $\Delta M = M_{\rm match} - M_{\rm prev,step} $, is typically better than $0.2\,\Msun$. He-core masses are matched even more precisely, to within 0.05\,$\Msun$. Other parameters (we show $R$, $Y_{\rm center}$, $Y_{\rm surf}$ in Figure \ref{fig:matching}) are also closely matched, justifying our assumption that the non-degenerate star in a detached binary can be accurately represented by a single star model, at least so long as it remains detached.

A CO component in the binary system is treated as a point mass and does not need a matching process. 
We also keep the orbital separation and eccentricity constant in this transition, and thus any small difference in the matched star's mass from the previous step results in a small relative change in the orbital period. 
In addition, by conserving the spin angular momentum on non-degenerate stars during the matching step, we can determine their initial angular frequency $\Omega$ at the beginning of the detached step.

\subsubsection{Further evolution of an eccentric detached binary system}\label{sec:method:ODE}

Once matched with single star models, we evolve the stars in detached binaries as essentially single stars, accounting for their effects on the binary's orbit. For a non-degenerate star, its parameters (e.g., mass, radius, moment of inertia) are evolved according to its interpolated stellar track. 
At the same time, its spin $\Omega$ as well as the system's $a$ and $e$ are evolved solving a set of coupled ordinary differential equations that describe their rate of change due to wind mass loss, tides, magnetic braking, and gravitational radiation. That we assume the back reaction of each of these effects does not significantly impact the internal structure of either star so that our single-star models are sufficiently accurate. For instance, although we follow each star's spin using the moment of inertia of the single-star models, we cannot account during this phase for the star's  internal differential rotation and effects such as rotational mixing.

This approach can only handle scenarios where no RLO mass transfer takes place between the two stars; as soon as a H-rich star enters RLO, we stop the binary's evolution and transition to our grid of \mesa{} mass transfer simulations described in Section \ref{sec:CO-HMS_RLO}. Binaries are assumed to circularize instantaneously upon RLO with an orbital separation equal to the binary's separation at periastron. Alternatively,  we also allow for a user to choose to circularize the orbit assuming angular momentum is conserved.\footnote{This option results to circularized orbits where the star does not fill its Roche lobe anymore. In this case, we allow for the start to further evolve until it fills its Roche lobe again, but without changing the orbit.} 
Likewise, this step of evolution also ends if a non-degenerate star reaches the end of its life, in which case the binary is sent to a step that handles core collapse (Section~\ref{sec:core_collapse}).

A third stopping condition exists for binaries consisted of two COs, which  merge due to gravitational wave radiation. Note that when modeling two COs, only effects due to gravitational wave radiation contribute to orbital evolution. We calculate the orbital decay until the merger or the maximum simulation time is reached. 

Orbital evolution during the detached step is due to a combination of the relevant pieces of physics, which we assume have additive effect:
\begin{eqnarray}
    \dot{a} = \dot{a}_{\rm wind}  + \dot{a}_{\rm tides,1} + \dot{a}_{\rm tides,2}  + \dot{a}_{\rm GR},
\label{eq:a_dot_total}\\
        \dot{e} = \dot{e}_{\rm tides,1} + \dot{e}_{\rm tides,2} + \dot{e}_{\rm GR},
\label{eq:e_dot_total}\\
    \dot{\Omega}_{1} = \dot{\Omega}_{\rm wind,1} + \dot{\Omega}_{\rm intertia,1} + \dot{\Omega}_{\rm tides,1} + \dot{\Omega}_{\rm mb,1},
    \label{eq:omega_dot_total1}\\
    \dot{\Omega}_{2} = \dot{\Omega}_{\rm wind,2} + \dot{\Omega}_{\rm intertia,2} + \dot{\Omega}_{\rm tides,2} + \dot{\Omega}_{\rm mb,2}.
    \label{eq:omega_dot_total2}
\end{eqnarray}
The orbital separation, eccentricity and stellar spins are evolved using a set of self-consistent, coupled equations. We describe each of the terms in Eq.~\eqref{eq:a_dot_total}--\eqref{eq:omega_dot_total2} below.

{\bf Mass Loss:}
We ignore mass accretion onto a star (either non-degenerate or CO) from a  companion star's wind. So non-degenerate stars will only lose mass due to their own stellar winds, with the mass lost carrying away the specific orbital angular momentum of the mass-loosing star \citep[Jeans-mode mass loss; for a review, see][]{2006csxs.book..623T}:
\begin{eqnarray}
\dot{a}_{\rm wind} &=& 
    -a\, {\dot{M}_{\rm w,1} + \dot{M}_{\rm w,2} \over {M_1+M_2}}.
\end{eqnarray}
For CO binary components, $\dot{M}_{\rm w} = 0$, while in general for spherical, isotropic fast winds, the orbit-averaged $\dot{e}$ due to winds is zero. We discuss the effect of mass loss on stellar spin later in this section.

{\bf Tides:}
Changes in the orbit's period and eccentricity due to tidal forces are described by a set of ordinary differential equations, according to \citet{1981A&A....99..126H}. 
In order to be able to compute tidal spin--orbit coupling, we treat the donor star of mass $M$,  radius $R$ and of moment of inertia  $I$, as a solid body rotating with angular velocity $\Omega$. 
The initial angular-momentum budget of the non-degenerate star is assumed to be the same as from the end of the previous step.

The change of the orbital separation due to tidal forces on the first star (subscript 1) is given by:
  \begin{eqnarray}\label{eq:dotA_tides}
   \dot{a}_{\rm tides, 1} & = &
    -6\, \left( {k \over T}\right)_1 \,{M_1(M_1+M_2) \over M_1^2} \left( {R_1 \over a} \right)^8 {a \over {\left( 1 - e^2 \right)^{15/2}}} \nonumber \\
   & &  \times \left[ f_1 \left( e^2 \right) - \left( 1 - e^2 \right)^{3/2} f_2 \left( e^2 \right) {\Omega_1 \over \Omega_{\rm orb}} \right] \, ,
 \end{eqnarray}
where $\Omega_{\rm orb} = {2\pi  /  P_{\rm orb}}$ is the  mean orbital angular velocity. When both stars are non-degenerate, they each have their own contribution to the orbit's evolution. Therefore an analogous equation exists providing $\dot{a}_{\rm tides, 2}$, where $R_1$ is replaced with $R_2$, the $k / T$ term is calculated for the secondary star, and $M_1$ and $M_2$ are switched.

The $k / T$ term in Eq.~\eqref{eq:dotA_tides} depends on a star's structure and the associated physical process of tidal dissipation. We calculated them separately for dynamical and equilibrium tides, in the same way as in our detailed, binary-star model grids (Section~\ref{sec:Tides}) described in Eq.~\eqref{eq:kt_rad} and Eq.~\eqref{eq:kt_conv}, respectively. We apply the maximum of these two at each timestep. 

The change of the orbital eccentricity and the stellar spin from tidal forces is also calculated as
  \begin{eqnarray}\label{eq:dote_tides}
   &\dot{e}_{\rm tides, 1} & = 
    -27\, \left({k \over T}\right)_1 \,{M_2(M_1+M_2) \over M_1^2} \,  \left( {R_1 \over a} \right)^8  {e \over {\left( 1 - e^2 \right)^{13/2}}}  \nonumber \\
   & & \times \left[ f_3 \left( e^2 \right) - {11 \over 18} \left( 1 - e^2 \right)^{3/2} f_4 \left( e^2 \right) {\Omega_1 \over \Omega_{\rm orb}} \right] \, ,
 \end{eqnarray}
 \begin{eqnarray}\label{eq:dotomega_tides}
   \dot{\Omega}_{\rm tides, 1} & = &
    3\, \left({k \over T}\right)_1 \, \left ({M_2 \over M_1} \right)^2 \left ({M_1 R_1^2 \over I_1} \right) \left( {R_1 \over a} \right)^6 {\Omega_{\rm orb} \over {\left( 1 - e^2 \right)^{6}}}  \nonumber \\
   & & \times \left[ f_2 \left( e^2 \right) - \left( 1 - e^2 \right)^{3/2} f_5 \left( e^2 \right) {\Omega_1 \over \Omega_{\rm orb}} \right] \, .
 \end{eqnarray}
As in the Eq.~\eqref{eq:dotA_tides} for $\dot{a}_{\rm tides}$, when the companion star is non-degenerate,  $\dot{e}_{\rm tides,2}$ and $\dot{\Omega}_{\rm tides,2}$ terms exist, which can be calculated by switching subscripts $1$ and $2$ in Eq.~\eqref{eq:dote_tides} and \eqref{eq:dotomega_tides}.
The $f_i(e^2)$, $i=1-5$ terms in Eq.~\eqref{eq:dotA_tides}, \eqref{eq:dote_tides}, and \eqref{eq:dotomega_tides} can all be found in \citet{1981A&A....99..126H}.

{\bf Stellar Evolution:}
During the detached orbital evolution, we also take into account the change of stellar spin due to the evolution of the stars themselves. 
This includes spin down because of wind mass loss that carries away the specific angular momentum of the star's surface,  as well as changes in its spin due to the evolution of its moment of inertia due to changes in its internal structure, 
\begin{equation}\label{eq:dotomega_wind_deform}
   \dot{\Omega}_{\rm wind+inertia,1} = \frac{2}{3} {R_1^2 \, \Omega_1  \over I_1} \dot{M_1}  - {\Omega_1 \over I_1} \, \dot{I}_1  \, ,
 \end{equation}
where 
$\dot{I}$ the rate of change of its moment of inertia. For binaries in which both stars are non-degenerate, an equation equivalent to Eq.~\eqref{eq:dotomega_wind_deform} exists for $\dot{\Omega}_{\rm wind+inertia,2}$.

Non-degenerate stars tend to spin down, due to their expansion and their wind mass loss. However, they may also be spun up in phases where they contract. 
Therefore, for numerical-stability reasons we artificially limit the second term in Eq.~\eqref{eq:dotomega_wind_deform} to  $+100\,{\rm rad\,yr^{-2}}$ (usually reached during a sudden contraction to form a WD). 
Although we take into account the effect of stellar spin on the orbit via tidal spin--orbit coupling, we do not include effects of spin on the stellar structure, such as stellar deformation, rotational mixing or rotationally-enhanced winds.

{\bf Magnetic Braking:}
In case the binary contains low-mass non-degenerate stars, spin-down due to magnetic braking can become important i.e., the loss of  spin angular momentum due to ionized material ejected from the star that is trapped in its own radial magnetic field. 
The spin-down rate is given by
\begin{equation}\label{eq:dotomega_magnetic}
\dot{\Omega}_{\rm mb,1}  = {\tau_{\rm mb,1} \over I_1} \, ,
\end{equation}
where $\tau_{\rm mb,1}$ is the torque calculated as in Eq.~(36) of \citet{1983ApJ...275..713R},
\begin{equation}\label{eq:tau_magnetic}
\tau_{\rm mb,1} = - 6.82 \times 10^{34} {\rm dyn\,  cm} \left({M_1 \over \Msun}\right) \left({R_1 \over \Rsun}\right)^{\gamma_{\rm mb}} \left({\Omega_1 / 2\pi \over 1/ {\rm day}}\right)^3,
\end{equation}
with $\gamma_{\rm mb}=4$ \citep{1981A&A...100L...7V}. 
We apply the full torque to all non-degenerate stars below $1.3\,\Msun$ and assume no magnetic braking for stars above $1.5\,\Msun$, with linear interpolation in-between. Again, for binaries with two low-mass, non-degenerate stars, an equation equivalent to Eq.~\eqref{eq:dotomega_magnetic} exists for $\dot{\Omega}_{\rm mb,2}$.

{\bf Gravitational Wave Radiation:}
Finally, we take into account orbital changes due to gravitational radiation \citep{1964PhRv..136.1224P,1992MNRAS.254..146J},
\begin{eqnarray}\label{eq:dotA_GR}
\dot{\alpha}_{\rm GR}  &=& 
  - {{2} \over {15}}\,{\nu c \over {\left( 1-e^2 \right)^{9/2}}} \left[ {{G \left( M_1 + M_2 \right)} \over 
  {a\, c^2}} \right]^3 \nonumber \\
  & & \left[ g_1\left(e^2\right) - {1 \over 28}{{G \left( M_1 + M_2 \right)} \over {a\, c^2}}g_2\left(e^2\right) \right] \, ,
\end{eqnarray}
\begin{eqnarray}\label{eq:dote_GR}
\dot{e}_{\rm GR}  &=& 
  - {1 \over {15}}\, {{\nu\, c^3} \over {G \left( M_1 + M_2 \right)}} \left[ {{G \left( M_1 + M_2 \right)} \over 
  {a\, c^2}} \right]^4 {e \over {\left( 1-e^2 \right)^{7/2}}} \nonumber \\
& & \left[ g_3\left(e^2\right) - {1 \over 56}{{G \left( M_1 + M_2 \right)} \over {a\, c^2}}g_4\left(e^2\right) \right] \, ,
\end{eqnarray}
where 
$\nu = M_1\,M_2/(M_1 + M_2)^2$, and the functions $g_i(e^2)$, $i=1-4$, are defined as:

\begin{eqnarray}
g_1 \left( e^2 \right) &=& \left(96 + 292e^2 + 37e^4 \right)\left(1-e^2 \right),\\
g_2 \left( e^2 \right) &=& \left(14008 + 4704\nu \right) + \left(80124 + 21560 \nu \right)e^2 + \nonumber \\
& & \left(17325 + 10458 \nu\right)e^4 - \nonumber \\
& &{1\over 2}\left(5501 - 1036 \nu \right)e^6,\\
g_3 \left( e^2 \right) &=& \left(304 + 121e^2\right)\left(1-e^2 \right),\\
g_4 \left( e^2 \right) &=& 8\left(16705 + 4676\nu \right) + 12\left(9082 + 2807 \nu \right)e^2 - \nonumber \\
& &\left(25211 + 3388 \nu\right)e^4 \, .\label{eq:g_GR}
\end{eqnarray}

These general-relativistic terms of orbital evolution are usually negligible apart from cases of close binaries. For binaries consisted of two COs, only Eq.~\eqref{eq:dotA_GR}--\eqref{eq:g_GR} govern the evolution of the binary's orbit.

\subsection{Common-envelope evolution}\label{sec:common_env}
\label{sec:common_envelope}

Binary interactions can lead to a dynamically unstable mass-transfer phase \citep[][]{2013A&ARv..21...59I, 2020cee..book.....I}. 
We have described in Section~\ref{sec:binary_physics_instability} all the conditions that are assumed to trigger an unstable mass-transfer episode: a maximum mass-transfer rate of $0.1\,\rm M_{\odot}\,yr^{-1}$, L$_2$ overflow, a contact phase with a post-MS star, or exceeding the threshold of the trapping radius during accretion onto a CO. 

If the donor star that triggered the unstable phase is in its MS phase or is a stripped He star during its He core-burning phase, we assume that the two stars promptly merge, as no distinct core has formed yet in its interior.
In v1.0 of \posydon{} we do not follow the further evolution of stellar merger  products. 

For all the other donor stellar states, a trigger of unstable mass transfer is assumed to lead to a CE phase. 
If the donor has a H-rich envelope at the beginning of the phase, this envelope is considered to form the CE, inside of which the  donor's He-rich core and its binary companion will spiral-in. For stripped donors, the He-rich envelope engulfs the companion which spirals in around the donor's C/O core. 
In case the companion star also has a giant-like structure with a distinct core-envelope separation, (i.e., anything but a MS star, a He star in its He-MS, or a CO), then its envelope also may contribute in a (double) CE.

The outcome of the CE phase is calculated using the $\alpha_{\rm CE}$--$\lambda_{\rm CE}$ prescription \citep{1984ApJ...277..355W, 1988ApJ...329..764L}, which equates a fraction $\alpha_{\rm CE}$ of the orbital energy released during the spiral-in with the binding energy of the CE. 
The $\alpha_{\rm CE}$ parameter is set equal to $1$ in the example population runs shown in Section~\ref{sec:populations}, following previous population synthesis works \citep[e.g.,][]{2002MNRAS.329..897H}, but is, in general, a free parameter in \posydon{}. 

\begin{figure}[t]\center
\includegraphics[width=0.48\textwidth]{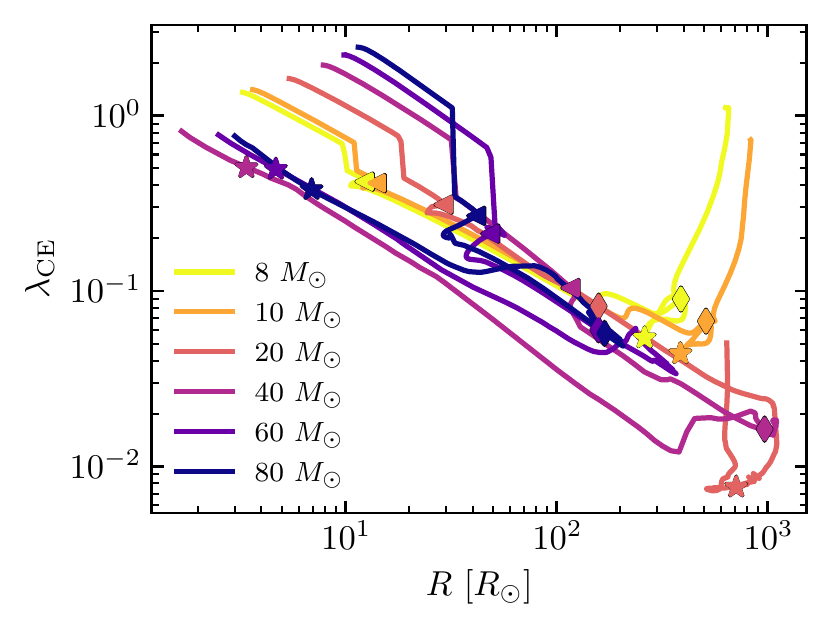}
\caption{Evolution of $\lambda_{\rm CE}$ parameter of \posydon{} single-star models of different initial masses. 
For these calculations, the assumed core--envelope boundary is located at the point where the H mass fraction drops below 10\%. 
The triangle, diamond, and star markers represent the start of shell H burning, the start of core He burning, and the end of core He burning, respectively.}
\label{fig:lambda_ce}
\end{figure}  

The parameter $\lambda_{\rm CE}$ has been introduced to parametrize the binding energy of the envelope using the total stellar radius and mass \citep{1990ApJ...358..189D}. 
In \posydon, $\lambda_{\rm CE}$ values are calculated from the detailed stellar profile of the donor star at the beginning of CE (or of both stars, in case of a double CE). 
This is an important quantitative improvement of \posydon,  compared to pBPS codes. 
The latter need to adopt $\lambda_{\rm CE}$-value fits from the literature, based on single-star models with often inconsistent stellar-physics, and apply them to post-interacting stars. In our common CE energy  calculation, we integrate both the gravitational and the internal  energy of the envelope from the detailed stellar structure profiles of our binary models,
subtracting the recombination energy from the internal energy. 
In Figure~\ref{fig:lambda_ce} we show the $\lambda_{\rm CE}$ for a few example single-star \posydon{} models. 
The parameter $\lambda_{\rm CE}$ tends to decrease as the star evolves and expands as a giant. However, for initially very massive ($\gtrsim$ 20 $\Msun$) stars that strip their H-rich layers due to their own wind mass loss,  $\lambda_{\rm CE}$ increases again. 
We find comparable trends and values with other works that study the detailed stellar structures of giant stars \citep[e.g.,][]{2016A&A...596A..58K, 2021A&A...645A..54K}.

The binding energy of the envelope (and thus the outcome of the CE phase) is sensitive to the exact assumed core-envelope boundary of the donor, as the deeper envelope layers tend to be the most tightly bound \citep{2000A&A...360.1043D, 2013A&ARv..21...59I, 2019ApJ...883L..45F}. 
For this reason, we allow for different core-envelope boundaries, defined for H-rich stars as the outermost layer where the H mass fraction drops below $0.3$, $0.1$ (default), and $0.01$ and for stripped-stars when the sum of H and He drops below $0.1$. 

Given the properties of the binary at the onset of the CE, the assumed $\alpha_{\rm CE}$ value and the estimated $\lambda_{\rm CE}$, one can calculate how much a binary's orbit shrinks in order for the released orbital energy to fully unbind the CE. 
The final post-CE orbital separation $a_{\rm post,CE}$ is given by solving \citep[][]{1984ApJ...277..355W}:
\begin{equation}
    \frac{GM_{\rm don,core}M_{\rm acc}}{2a_{\rm post,CE}}-\frac{GM_{\rm don}M_{\rm acc}}{2a_{\rm pre,CE}} = \frac{GM_{\rm don}M_{\rm don,env}}{\alpha_{\rm CE} \lambda_{\rm CE}R_{\rm don}},    
\end{equation}
where $M_{\rm don}$, $M_{\rm don,core}$, and  $M_{\rm don,env}$ are the total, core and envelope masses of the donor star,  $R_{\rm don}$ the donor star's radius, $M_{\rm acc}$ the mass of the accreting star, and $a_{\rm pre,CE}$ the orbital separation at the onset of the CE.
If the final estimated $a_{\rm post,CE}$ is such that neither the accreting star nor the stripped core of the primary star are filling their respective Roche lobes, then the CE is consider successful and results in a detached, circular, tight binary. 
Alternatively, the binary is assumed to merge, and its evolution is not further followed in v1.0 of \posydon{}.

One complication with the flexibility we offer regarding the core--envelope boundary definition is that the post-CE stripped donor star might still contain some H in its outer layers, while in the next evolutionary steps we assume that the H envelope is fully removed. Exactly how much H remains depends on a user's choice of 0.01, 0.1, or 0.3 for a fractional H abundance when defining the core-mass boundary.
We account for this inconsistency by assuming that either the remaining H-rich layers are either removed by stellar winds or these layers re-expand after the CE and are removed via stable mass transfer \citep[e.g.,][]{2019ApJ...883L..45F}. 
Both assumptions result in slight corrections to the post-CE donor masses and orbital separation. 
Although the former assumption is the default one in \posydon{}, we find they both lead to correction at the level of only a few percent, and thus the choice between the two is in practice inconsequential.

\subsection{Core-collapse and compact-object formation \label{sec:core-collapse}}
\label{sec:core_collapse}

The end fate of stars primarily depends on their masses. 
The most massive stars undergo all nuclear burning phases (hydrogen, helium, carbon, neon, oxygen, silicon) up to the formation of an iron core . 
The iron core keeps growing by silicon shell burning to a mass of around the Chandrasekhar mass limit $\sim 1.44\,\Msun$ when electron degeneracy pressure can no longer stabilize the core and it collapses. 
This runaway process can lead to the explosion of a star in a SN or to a direct collapse into a BH and it is known as core-collapse SN (CCSN) \citep[see][for a review]{2007PhR...442...38J}. 

Lower mass stars do not complete all nuclear burning phases. 
For stars which do not ignite oxygen but for which their He-cores masses are $1.4\,\Msun \lesssim M_\mathrm{He\mbox{-}core} \lesssim 2.5\,\Msun$ \citep{2004ApJ...612.1044P} we assume a star collapses into a NS in an electron-capture SN (ECSN). \posydon{} alternatively includes the option to determine whether a star undergoes an ECSN based on its C/O core mass: $1.37\,\Msun \lesssim M_\mathrm{C/O\mbox{-}core} \lesssim 1.43\,\Msun$ \citep{2015MNRAS.451.2123T}. Stars with core masses below the lower limit for ECSN evolve into white dwarfs.

\begin{figure}[t]\centering
\includegraphics{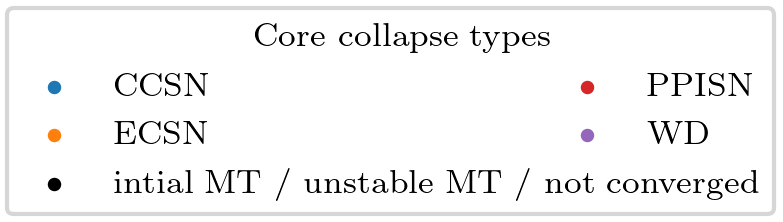}\\
\includegraphics{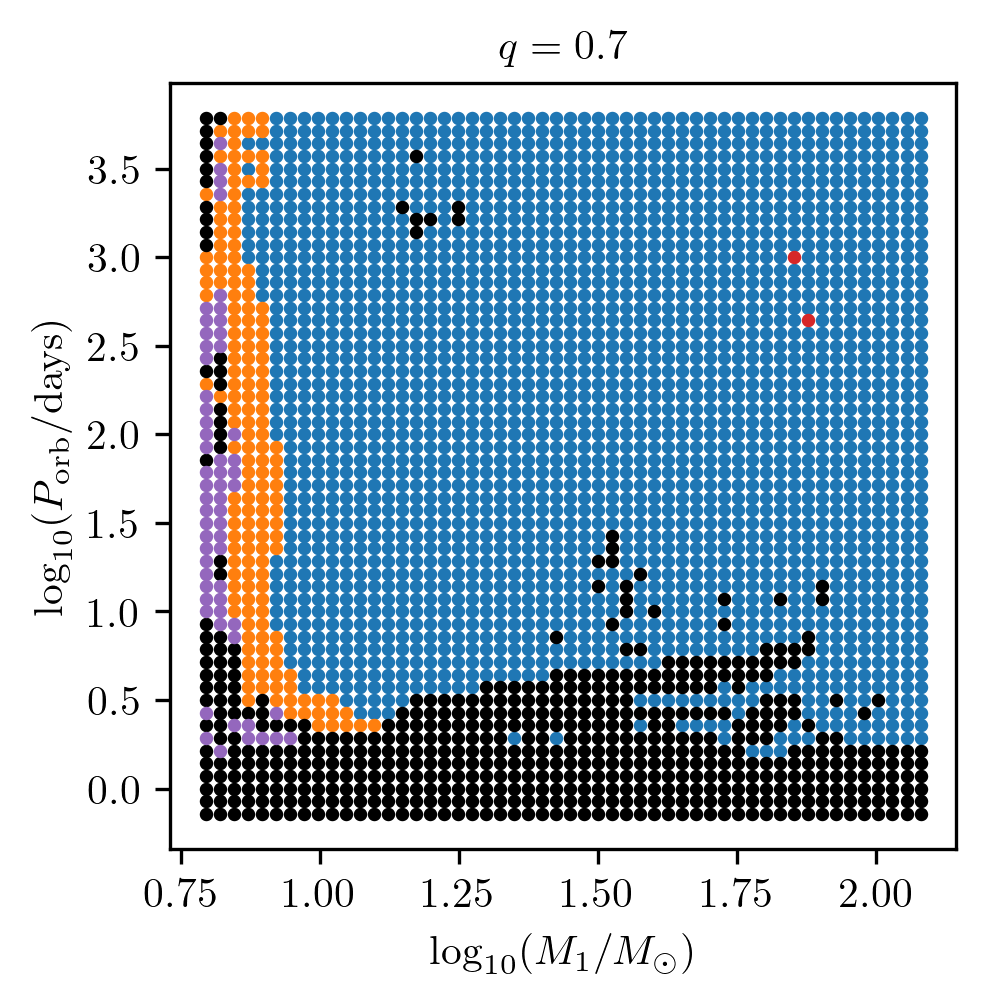}
\caption{The core-collapse type for the $q=0.7$  slice of our binary-star model grid composed of two H-rich stars. 
We distinguish between WD formation, electron caption SN (ECSN) following \citet{2005MNRAS.361.1243P}, core-collapse SN (CCSN) and pair-pulsational instability SN (PPISN) following \citet{2019ApJ...882...36M}. 
Models that did not reach the end of stellar evolution are indicated in black.}
\label{fig:HMS-HMS_collpase_SN_type}
\end{figure}

In Figure~\ref{fig:HMS-HMS_collpase_SN_type}, we show, for a slice at a fixed initial mass-ratio $q=0.7$ of the binary-star model grid composed of two H-rich stars, the core-collapse type as a function of initial  orbital period primary star mass. The transition region between the ECSN and CCSN, occurs at ZAMS masses of $\simeq 8\, \Msun$ \citep[consistent with previous studies, e.g.,][]{1984ApJ...277..791N, 2014ApJ...797...83J}, but depends somewhat on the initial $P_{\rm orb}$.

\subsubsection{Pulsational pair-instability SN}

During the post-carbon burning phase of massive stars (not modelled here), photons produced in the core can be energetic enough to produce electron--positron pairs, softening the equation of state and diminishing the pressure support of the core \citep[][and references therein]{2007Natur.450..390W}. 
In such stars the core rapidly contracts and the temperature increases, leading to explosive oxygen burning \citep[e.g.,][]{2015ASSL..412..199W} that creates a series of energetic pulses which eject material from the star surface. 
This phenomenon of material ejection due to pulses is known as pulsational pair instability SN (PPISN) and occurs for stars with He-core masses in the range $\sim [32,64] \, \Msun$,  \citep[][]{2016MNRAS.457..351Y,2017ApJ...836..244W,2019ApJ...882...36M,2020MNRAS.493.4333R}.
For more massive stars with He-core mass in $\sim [61,124] \, \Msun$, the first pulse is so energetic that can unbind and destroy the whole star in a so-called pair instability SN \citep[PISN;][]{1964ApJS....9..201F,1967ApJ...148..803R,1967PhRvL..18..379B}, leaving no remnant behind. 

To identify systems that will undergo PPISN and PISN, we adopt a polynomial fit, as implemented in  \citet[][see Eq.~4 therein]{2020ApJ...898...71B}, to \mesa{} single-star simulations (at $Z=0.1Z_\odot$) by \citet[][see Table~1 therein]{2019ApJ...882...36M}. This fitting formula is used to map the He core mass at carbon depletion in the range $31.99\,\Msun \leq M_\mathrm{He\mbox{-}core} \leq 61.10\,\Msun$ to the stellar mass collapsing to form the CO. 
We use PPISN models computed at 1/10-th solar metallicity, as it was shown that such a limit is independent of metallicity \citep{2019ApJ...887...53F}, while highly dependent on the uncertain ${}^{12}\mathrm{C}(\alpha,\gamma){}^{16}\mathrm{O}$ reaction rates \citep{2020ApJ...902L..36F}. 
In our case, these reaction rates follow \citet{cyburt2010}, consistent with the rates used by \citet{{2019ApJ...882...36M}}. 

In Figure~\ref{fig:HMS-HMS_collpase_SN_type}, we can identify two systems that enter the regime of PPISN as they possess a He-core with mass slightly larger than $31.99\,\Msun$ at carbon depletion. 
Other mass ratio slices show a few more similar systems but without becoming statistically relevant. We expect PPISNe to be more present at sub-solar metallicities, as stellar wind-mass loss at $Z_\odot$ prevent the stars from reaching carbon depletion with a He-core mass in the relevant mass range for PPISNe.

\begin{figure*}\center
\includegraphics{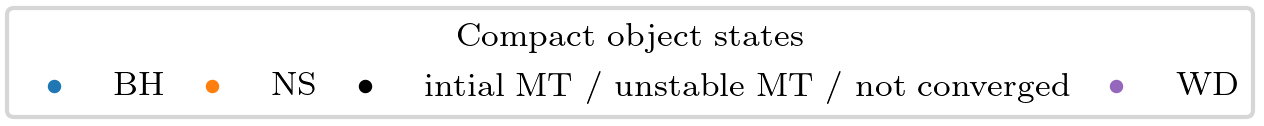}\\
\includegraphics{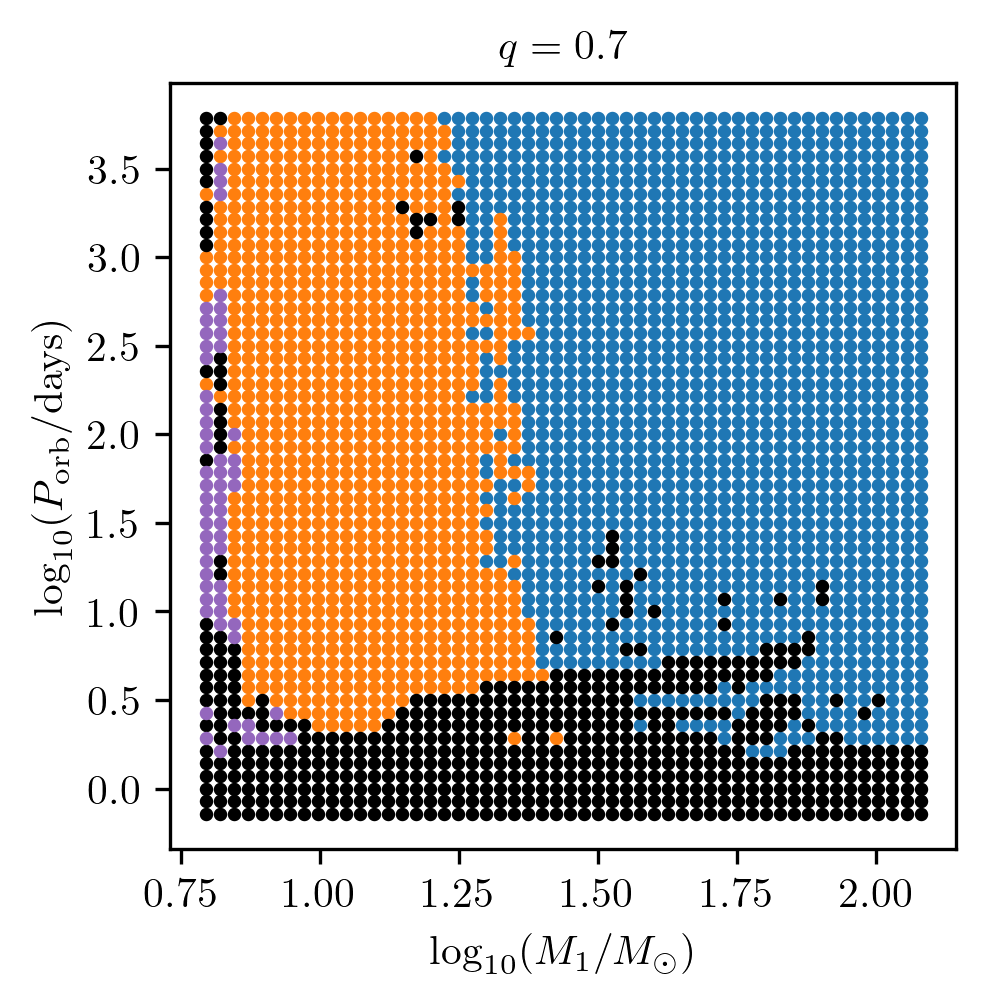}
\includegraphics{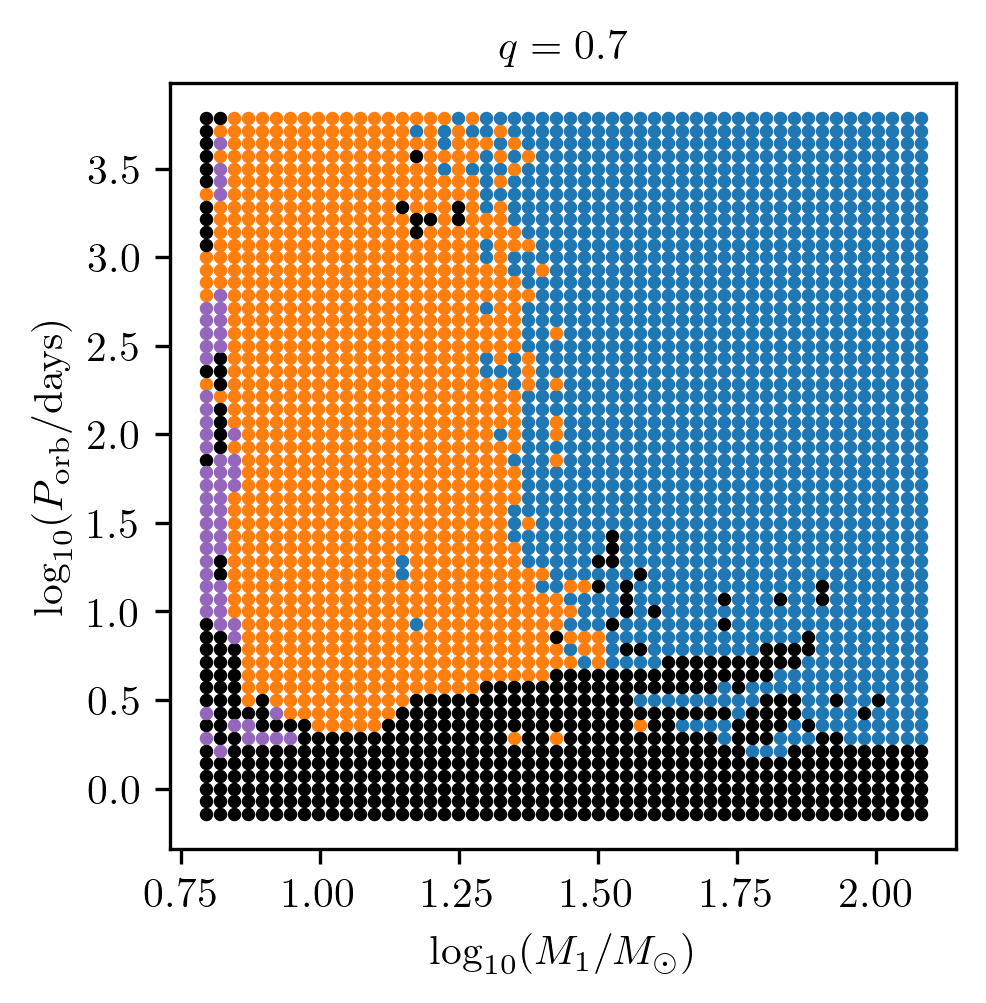}
\caption{The CO state for the $q=0.7$ grid slice of HMS--HMS \mesa{} simulations in the initial primary mass--orbital period plane. 
We distinguish between the COs: white dwarf (WD), neutron star (NS) and black hole (BH) according to the legend. 
Models that did not reach the end of the stellar evolution are indicated in black. The two panels compare the \citet{2012ApJ...749...91F} delayed core-collapse mechanism (left) with the outcome of \citet{2020MNRAS.499.2803P} N20 core-collapse engine (right).}
\label{fig:HMS-HMS_collpase_state}
\end{figure*} 

\subsubsection{Remnant baryonic mass}\label{sec:rembar}

In this version of \posydon{}, we calculate the mass left behind by the collapse using different models: (i) direct collapse where all the stellar mass is conserved; (ii) fits to the results of two-dimensional core-collapse models of \citet{2012ApJ...749...91F}; (iii) nearest neighbor interpolations of the results of the detailed one-dimensional core-collapse models of \citet{2016ApJ...821...38S}, or (iv) with the explodability criteria of \citet{2020MNRAS.499.2803P}. 
The last is our default option.

\citet{2012ApJ...749...91F} presents two mechanisms which are known as \emph{rapid} and \emph{delayed} based on how quickly convective instabilities are expected to grow after core bounce. 
The rapid prescription produces a mass gap between BHs and NSs by assuming strong convection which allows instabilities to grow quickly after core bounce, producing a more energetic SN explosion. 
In contrast, the delayed mechanism produces a continuous spectrum of compact remnant masses. 
Both prescriptions determine the baryonic mass of the compact remnant $M_\mathrm{rembar}$ given the pre-SN C/O core mass, $M_\mathrm{CO-core}$. 
More precisely, $M_\mathrm{C/O-core}$ determines whether the star explodes into a SN, and what fraction, $f_\mathrm{fb}$, of the ejected mass falls back onto the CO. 
In the case that the star directly collapses to form a BH, $f_\mathrm{fb}=1$. 
For the rapid prescription, direct collapse occurs for $M_\mathrm{C/O-core} \geq 7.6\,\Msun$ while for the delayed prescriptions direct collapse occurs for $M_\mathrm{C/O-core} \geq 11\,\Msun$.

In \citet{2016ApJ...821...38S}, the outcome of the collapse of their pre-SN models has been calibrated against the well-studied SN 1987A progenitor. 
We have implemented several of their SN engine calibrations,  namely N20, S19.8, W15, W20 and W18, although for the training of the initial--final interpolation  we only consider the default N20 option, which is the most optimistic option for successful explosions. 
In contrast to \citet{2012ApJ...749...91F} results, \citet{2016ApJ...821...38S} finds sharply varying behavior between the initial star mass and the final core properties, linked to convective carbon-burning episodes occurring in the later evolutionary phases. 
This results in a region of the parameter space where the outcome of the collapse, i.e., NS and BH formation, appears stochastic in its nature. 
To mitigate interpolation errors, we determine the remnant baryonic mass of a collapsing star using a nearest neighbor technique on the He-core mass at carbon depletion to map our stars to the \citet{2016ApJ...821...38S} simulation results. 

In the \citet{2020MNRAS.499.2803P} prescription, the C/O core mass and the average carbon abundance of the core at carbon ignition are used to determine the explodability of the pre-SN core. For every single- and binary-star model in our grids, we store these two values, and by applying a $k$-nearest neighbour interpolation, with $k=5$, we map to the explodability parameters $M_4$ and $\mu_4$ from \citet{2016ApJ...818..124E}, as described in \citet{2020MNRAS.499.2803P}. 
These two explodability parameters allow us to infer whether a SN is successful, and, if so, we estimate the resulting NS mass to be approximately equal to $M_4$.
We assume that BHs are produced only from failed explosions which result in a direct collapse.  
Finally, for the \citet{2020MNRAS.499.2803P} prescription, we have implemented the same SN engine options as for \citep{2016ApJ...821...38S} with N20 as our default option, using the updated calibration from \citet{2020ApJ...890...51E}.

In Figure~\ref{fig:HMS-HMS_collpase_state}, we show a comparison between the final CO state for the same grid slice as Figure~\ref{fig:HMS-HMS_collpase_SN_type}, as predicted by \citet{2012ApJ...749...91F} \textit{delayed} prescription compared and the \citet{2020MNRAS.499.2803P} prescription, based on the N20 engine. The differences between our choice of SN prescription are slight, but noticable when focusing on the NS/BH boundary. The \citet{2012ApJ...749...91F} \textit{delayed} prescription produces BHs for somewhat less massive stars, while the \citet{2020MNRAS.499.2803P} prescription shows a more variable boundary between NSs and BHs.

In both \citet{2016ApJ...821...38S} and \citet{2020MNRAS.499.2803P} prescriptions we assume fallback fractions of $f_\mathrm{fb}=1$ for BHs and $f_\mathrm{fb}=0$ for NSs. 
For \citet{2012ApJ...749...91F} prescriptions the fallback fractions are computed explicitly, with the exception of NS ECSN where we assume $f_\mathrm{fb}=0$.

\begin{figure*}[t]\center
\includegraphics{figures/TF1_legend.png}\\
\includegraphics{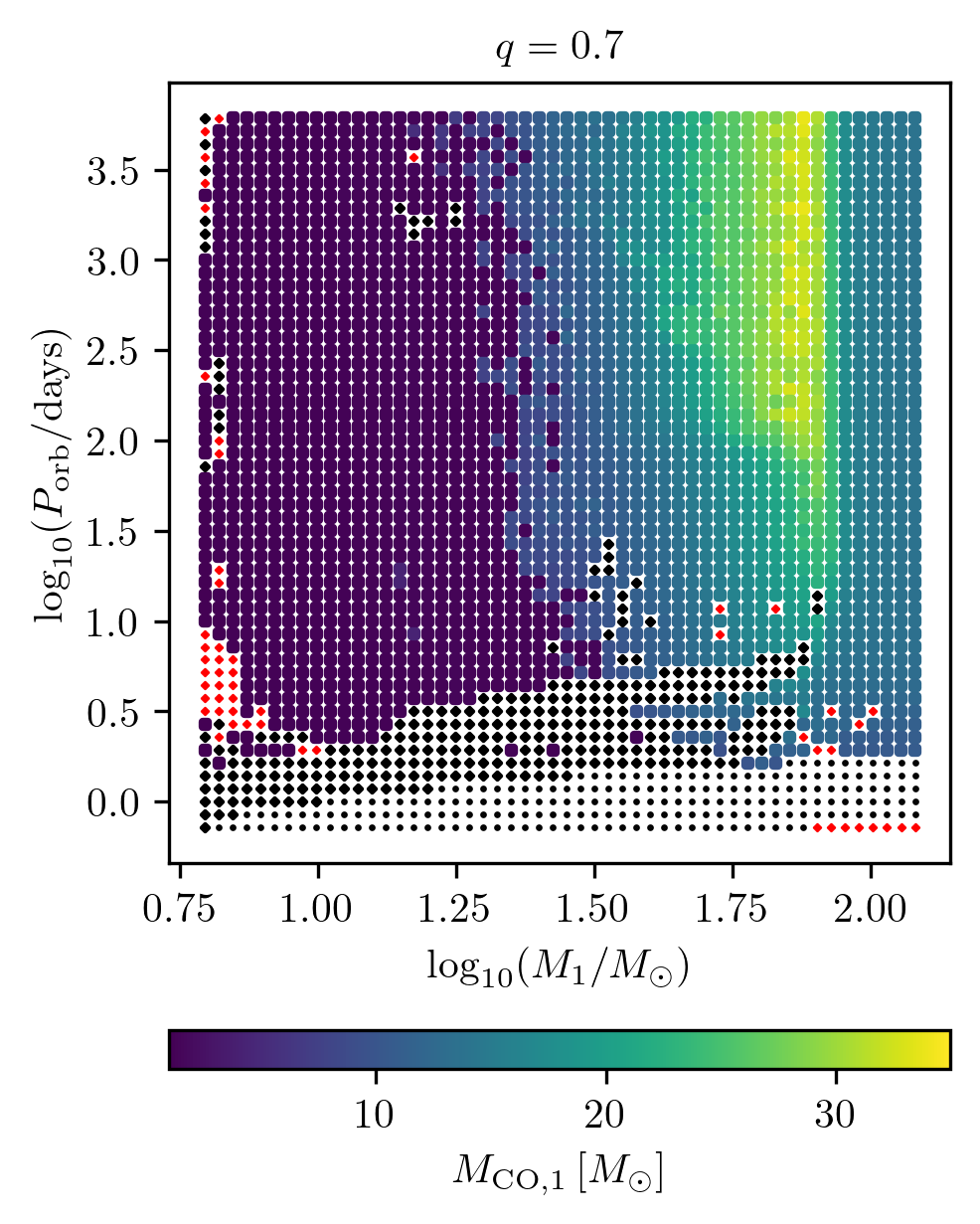}
\includegraphics{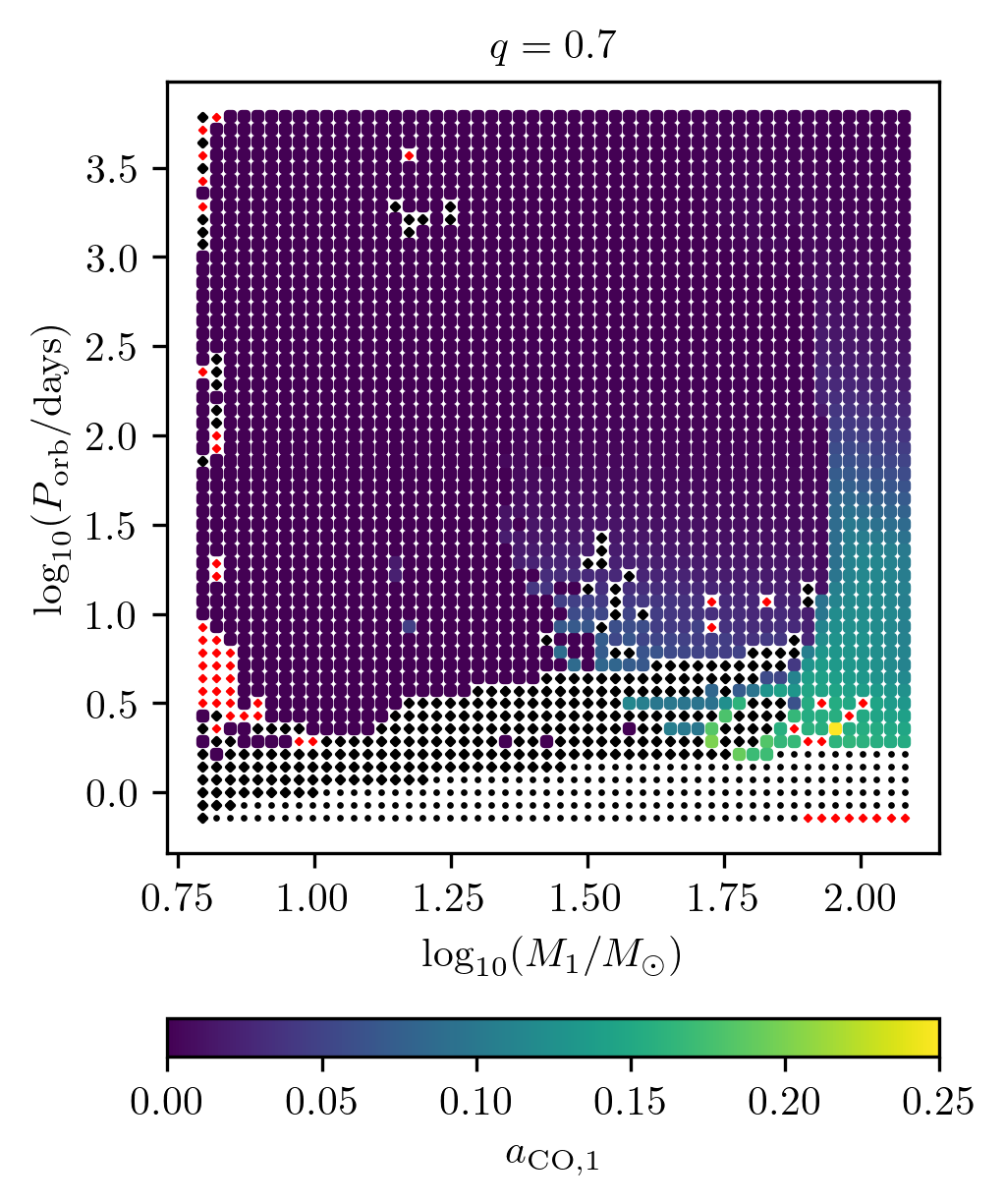}
\caption{The CO mass $M_\mathrm{CO1}$, and spin $a_\mathrm{CO1}$ for the $q=0.7$ grid slice of HMS--HMS \mesa{} simulations in the initial primary mass--orbital period plane as predicted by the \citet{2020MNRAS.499.2803P} N20 engine. 
We assume that both the neutrino mass loss up to $0.5\,\Msun$ and the ejected mass during core collapse carries away the corresponding angular momentum. 
Spinning BHs are formed in binary systems avoiding mass transfer or undergoing stable mass transfer during contact phase or Case A mass transfer, see Figure~\ref{fig:HMS-HMS_MESA_grid_TF12}.}
\label{fig:HMS-HMS_collpase_properties}
\end{figure*} 

\subsubsection{CO gravitational mass}\label{sec:grab_mass}

To convert the remnant baryonic mass to gravitational mass, we use the prescription by \citet{2020ApJ...899L...1Z}, which is an updated version of the one by \citet{1989ApJ...340..426L} based on the neutrino observations of SN\,1987A. 
This new conversion caps the maximum neutrino mass loss to $0.5\,\Msun$ (C.~Fryer, private communication) and removes any artificial discontinuity in the mass spectrum between NS and BH formation (in the case of direct collapse or the \citet{2012ApJ...749...91F} delayed mechanism) as
\begin{equation}
     M_\mathrm{grav} =
     \begin{cases}
     \dfrac{20}{3}\left(\sqrt{1 + 0.3 \dfrac{M_\mathrm{rembar}}{\Msun}} - 1\right){}\Msun,\, \\ \,\,\,\,\,\,\,\,\,\,\,\,\,\,\,\, 
     \,\,\,\,\,\,\,\,\,\,\,\,\,\,\,\,
     \,\,\,\,\,\,\,\,\,\,\,\,\,\,\,\,
     \,\,\,\,\,\,\,
     M_\mathrm{rembar}-M_\mathrm{grav} < 0.5 \, \Msun \\
     M_\mathrm{rembar} - 0.5{}\Msun \,\,\,\,\,\,\,\,\,\,\,\,\,\,\,\,\, \text{otherwise}. \\
   \end{cases}
\end{equation}
If $M_\mathrm{grav} < 2.5\,\Msun$ we classify the CO as a NS otherwise as a BH. 
There is a large uncertainty in the exact maximum NS mass and this range spans $2.0$--$2.7\,\Msun$ \citep{2010arXiv1012.3208L, 2017ApJ...850L..19M, 2018ApJ...852L..25R,2020ApJ...893..146A,2020PhRvD.102f3006S, 2021PhRvC.104c2802L, 2021ApJ...918L..28M, 2021ApJ...918L..29R}. 
In \posydon{}, the maximum neutron star mass is set to $M_\mathrm{NS}^\mathrm{max}=2.5\,\Msun$ \citep[see discussion in][and references therein]{2020ApJ...896L..44A}. 

In the left panel of Figure~\ref{fig:HMS-HMS_collpase_properties}, we show the gravitational mass of the CO as predicted by \citet{2020MNRAS.499.2803P} N20 prescription for the same grid slice as Figure~\ref{fig:HMS-HMS_collpase_state}.

Finally, in the case of BH formation, \posydon{} also allows us to take into account the detailed internal structure of the star at the moment of collapse. As illustrated in the next section, we can then make an estimate of the spin of the CO taking into account the angular momentum profile of collapsing star.

\subsubsection{Birth spins of COs}\label{sec:birth_spin}

We estimate the spin of the resulting BH following the collapse of the stellar profile as presented in \citet[][Appendix~D]{2021A&A...647A.153B}. 
For convenience we summarize here the key assumption of this procedure.  
The final mass and spin of the BH resulting from the collapse is calculated by following the accretion history of $M_\mathrm{rembar}$ soon after the direct collapse of the central core of the star which forms a proto-BH of mass $2.5\,\Msun$. The mass lost in neutrinos during the formation of this proto-BH, also carry away specific  angular momentum equal of that collapsing central part of the core that forms the proto-BH. 
If $M_\mathrm{rembar}<M_\mathrm{star}$ we assume the ejected mass takes away the outer layers of the star. 

The angular momentum content of the infalling material can in principle support the formation of an accretion disk. We consider a collapsing star to be a collection of shells with radius $r$, mass $m_\mathrm{shell}$ and angular velocity $\Omega_\mathrm{shell}$ that falls one by one onto the central BH. A shell of mass is accreted by the BH once it reaches the BH’s event horizon. 
The specific angular momentum of the in-falling material, $j(r,\theta) = \Omega_\mathrm{shell}(r) r^2 \sin(\theta)$, where $\theta$ is the polar angle, determines the properties of the accretion flow.
Disk formation occurs when the specific angular momentum of the shell $j(r,\theta)$ exceeds the specific angular momentum of the ISCO $j_\mathrm{ISCO}$. 
This condition can be redefined as the polar angle at which disk formation occurs as
\begin{equation}
    \theta_\mathrm{disk} = \arcsin{\left( \frac{j_\mathrm{ISCO}}{\Omega_\mathrm{shell}(r)r^2} \right)^{1/2}} \, . 
\end{equation}
The portion of the shell with $\theta < \theta_\mathrm{disk}$ will collapse directly onto the BH on a dynamical timescale transferring $j(r,\theta)$ to the hole, while the portion of the shell with $\theta \geq \theta_\mathrm{disk}$ will form a disk and transfer only $j_\mathrm{ISCO}$ to the BH. 
The disk will be accreted on a viscous timescale which is assumed to be much smaller than the fallback timescale of the following shell \citep{2019arXiv190404835B}. 

Therefore each collapsing shell contributes to the angular momentum of the BH by
\begin{equation}
\begin{split}
    J_\mathrm{shell} & \equiv J_\mathrm{direct} + J_\mathrm{disk} = \\
    & = \int_0^{\theta_\mathrm{disk}} M_\mathrm{shell} \Omega_\mathrm{shell}(r) \, r^2 \sin^3(\theta) \, \mathrm{d}\theta \, + \\
    & \,\,\,\,\,\, + \int_{\theta_\mathrm{disk}}^{\pi/2} M_\mathrm{shell} j_\mathrm{ISCO} \sin (\theta) \, \mathrm{d} \theta \, .
\end{split}
\end{equation} 
The mass-energy accreted onto the BH from the disk is $ M_\mathrm{disk} = \varepsilon \,M_\mathrm{shell} \cos (\theta_\mathrm{disk})$ while the fraction $\eta\equiv1- \varepsilon =1-[1 - 2G M_\mathrm{BH} /(3 c^2 r_\mathrm{ISCO})]^{1/2}$ is radiated away \citep{1970Natur.226...64B,1974ApJ...191..507T}. 
Here, $r_\mathrm{ISCO}$ is the radius of the ISCO of the accreting BH. This means that the resultant BH will have mass smaller than $M_\mathrm{grav}$ as a fraction of the disk will be radiated away.  The dimensionless spin parameter of the BH is updated after each shell is accreted onto the BH with the following relation
\begin{equation}
    a = \frac{cJ_\mathrm{BH}}{GM^{2}_\mathrm{BH}} \, ,
\end{equation}
where $J_\mathrm{BH}$ is the angular momentum of the BH and $M_\mathrm{BH}$ its mass after accreting the directly infalling part of the shell and, if formed, the thin disk.

The presented treatment is applicable only to the case of BH formation. 
For simplicity and the lack of firm alternatives in this version of \posydon{}, we assume a zero spin for NSs.

 In the right panel of Figure~\ref{fig:HMS-HMS_collpase_properties}, we show the CO spin for the $q=0.7$ mass ratio slice of the binary-star grid composed of two H-rich stars, as predicted by the stellar profile collapse assuming the remnant baryonic mass is determined with  \citet{2020MNRAS.499.2803P} N20 prescription. 
 
\subsubsection{SN kicks}\label{sec:SN_kicks}

During a SN, the binary system experiences abrupt mass loss, away from the center of mass, affecting its orbital parameters \citep{1961BAN....15..265B,1961BAN....15..291B}. 
Furthermore,  asymmetric ejection of matter \citep{1994A&A...290..496J,1996PhRvL..76..352B,2013MNRAS.434.1355J} or asymmetric emission of neutrinos \citep{1993A&AT....3..287B,2005ApJ...632..531S} can provide a momentum kick to the newly formed CO. 
Here we assume that the magnitudes of the asymmetric kicks ($v_{\rm k}$) are drawn from a Maxwellian distribution with dispersion $\sigma$:
\begin{equation}
    f(v_\mathrm{k}) = \sqrt{\frac{2}{\pi}}\frac{v^2_\mathrm{k}}{\sigma^3}\exp\left(-\frac{v^2_\mathrm{k}}{2\sigma^2}\right) \, .
\end{equation}
As our fiducial assumption we take $\sigma_\mathrm{CCSN}=265\,\mathrm{km\,s^{-1}}$ \citep{2005MNRAS.360..974H} and $\sigma_\mathrm{ECSN}=20\,\mathrm{km\,s^{-1}}$ \citep{2019MNRAS.482.2234G} for CCSN and ECSN, respectively. However, these velocities are free parameters.
\posydon{} supports multiple kicks rescaling options, e.g., if the prescription used to calculate the remnant baryonic mass assume mass loss, i.e., the fallback mass fraction $f_\mathrm{fb} <1$, the kick is then rescaled by $1-f_\mathrm{fb}$ \citep{2012ApJ...749...91F}. 
Alternatively BH kicks are rescaled by a factor $1.4\,\Msun/M_\mathrm{BH}$ (using the gravitational mass) while NS kicks are not rescaled (our default option). Finally a user can opt to either not rescale any kicks or turn off SN kicks altogether.

These kicks can tilt the orbit of the binaries, add eccentricity or disrupt it.
We take into account all these orbital changes including orbital changes for eccentric binaries following the analytical calculations of \citet{1996ApJ...471..352K,2012ApJ...747..111W}. 

We assume the collapsing star to lie on the origin of the coordinate system moving in direction of positive $y$-axis. 
The companion lies on the negative $x$-axis and $x$-axis completes the right-handed coordinate system \citep[cf.\ ][Figure~1]{1996ApJ...471..352K}. 
The semi-major axis after the kick $a_f$ is computed given the instantaneous orbital separation $r_\mathrm{i}= a_\mathrm{i}[1 - e_\mathrm{i} \cos(E)]$ pre-SN, where $E$ is the eccentric anomaly,  as
\begin{equation}
    a_\mathrm{f} = \left(
            \frac{2}{r_\mathrm{i}}  - \frac{v_\mathrm{k}^2 + v_\mathrm{r}^2 + 2v_\mathrm{k}^yv_\mathrm{r}}{GM^\mathrm{f}_\mathrm{tot}} \right)^{-1},
\end{equation}
where $M^\mathrm{f}_\mathrm{tot}$ is the binary total stellar mass after the core collapse, $v_\mathrm{r}$ is the pre-SN velocity of the collapsing star relative to the companion directed along the positive $y$-axis and $v_\mathrm{k}^y$ the $y$-axis component of the kick.
The eccentricity after the kick is then
\begin{equation}
        e_\mathrm{f} = \left\{1 - 
                    \frac{(v_\mathrm{k}^{z})^2 + [\sin(\psi) (v_\mathrm{r} + v_\mathrm{k}^y) - \cos(\psi) v_\mathrm{k}^x]^2}
            {G M^\mathrm{i}_\mathrm{tot} a_\mathrm{f}} r_\mathrm{i}^2
            \right\}^{{1}/{2}} \, ,
\label{eq:kickef}
\end{equation}
where  $M^\mathrm{i}_\mathrm{tot}$ is the binary total stellar mass before the core collapse, $\psi$ is the polar angle of the position vector of the collapsed star with respect to its pre-SN orbital velocity in the companion's reference frame and
\begin{equation}
    \sin(\psi) = \frac{\sqrt{G M^\mathrm{i}_\mathrm{tot} (1 - e_\mathrm{i}^2) a_\mathrm{i}}}{v_\mathrm{r}r_\mathrm{i}} \, .
\end{equation}
In the above equations, in the case the CO receive no natal kick $v_k=0\,\mathrm{km\,s^{-1}}$ but the star loses some mass, the orbit is still readjusted to conserve Kepler's third law.

We consider a binary is disrupted if it does not satisfy the condition that demands the post-SN orbit passes through the pre-SN position \citep{2005ApJ...625..324W,1975A&A....39...61F},
\begin{equation}
    1 - e_\mathrm{f}  \leq \frac{r_\mathrm{i}}{a_\mathrm{f}} \leq 1 + e_\mathrm{f} \, ,
\end{equation}
or if it is outside the limits of the amount of orbital contraction or expansion that can take place for a given amount of mass loss and a given magnitude of
the kick velocity \citep{2000ApJ...530..890K,2005ApJ...625..324W}
\begin{equation}
    2 - \frac{M^\mathrm{i}_\mathrm{tot}}{M^\mathrm{f}_\mathrm{tot}}\left(\frac{v_\mathrm{k}}{v_\mathrm{r}} + 1\right)^2 < \frac{r_\mathrm{i}}{A_\mathrm{f}}  < 2 - \frac{M^\mathrm{i}_\mathrm{tot}}{M^\mathrm{f}_\mathrm{tot}}\left(\frac{v_\mathrm{k}}{v_\mathrm{r}} - 1\right)^2 \, .
\end{equation}
Finally, we also verify that $e_\mathrm{f}$ does not exceed 1 or that the argument of the square root in Eq.~\eqref{eq:kickef} does not become negative, if this is the case the binary is considered to be disrupted.

\section{How \posydon{} Evolves an Individual Binary System}
\label{sec:flow}

To evolve a single binary within \posydon{}, we use a hierarchy of classes. Every binary system is represented as a {\tt BinaryStar} class containing two {\tt SingleStar} classes, each with attributes that define their current state. 
To evolve the binary through each step, we have implemented a Pythonic flow, which takes the combination of a binary's state and event, and each stars' state to direct a particular binary to its next step. 
We have a complete flow set as default, which can self-consistently track the evolution of binaries from their ZAMS state comprised of two H-stars, through all parts of the evolutionary tree shown in \autoref{fig:structure}.

All steps in \posydon{} are Python classes that update a binary via the user-defined {\tt call} method. 
Steps are  implemented based on on-the-fly calculations for evolutionary phases such as the CE or the core collapse, or based on pre-calculated grids of detailed binary-star models, which is a novel component of \posydon{}. 
In the latter case, to estimate the evolution of systems for which no detailed model exists for the exact initial binary properties, we use initial--final classification and interpolation algorithms, trained on the grids of detailed binary-star models (Section~\ref{sec:machine_learning}) or alternatively a nearest neighbor matching scheme. 
As the binary and its component stars evolve through these steps, the {\tt BinaryStar} and {\tt SingleStar} characteristics are appended to the objects, so every binary maintains a historical record of its evolution.

If any system enters a phase that does not require further evolution, we use an end step to halt the evolution.
This is used for binaries that either merge, disrupt, or reach the maximum physical time. 
For binaries whose evolution time ends in the middle of a step based on pre-calculated grid of models (e.g., if our star formation history randomly generates an initialization time for a binary within a few megayear of the end run time), we can no longer use our pre-trained classification and interpolation algorithms as these only apply to the end state of the binary. 
In these cases, we instead use the system's nearest-neighbor pre-computed track directly (interpolating full binary tracks is non-trivial and is being investigated for future versions of \posydon{}). 
This is the default behavior of all \mesa{} grid steps, with various classification and interpolation algorithms ready to use.

\begin{figure*}\center
\includegraphics[width=\textwidth]{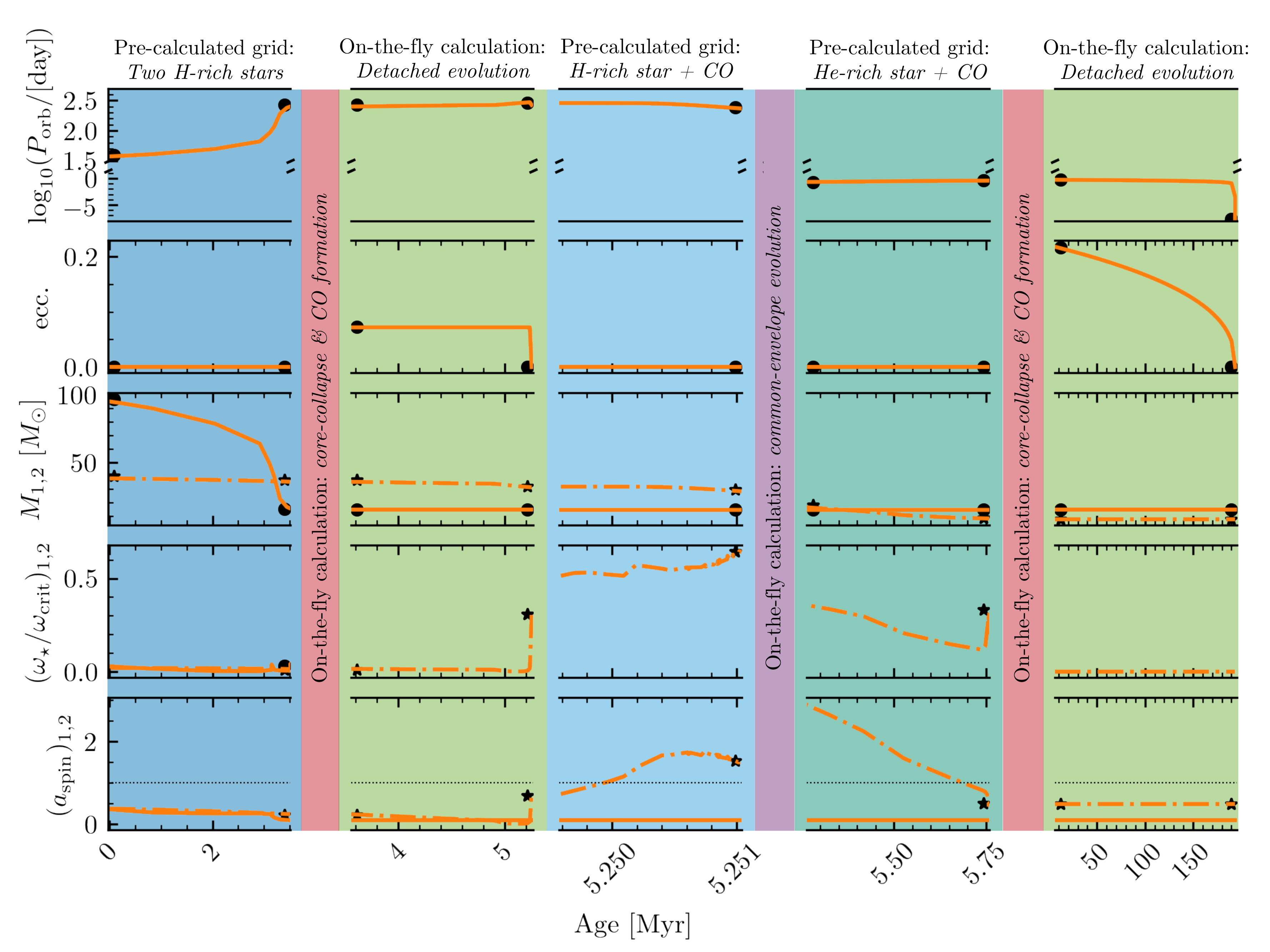}\\
\caption{Evolution over time, from ZAMS to binary BH formation, of one example binary system with initial  properties $M_{1} = 97.07\,\Msun$, $M_{2} = 39.86$\,\Msun\,in a $P_\mathrm{orb}=39.45$\,days. 
The top two rows of panels show the evolution of the binary's $P_\mathrm{orb}$ and $e$ for our nearest neighbor matching scheme (orange lines) and our initial--final interpolation method (black, circle markers). In the bottom three rows, we show the primary star's (1; solid lines) and secondary star's (2; dashed-dotted lines) properties using the nearest neighbor interpolation method and compare them against the same binary using the initial--final interpolation method (circle markers for primary, star markers for secondary). The binary's evolution is followed across its different evolutionary steps (note that the time scale varies for each step), through both \mesa{} grids (Section~\ref{sec:grids}) and on-the-fly calculations (Section~\ref{sec:other_physics}).
}
\label{fig:one_system}
\end{figure*}

Figure~\ref{fig:one_system} depicts the complete evolution of one particular binary from ZAMS to the formation of a BBH system that merges within a Hubble time. 
Each vertical, colored band indicates an extended stage of a binary's evolution, whereas the two core-collapse events and the CE phase are essentially instantaneous processes, occurring in between the other, extended phases. 

The initial masses of the system are $M_{1} = 97.07\,\Msun$, $M_{2} = 39.86$\,\Msun\,in a $P_\mathrm{orb}=39.45$\,days circular orbit. The first part of the evolution of the system is based on the HMS--HMS binary grid (Section~\ref{sec:HMS-HMS}), and its subsequent evolution is followed either through the nearest neighbor interpolation (orange lines) or through initial--final interpolation (black dots; using our classification and interpolation methods) for the same initial configuration. 
In both cases we adopt the same SN kick after the two core-collapse episodes in order to compare them as close as possible.

The system does not experiences mass transfer before the first supernova, however the primary experience strong stellar wind mass loss during the Wolf--Rayet phase which widens the orbit .
Shortly after the primary collapses into a $\sim 14.96\,\Msun$ BH with a low $\left(\alpha_{\mathrm{spin}}\right)_1$. 
The subsequent detached evolution (Section~\ref{sec:detached}) of the mildly eccentric system (due to the BH natal kick), after matching the companion of the BH to a single star grid, leads to a small increase of the period predominantly due to winds. Eventually the secondary star fills its Roche lobe at periastron, where we assume that the system circularizes. 
Mass transfer onto the BH is interpolated through the binary grid of COs with H-star companions (Section~\ref{sec:CO-HMS_RLO}) and  lasts for a few thousand years, becoming unstable and leading to a CE episode (Section~\ref{sec:common_envelope}). 
The system survives the process, forming a tight binary comprised of a BH with a stripped He star on a $\sim 0.2$\,day orbit. 
Tidal forces become important in this tight orbit, spinning up the He star (Section~\ref{sec:CO-HeMS}), which eventually also forms a mildly spinning ($a_{\mathrm{spin}, 2} \sim 0.48$) BH of $7.90\,\Msun$. 
The two BHs merge after 183\,Myr from birth, due to gravitational wave radiation. 

We have specifically chosen a binary where the differences between the nearest neighbor and initial--final interpolation schemes are relatively small, so that we can accurately display how the binary evolves through each step. Differences between the two evolution options for binaries in general are significantly larger, and the initial--final interpolation scheme is our default choice.

The binary shown in Figure~\ref{fig:one_system} was evolved using our default configuration, although we have purpose-built \posydon{} to be modular. 
Throughout the previous sections we have described possible changes to physical prescriptions that a user can make. 
However, a user can also easily supplement their own functions for specific steps, or even define the entire binary flow. 

Finally, for debugging we keep track of any errors or warnings raised throughout a binary's evolution. 
This allows us to isolate the problematic step for a binary, or a stellar and binary state--event combination that our flow structure cannot handle. 
This error tracking can be especially useful for user-defined steps and flow structures.

\section{How \posydon{} Evolves a Binary Population}
\label{sec:populations}
Generating a model binary population for comparison to observations 
requires two separate steps: initializing individual binaries and then evolving those binaries, which we describe in Section~\ref{sec:initialization} and Section~\ref{sec:evolution} below. 
Finally, in Section~\ref{sec:example_population} we describe a sample binary population evolved with \posydon{}.

\begin{figure*}[t]\center
\includegraphics[width=\textwidth]{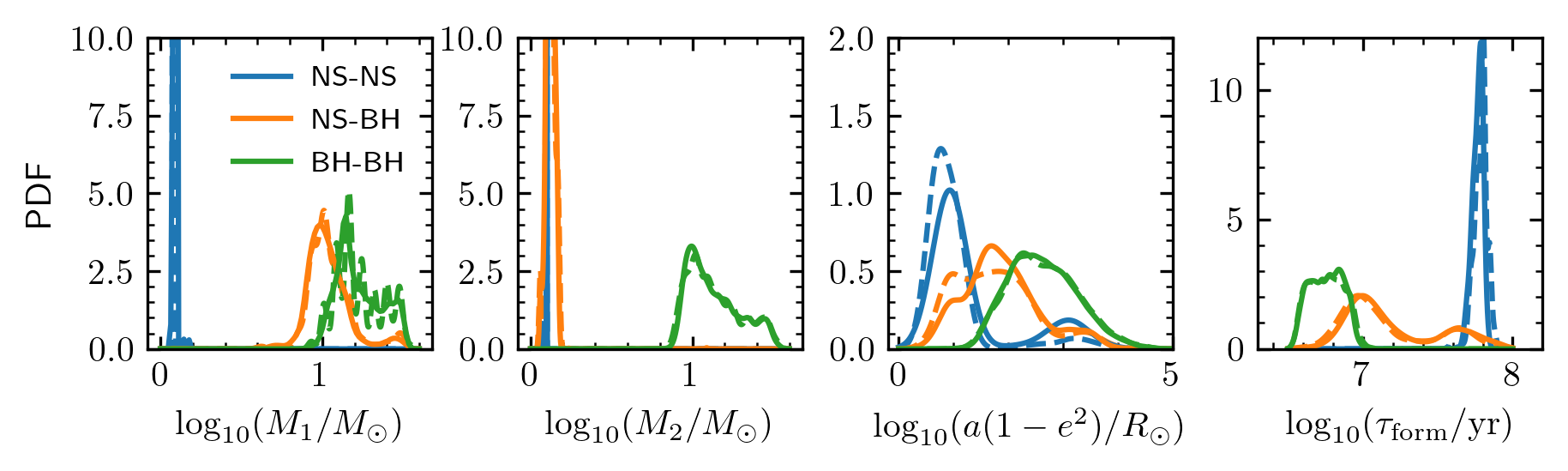}
\caption{
Example of the double compact object populations produced by \posydon{}. Quantities shown are the final distributions, as the binary populations appear today. 
We separately indicate the parameters for NS--NS, BH--NS, and BH--BH systems. \change{}{Results from a population synthesis using linear interpolation is indicated with solid lines, while results using the first nearest neighbor approach with dashed lines.} Comparison with {\tt COSMIC} \citep[Fig.\,3 from][]{2020ApJ...898...71B} shows morphological similarities, but important quantitative differences between the same binary CO populations. }
\label{fig:population}
\end{figure*}

\subsection{Initialization}\label{sec:initialization}
The primary function of \posydon{} is to produce synthetic populations of binaries which requires evolving a random distribution of binaries from ZAMS. 
We generate an initial population by sampling binary parameters from standard distributions. 

{\bf Component Masses:}
For the primary component mass of a binary, we implement initial mass functions (IMF) from \citet{1955ApJ...121..161S,1993MNRAS.262..545K,2001MNRAS.322..231K}. By default we use the \citet{2001MNRAS.322..231K} IMF in the range $M_1 \in [7,150] \; M_\odot$ with $\alpha=2.3$. 
Then the secondary component mass is given by drawing the mass ratio $q = M_2/M_1$ from a flat distribution $[0,1]$ with $M_2 \in [0.35,150] \; M_\odot$.

{\bf Orbital Parameters:}
For a binary population's initial orbital period, \citet{2013A&A...550A.107S} describe a power-law in log-space, while Opik's Law describes a log-flat distribution in orbital separation rather than orbital period space \citep{1983ARA&A..21..343A}. 

In \posydon{} we allow for both models to be adopted, with our default being \citet{2013A&A...550A.107S}.
The minimum orbital period is set by systems that undergo RLO at ZAMS, and a default maximum orbital period to be $6000\,\mathrm{days}$. 
The orbital period distribution from \citet{2013A&A...550A.107S} is undefined for $P_{\rm orb} < 1$\,day; therefore, we extend this distribution so that it is uniform in $\log_{10} P_{\rm orb}$ space down to a $P_{\rm orb}$ of 0.35\,days. 
We further set the maximum $P_{\rm orb}$ to $10^{3.5}\,\mathrm{days}$.
Although binaries may form at wider separations, we choose this upper limit to account for all interacting binary models (e.g., Figure~\ref{fig:HMS-HMS_MESA_grid_TF12}).

Since we only model circular binaries in our detailed binary-star models with the \mesa{} code, the binary eccentricity $e$ at ZAMS is set to zero (see Section~\ref{sec:binary_physics} and Section~\ref{sec:detached}). 
Since we plan to generalize this assumption in future work, we also add the option to generate $e$ from a thermal distribution \citep{1991A&A...248..485D}.

{\bf Star-formation History:}
To account for the star-formation history (SFH) of a binary population, we assign to each binary in the population  the same maximum age. 
Then different star-formation rates (SFRs) can be modeled by modifying the distribution of birth times.
In \posydon{} we offer two options for the SFR: a burst of star formation or a constant SFR. 
For the burst model, all binaries have identical birth times some number of years prior to the end of the simulation.
For a constant SFR, we randomly generate birth times from a uniform distribution within a user-defined range. 
The average metallicity of the Galaxy and the greater Universe evolve over time, but this is something that we cannot currently model accurately, as our grids of models presented in v1.0 of \posydon{} are only calculated for stars at solar metallicity.

\subsection{Evolution}
\label{sec:evolution}

Evolving a population of ZAMS binaries initialized using the distributions described in Section~\ref{sec:initialization} requires implementing the procedure outlined in Section~\ref{sec:flow} for each binary.
In \posydon{} we have created an overarching {\tt BinaryPopulation} class which is a container for a list of individual {\tt BinaryStar} instances. Each {\tt BinaryStar} instance is then iteratively evolved until the entire population has been processed.

The {\tt BinaryPopulation} class contains a number of additional capabilities to efficiently and easily evolve populations of binaries. 
First, evolved populations are automatically saved in an efficient {\tt hdf5} file format with two datasets: one that contains each binary as a single line providing both its initial and final states, and a second dataset that contains the entire evolutionary history of each binary. 
Second, populations can be evolved either serially or in parallel, so that large ($>10^6$ binaries) populations can be run quickly on a high-performance computing cluster. 
Third, we have implemented routines that catch and keep warnings and errors from each binary, so that the code does not crash when a single binary fails to complete its evolution due to a bug. 
We have found this to be  useful for identifying and resolving coding bugs when implementing new physical prescriptions. 
Last, we have implemented various routines that allow a user to easily select only certain types of binaries (e.g., only BBHs or only double COs).

We envision that a typical user will interact primarily with the {\tt BinaryPopulation} class and its associated {\tt SimulationProperties} class, which when combined provide the interface for customizing a particular user's binary population synthesis needs.

\subsection{Example Population}
\label{sec:example_population}

To demonstrate the results of a binary population synthesis run with \posydon{}, we construct a basic population of $10^6$ binaries generated with the default initial conditions described in Section~\ref{sec:initialization}. We choose a constant star formation history over the past 10\,Gyr. The simulation on our high performance computing cluster, {\tt Trident}, takes approximately 2 hours of wall time using 5 nodes, each with 20 cores, i.e., less than 1\,s of CPU time per binary. 

From the resulting binaries, we select those that evolve into bound NS--NS, NS--BH, or BH--BH systems. 
The present-day properties of these binaries are provided in Figure~\ref{fig:population}; this can be compared with the results of other studies, e.g., Figure 3 from \citet{2020ApJ...898...71B} with similar initial conditions at solar metallicity. Overall we find good agreement reaffirming that the code is producing reasonable results. 
In the left two panels, we show distributions of the component compact object masses. Our NS masses are in very close agreement with those of \citet{2020ApJ...898...71B}, although the BHs we produce extend to larger masses. The reason is that current stellar models with stellar structure and evolution parameters calibrated to the latest observations (e.g., overshoorting, etc) produce cores more massive compared to those in the late-nineties models used in rapid BPS codes. The third panel from the left shows the semi-latus rectum of the population. Again, the distribution results are qualitatively similar, with a preference for NS--NS systems at smaller $a(1-e^2)$, and NN--BH and BH--BH systems having larger $a(1-e^2)$. Finally, the rightmost panel of Figure~\ref{fig:population} shows an increasing formation timescale (from ZAMS to the second SN) as we move from BH--BH to NS--BH to NS--NS systems. This is expected since BHs tend to form from more massive systems that complete their evolution more quickly. 
Our binary CO populations appear morphologically similar to those produced by {\tt COSMIC} and described in \citet{2020ApJ...898...71B}. 
We will undertake detailed descriptions of the specific binary populations of interest to separate, science studies.

\section{Summary and Future work}
\label{sec:future}

Here we present \posydon{}, a new, next-generation
computational tool for general population synthesis of single and binary stars. \posydon{} incorporates full stellar structure and evolution sequences for interacting binaries, using the \mesa{} code. Compared to other existing, binary population synthesis code there are significant advances: (i) binary evolution is treated self-consistently without analytical fits of single-star evolutionary tracks and the need for simplified or artificial recipes to emulate the behavior of stars in interacting binaries; (ii) initial-final classification and interpolation methods trained on the pre-calculated grids of binary evolution models, allowing general synthetic simulations of binary populations. The code base along with the existing evolutionary-track grids are publicly available through the \posydon{} collaboration's web portal (\href{https://posydon.org}{https://posydon.org}) along with full documentation and tutorials for how to use the code. An advanced query system is also available for users to be able to mine the grids of single- and binary-star evolutionary tracks and download relevant data using pre-programmed and customized queries \citep[e.g.,][]{10.1145/3469830.3470916,10.1145/3474717.3483989}. Finally, we provide a user-friendly web-application that allows a user to perform small-scale simulations with \posydon{} online, without the need of code installation and configuration.

Compared to current rapid population synthesis codes \posydon{} has a smaller set of free parameters, for many of which there are already multiple options for the user to choose from. The code structure is modular and an advanced user is able to implement their own choices of evolutionary parameters; from as simple as changing the initial properties of the binaries to as complex as incorporating their own custom-made evolutionary-track grids. In this first instrument \posydon{} paper we describe in detail the first version of the code, but technical and astrophysical advancements are ongoing and improved code versions will be released in the near and long-term future.

Our focus on the technical front is on classification and interpolation methods. Our current process of  first classifying the grids and then performing interpolation can lead to errors propagating throughout the pipeline. Instead these two could be combined into a joint treatment to reduce errors, making additional use of covariances between the different grid types \citep{errorProp}. We will also explore adopting kernelized-interpolation approaches \citep[i.e.,][]{interpol1,interpol2,interpol3}. To illustrate the use of such an approach, it is observed in Figure~\ref{fig:classif_confusion} that the confusion matrix for our grid of He-rich stars with compact object companions has a lower accuracy for unstable mass transfer. This is perhaps due to the non-linearly separable decision boundary, as visible in Figure~\ref{fig:classif_dec_boundary}. Adopting a kernelized-interpolation technique has the potential to increase the class-specific interpolation accuracy. In particular, kernel-based approaches such as support vector machines and Gaussian processes may prove fruitful. In principle, we may also use a neural networks.
Also, improvements can be made by adopting non-Euclidean metrics when defining our distance functions in our $k$-nearest neighbors classifiers in Section~\ref{sec:classification}. 

Apart from methods exploration for the existing classification and interpolation processes, we will focus on the next step of interpolations critical for astrophysical studies: interpolation \emph{between} whole evolutionary tracks along time. This is a challenging problem that is of particular interest for any study that requires tracking binary properties as a function of their age (e.g., X-ray binary luminosities). In parallel, we are already working on increasing the computational efficiency of building the pre-computed grids necessary for future \posydon{} versions. Specifically, for any future grid development, we will take advantage of a new active-learning method developed by our team \citep{2022arXiv220316683A} that allows us to achieve the same classification and interpolation accuracies with a significantly smaller \mesa{} tracks, by dynamically placing them at class boundaries in the parameter space. Such dynamic placement, informed by active learning, leads to grids with non-regular, but smart  placement of evolutionary tracks. This work becomes more critical as we expand to more metallicities and add eccentricity as another dimension in order to keep the \posydon{} package sizes appropriate for downloads. 

The first version of \posydon{} is fully functional as an astrophysical tool in the sense that it can be used for complete simulations of binary populations from ZAMS to either formation of a binary with two compact objects or binary distractions (stellar merger or binary disruption), but it is still limited in two ways: our current pre-computed grids are for primaries massive enough to likely form a NS or a BH and are calculated at solar metallicity only. Our next version will expand to a grid of metallicities appropriate for populations across the Universe instead of just the Milky Way. Challenges related to the convergence of single- and binary-star evolutionary models often depend on metallicity, e.g. because the different opacities will results in massive stars reaching the Eddinghton limit within their interiors at different mass ranges and evolutionary phases. Typically, however, such challenges become less severe with decreasing metallicity. Interpolation across metallicities will be a follow-up step as well, once grids at a sufficient number of metallicities have been computed. All the interpolation and classification methods we have developed scale naturally to additional dimensions in the initial conditions parameter space. Further we will expand our grids to low-mass primaries so populations with WDs can also be modeled. Expansion of the grids will inevitably increase data set sizes proportionally. The development of interpolation methods for not only the final properties of each grid but also the entire evolutionary tracks, which is one of our primary future objectives, will allow future \posydon{} releases to come only with with the pre-trained interpolation objects. The latter are expected to have a significantly smaller data footprint compared to the downsampled grids we currently ship with \posydon{}. 

We will also continue to improve the physics treatment of binary evolution. Specifically we are already working on three improvements: (i) in this first version our treatment of the CE phase, once dynamical instability is recognized (taking into account the full stellar structure of the RLO star), is similar to what is done in pBPS codes, apart from the self-consistent calculation of the CE's binding energy. However, since we model binaries with \mesa{}, we are able to treat the phase in a more physical way, either by following the CE inspiral self-consistently using one-dimensional hydrodynamic simulations \citep[e.g.,][]{2019ApJ...883L..45F} or by following the long-term response of the RLO star to losing its envelope on a very high rate and have its Roche lobe shrinking rapidly \citep[e.g.,][]{2021A&A...650A.107M, 2021ApJ...922..110G}. One of the objectives of the next version of \posydon{} will be to improve the physics of our CE treatment. (ii) like all binary population modeling to date, we assume that binaries circularize instantly upon RLO and for this reason we also assume that all ZAMS binaries are circular. However, the physics of secular binary evolution through mass transfer in eccentric orbits has been fully developed recently \citep{2007ApJ...667.1170S,2009ApJ...702.1387S, 2010ApJ...724..546S, 2016ApJ...825...70D, 2016ApJ...825...71D, 2021Hamers} and will be implemented in future \posydon{} versions. (iii) Most recently more physical models for magnetic braking have been developed and calibrated against single-star rotational-velocity data \citep[e.g.,][]{2019ApJ...886L..31V,2021ApJ...922..174V, 2021ApJ...912...65G} and we will use them to update the options for magnetic braking evolution in binaries.

\FloatBarrier   
\appendix
\newcommand{\downsymbol}{\textasteriskcentered}
\newcommand{\notdown}{\tablenotemark{\downsymbol}}

\startlongtable
\begin{center}
\begin{deluxetable}{l p{11cm} l}
\tablecaption{
    The variables in history tables in our grids, taken from the \mesa{} output. 
    The \texttt{star\_age} is not included since we provide the \texttt{age} variable in the binary history (cf.\ \autoref{table:binary_properties}) which refers to both the systems, and its components. 
    Since these quantities are undefined for compact objects, history tables are not provided for NSs and BHs; however, the evolution of the mass is given in the binary history tables.
    \label{table:MESA_singlestar_attributes}}
\tablehead{\colhead{Name}  & \colhead{Description} & \colhead{Unit}}
\startdata
he\_core\_mass                       & Helium core mass. & \Msun \\
c\_core\_mass                        & Carbon core mass. & \Msun \\
o\_core\_mass                        & Oxygen core mass. & \Msun \\
he\_core\_radius\notdown             & Helium core radius. & \Rsun \\
c\_core\_radius\notdown              & Carbon core radius. & \Rsun \\
o\_core\_radius\notdown              & Oxygen core radius. & \Rsun \\
center\_h1                           & Center \isotope[1]{H} mass fraction. & \\
center\_he4                          & Center \isotope[4]{He} mass fraction. & \\
center\_c12\notdown                  & Center \isotope[12]{C} mass fraction. & \\
center\_n14\notdown                  & Center \isotope[14]{N} mass fraction. & \\
center\_o16\notdown                  & Center \isotope[16]{O} mass fraction. & \\
surface\_h1\notdown                  & Surface \isotope[1]{H} mass fraction. & \\
surface\_he4\notdown                 & Surface \isotope[4]{He} mass fraction. & \\
surface\_c12\notdown                 & Surface \isotope[12]{C} mass fraction. & \\
surface\_n14\notdown                 & Surface \isotope[14]{N} mass fraction. & \\
surface\_o16\notdown                 & Surface \isotope[16]{O} mass fraction. & \\
c12\_c12\notdown                     & Decimal logarithm of the burning power from the ${}^{12}\mathrm{C} + {}^{12}\mathrm{C}$ reaction & $[\Lsun]$ \\
center\_gamma\notdown                & Plasma coupling parameter, ratio of the Coulomb to thermal energy. &\\
avg\_c\_in\_c\_core                  & Average \isotope[12]{C} abundance at carbon core. &\\
surf\_avg\_omega\notdown             & Average surface angular velocity. & $\rm yr^{-1}$\\
surf\_avg\_omega\_div\_omega\_crit\notdown  & Ratio of the average and critical surface angular velocity.  & \\
log\_LH\notdown                      & Decimal logarithm of the hydrogen burning power. &$[\Lsun]$ \\
log\_LHe\notdown                     & Decimal logarithm of the helium burning power. & $[\Lsun]$ \\
log\_LZ\notdown                      & Decimal logarithm of the total burning power excluding LH and LHe and photodisintegration. & $[\Lsun]$ \\
log\_Lnuc\notdown                    & Decimal logarithm of the total nuclear burning power. & $[\Lsun]$ \\
log\_Teff                            & Decimal logarithm of the effective temperature. & $\rm [K]$ \\
log\_L                               & Decimal logarithm of the luminosity. & $[\Lsun]$ \\
log\_R                               & Decimal logarithm of the radius. & $[\Rsun]$ \\
total\_moment\_of\_inertia           & Total momentum of inertia. & $\rm {g}\,{cm}^2$\\ 
spin\_parameter\notdown              & Dimensionless stellar spin parameter. & \\
log\_total\_angular\_momentum        & Decimal logarithm of the total angular momentum. & $\rm [g\,cm^2\,s^{-1}]$ \\
conv\_env\_top\_mass\notdown         & Mass coordinate of the top boundary of the outermost convective region. & \Msun \\
conv\_env\_bot\_mass\notdown         & Mass coordinate of the bottom boundary of the outermost convective region. & \Msun \\
conv\_env\_top\_radius\notdown       & Radial coordinate of the top boundary of the outermost convective region. & \Rsun \\
conv\_env\_bot\_radius\notdown       & Radial coordinate of the bottom boundary of the outermost convective region. & \Rsun \\
conv\_env\_turnover\_time\_g\notdown        & Global convective turnover time. & $\rm yr$ \\
conv\_env\_turnover\_time\_l\_b\notdown     & Local convective turnover time half of a scale height above the outermost convective zone bottom boundary. & $\rm yr$ \\
conv\_env\_turnover\_time\_l\_t\notdown     & Local turnover time one scale height above the outermost convective zone bottom boundary. & $\rm yr$ \\
envelope\_binding\_energy\notdown    & Binding energy of the envelope. & $\rm erg$ \\
mass\_conv\_reg\_fortides\notdown    & Mass of the most important convective region for equilibrium tides, as defined in Eq.~\eqref{eq:conv_timescale_tides}. & $\Msun$ \\
thickness\_conv\_reg\_fortides\notdown     &  Thickness of the most important convective region for equilibrium tides, as defined in Eq.~\eqref{eq:conv_timescale_tides}. & $\Rsun$ \\
radius\_conv\_reg\_fortides\notdown & Radial coordinate of the most important convective region for equilibrium tides, as defined in Eq.~\eqref{eq:conv_timescale_tides}. & $\Rsun$ \\
lambda\_CE\_1cent\notdown           & Common-envelope parameter of the envelope binding energy for core-envelope boundary where hydrogen mass fraction becomes lower than 1\%. & \\
lambda\_CE\_10cent\notdown          &  Common-envelope parameter of the envelope binding energy for core-envelope boundary where hydrogen mass fraction becomes lower than 10\%.  & \\
lambda\_CE\_30cent\notdown          &  Common-envelope parameter of the envelope binding energy for core-envelope boundary where hydrogen mass fraction becomes lower than 30\%.  & \\
co\_core\_mass                      & Carbon--oxygen core mass. & \Msun\\
co\_core\_radius\notdown            &  Carbon--oxygen core radius. & \Rsun\\
lambda\_CE\_pure\_He\_star\_10cent\notdown  & Common-envelope parameter of the He-rich envelope binding energy for core-envelope boundary where the sum of hydrogen and helium mass fraction becomes lower than 10\%. &\\
log\_L\_div\_Ledd\notdown           & Decimal logarithm of the ratio of the luminosity and Eddington luminosity.  & \\
\enddata
\tablenotetext{$\downsymbol$}{Property not accounted for when downsampling the grids (Section~\ref{sec:downsampling}).}
\end{deluxetable}
\end{center}
\startlongtable
\begin{center}
\begin{deluxetable*}{l p{11.0cm} l}
\tablecaption{
    The variables in binary history tables, taken from the \mesa{} output.
    \label{table:binary_properties}}
\tablehead{\colhead{Name} & \colhead{Description} & \colhead{Unit}}
\startdata
model\_number\notdown                   & The model number of the final state & \\
age\notdown                             & Binary age & $\rm yr$\\
star\_1\_mass                           & Mass of the first star & $\Msun$ \\
star\_2\_mass                           & Mass of the second star & $\Msun$ \\
period\_days                            & Orbital period in days & $\rm d$ \\
binary\_separation                      & Binary separation & $\Rsun$ \\
lg\_system\_mdot\_1                     & Decimal logarithm of rate of mass loss from the system from around the first star due to inefficient mass transfer & $[\Msun\,\rm yr^{-1}]$ \\
lg\_system\_mdot\_2                     & Decimal logarithm of rate of mass loss from the system from around the second star due to inefficient mass transfer & $[\Msun\,\rm yr^{-1}]$ \\
lg\_wind\_mdot\_1\notdown               & Decimal logarithm of rate of mass loss of the first star due to wind & $\rm [\Msun\, yr^{-1}]$ \\
lg\_wind\_mdot\_2\notdown               & Decimal logarithm of rate of mass loss of the second star due to wind & $\rm [\Msun\, yr^{-1}]$ \\
lg\_mstar\_dot\_1\notdown               & Decimal logarithm of rate of mass loss of the first star & $\rm [\Msun\,yr^{-1}]$ \\
lg\_mstar\_dot\_2\notdown               & Decimal logarithm of rate of mass loss of the second star & $\rm [\Msun\,yr^{-1}]$ \\
lg\_mtransfer\_rate                     & Decimal logarithm of mass-transfer rate & $\rm [\Msun\,yr^{-1}]$ \\
xfer\_fraction\notdown                  & Mass-transfer fraction & \\
rl\_relative\_overflow\_1\notdown       & Roche lobe overflow of the first star in units of donor Roche lobe radii & \\
rl\_relative\_overflow\_2\notdown       & Roche lobe overflow of the second star in units of donor Roche lobe radii & \\
trap\_radius\notdown                    & Trapping radius & $\Rsun$ \\
acc\_radius\notdown                     & Radius of the compact object & $\rm cm$ \\
t\_sync\_rad\_1\notdown                 & Tidal synchronization time-scale of the first star for stars with radiative envelopes & $\rm s$ \\
t\_sync\_conv\_1\notdown                & Tidal synchronization time-scale of the first star for stars with convective envelopes & $\rm s$ \\
t\_sync\_rad\_2\notdown                 & Tidal synchronization time-scale of the second star for stars with radiative envelopes & $\rm s$ \\
t\_sync\_conv\_2\notdown                & Tidal synchronization time-scale of the second star for stars with convective envelopes & $\rm s$ \\
\enddata
\tablenotetext{$\downsymbol$}{Property not accounted for when downsampling the grids (Section~\ref{sec:downsampling}).}
\end{deluxetable*}
\end{center}
\startlongtable
\begin{center}
\begin{deluxetable*}{l p{11cm} l}
\tablecaption{
    Quantities of final profiles of the stars, taken from the \mesa{} output. 
    \label{table:profile_columns}
    }
\tablehead{\colhead{Name} & \colhead{Description} & \colhead{Unit}}
\startdata
radius                              &  Radius at the outer boundary of the zone. & $[\Rsun]$ \\
mass\notdown                        &  Mass coordinate of the outer boundary of the zone. & $[\Msun]$\\
logRho                              &  Decimal logarithm of the density at the center of the zone. & $\rm[g\,cm^{-3}]$\\
omega                               &  Angular velocity. & $\rm [rad\,s^{-1}]$\\
energy\notdown                      &  Specific internal energy. & $\rm [erg\,g^{-1}]$\\
x\_mass\_fraction\_H\notdown        &  Mass fraction of all isotopes with atomic number 1.
& \\
y\_mass\_fraction\_He\notdown       &  Mass fraction of all isotopes with atomic number 2.
& \\
z\_mass\_fraction\_metals\notdown   &  Mass fraction of all elements except for those in x\_mass\_fraction\_H and y\_mass\_fraction\_He  & \\
neutral\_fraction\_H\notdown        &  Fraction of  neutral hydrogen (HI) of all the \isotope[1]{H}. & \\
neutral\_fraction\_He\notdown       &  Fraction of  neutral helium (HeI) of all the \isotope[4]{He}. & \\
avg\_charge\_He\notdown             &  Average charge of all  the \isotope[4]{He} isotopes. & electron charge [$e$]\\
\enddata
\tablenotetext{$\downsymbol$}{Property not accounted for when downsampling the grids (Section~\ref{sec:downsampling}).}
\end{deluxetable*}
\end{center}
\startlongtable
\begin{center}
\begin{deluxetable*}{l p{10cm} l}
\tablecaption{
    Post-processed variables referring to the final state of each binary in our grids. 
    Quantities referring to single-star quantities (i.e., all except termination flags and interpolation class) are prefixed with {\tt S1\_} and {\tt S2\_} to distinguish the corresponding star in the grid (e.g., \texttt{S1\_surface\_other} or \texttt{S2\_direct\_mass}). 
    These variables, along with the last values of the single and binary history variables (cf.\ Table~\ref{table:MESA_singlestar_attributes} and Table~\ref{table:binary_properties}; e.g., 
    \texttt{S1\_log\_L}), comprise the final values tables stored in the grids.
    \label{table:post_processed}
    }
\tablehead{\colhead{Name} & \colhead{Description} & \colhead{Unit}}
\startdata
termination\_flag\_1                    & Termination reason from \mesa output, or reach cluster timelimit.  & \\
termination\_flag\_2                    & RLO state (indicating which star is the donor), or \texttt{contact\_during\_MS} in case of stellar merger. &\\
termination\_flag\_3                    & State of primary star. &\\
termination\_flag\_4                    & State of secondary star. &\\
interpolation\_class                    & Classification based on termination flags 1 and 2, indicating broad groups based on mass transfer. & \\
surface\_other          &  Surface abundance fraction of elements excluding \isotope[1]{H}, \isotope[4]{He}, \isotope[12]{C}, \isotope[14]{N}, \isotope[16]{O}.  & \\
center\_other           &  Central abundance fraction of elements excluding \isotope[1]{H}, \isotope[4]{He}, \isotope[12]{C}, \isotope[14]{N}, \isotope[16]{O}.  & \\
direct\_state                       & CO state for the direct collapse prescription of the star.&\\
direct\_SN\_type                    & SN type for the direct collapse prescription of the star.&\\
direct\_f\_fb                       & Fallback mass fraction for the direct collapse prescription of the star.&\\
direct\_mass                        & CO mass for the direct collapse prescription of the star.& \Msun \\
direct\_spin                        & CO spin for the direct collapse prescription of the star.&\\
Fryer+12-rapid\_state               & CO state for the \citet{2012ApJ...749...91F} \textit{rapid} prescription of the star.&\\
Fryer+12-rapid\_SN\_type            & SN type for the \citet{2012ApJ...749...91F} \textit{rapid} prescription of the star.&\\
Fryer+12-rapid\_f\_fb               & Fallback mass fraction for the \citet{2012ApJ...749...91F} \textit{rapid} prescription of the star.&\\
Fryer+12-rapid\_mass                & CO mass for the \citet{2012ApJ...749...91F} \textit{rapid} prescription of the star.& \Msun\\
Fryer+12-rapid\_spin                & CO spin for the \citet{2012ApJ...749...91F} \textit{rapid} prescription of the star.&\\
Fryer+12-delayed\_state             & CO state for the \citet{2012ApJ...749...91F} \textit{delayed} prescription of the star.&\\
Fryer+12-delayed\_SN\_type          & SN type for the \citet{2012ApJ...749...91F} \textit{delayed} prescription of the star.&\\
Fryer+12-delayed\_f\_fb             & Fallback mass fraction for the \citet{2012ApJ...749...91F} \textit{delayed} prescription of the star. &\\
Fryer+12-delayed\_mass              & CO mass for the \citet{2012ApJ...749...91F} \textit{delayed} prescription of the star.& \Msun\\
Fryer+12-delayed\_spin              & CO spin for the \citet{2012ApJ...749...91F} \textit{delayed} prescription of the star.&\\
Sukhbold+16-engineN20\_state        & CO state for the \citet{2016ApJ...821...38S} \textit{N20 engine} prescription of the star.&\\
Sukhbold+16-engineN20\_SN\_type     & SN type for the \citet{2016ApJ...821...38S} \textit{N20 engine} prescription of the star.&\\
Sukhbold+16-engineN20\_f\_fb        & Fallback mass fraction for the \citet{2016ApJ...821...38S} \textit{N20 engine} prescription of the star. &\\
Sukhbold+16-engineN20\_mass         & CO mass for the \citet{2016ApJ...821...38S} \textit{N20 engine} prescription of the star.& \Msun\\
Sukhbold+16-engineN20\_spin         & CO spin for the \citet{2016ApJ...821...38S} \textit{N20 engine} prescription of the star.&\\
Patton\&Sukhbold20-engineN20\_state     & CO state for the \citet{2020MNRAS.499.2803P} \textit{N20 engine} prescription of the star.&\\
Patton\&Sukhbold20-engineN20\_SN\_type  & SN type for the \citet{2020MNRAS.499.2803P} \textit{N20 engine} prescription of the star.&\\
Patton\&Sukhbold20-engineN20\_f\_fb     & Fallback mass fraction for the \citet{2020MNRAS.499.2803P} \textit{N20 engine} prescription of the star. &\\
Patton\&Sukhbold20-engineN20\_mass      & CO mass for the \citet{2020MNRAS.499.2803P} \textit{N20 engine} prescription of the star.& \Msun\\
Patton\&Sukhbold20-engineN20\_spin      & CO spin for the \citet{2020MNRAS.499.2803P} \textit{N20 engine} prescription of the star.&\\
avg\_c\_in\_c\_core\_at\_He\_depletion  & Average carbon 12 abundance at carbon core at the state of He depletion of the star.&\\
co\_core\_mass\_at\_He\_depletion       & Carbon--oxygen core mass at the state of He depletion of the star.&\\
m\_core\_CE\_1cent                   & Mass of the hydrogen-deficient (i.e. helium) core, with the  core-envelope boundary defined as the outermost layer where the hydrogen mass fraction drops below 1\%. & \Msun\\
m\_core\_CE\_10cent                  & As m\_core\_CE\_1cent, but for hydrogen mass fraction of 10\%. & \Msun\\
m\_core\_CE\_10cent                  & As m\_core\_CE\_1cent, but for hydrogen mass fraction of 30\%. & \Msun\\
m\_core\_CE\_pure\_He\_star\_10cent  & Mass of the hydrogen- and helium-deficient  (i.e., carbon--oxygen) core, with the  core--envelope boundary defined as the outermost layer where the sum of hydrogen and helium mass fraction drops below 10\%. & \Msun\\ 
r\_core\_CE\_1cent                   & Radial coordinate of the core as defined in m\_core\_CE\_1cent. & \Rsun\\
r\_core\_CE\_10cent                  & Radial coordinate of the core as defined in m\_core\_CE\_10cent. & \Rsun\\
r\_core\_CE\_30cent                  & Radial coordinate of the core as defined in m\_core\_CE\_30cent. & \Rsun \\
r\_core\_CE\_pure\_He\_star\_10cent  & Radial coordinate of the core as defined in m\_core\_CE\_pure\_He\_star\_10cent.  & \Rsun\\
\enddata
\end{deluxetable*}
\end{center}
\FloatBarrier

\acknowledgments

We thank Corinne Charbonnel, Alex de Koter, Ilya Mandel, Pablo Marchant, Georges Meynet, Fred Rasio, Yorick Vink, Andreas Zezas, for valuable discussion on several aspects stellar- and binary-evolution physics; Aldo Batta, Monica Gallegos-Garcia, Samuel Imperato, Chase Kimball, and Maxime Rambosson for contributing to the code base of the project, and  Margaret Lazzarini and Mathieu Renzo for testing early development versions of the code and providing feedback.

The \posydon{} project is supported primarily by two sources: a Swiss National Science Foundation Professorship grant (PI Fragos, project number PP00P2 176868) and the Gordon and Betty Moore Foundation (PI Kalogera, grant award GBMF8477). 

The collaboration was also supported by the European Union's Horizon 2020 research and innovation program under the Marie Sklodowska-Curie RISE action, grant agreements No 691164 (ASTROSTAT) and No 873089 (ASTROSTAT-II). Individual team members were supported by additional sources:  
JJA acknowledges funding from Northwestern University through a CIERA Postdoctoral Fellowship, CPLB acknowledges support by the CIERA Board of Visitors Research Professorship, and SC through CIERA as a Computational Specialist. VK was partially supported through a CIFAR Senior Fellowship and a Guggenheim Fellowship. 
KK and EZ were partially supported by the Federal Commission for Scholarships for Foreign Students for the Swiss Government Excellence Scholarship (ESKAS No.\ 2021.0277 and ESKAS No.\ 2019.0091, respectively).
YQ acknowledges funding from the Swiss National Science Foundation (grant P2GEP2{\_}188242). DM and KR thank the LSSTC Data Science Fellowship Program, which is funded by LSSTCorporation, NSF Cybertraining Grant No. 1829740, the Brinson Foundation, and the Gordon and Betty Moore Foundation; their participation in the program has benefited this work.
 ZX was supported by the Chinese Scholarship Council (CSC).
MZ was supported as an IDEAS Fellow, through the NRT IDEAS program, a research traineeship program supported by the National Science Foundation (PI Kalogera, award DGE-1450006). 

The computations were performed at Northwestern University on the Trident computer cluster (funded by the GBMF8477 award) and at the University of Geneva on the Baobab and Yggdrasil computer clusters. This research was supported in part through the computational resources and staff contributions provided for the Quest high performance computing facility at Northwestern University which is jointly supported by the Office of the Provost, the Office for Research, and Northwestern University Information Technology.

\software{This manuscript has made use of the following Python modules: 
\texttt{numpy} \citep{2020NumPy-Array},
\texttt{scipy} \citep{2020SciPy-NMeth},
\texttt{pandas} \citep{mckinney2010data},
\texttt{matplotlib} \citep{hunter2007matplotlib},
\texttt{astropy} \citep{2013A&A...558A..33A,2018AJ....156..123A},
\texttt{scikit-learn} \citep{scikit-learn}.
}

\newpage

\bibliographystyle{aasjournal}
\bibliography{references}

\end{document}